\documentclass[11pt,a4paper,showkeys,showpacs]{article}
\usepackage{alpha}
\usepackage{defs}

\begin{document}

\title{
\begin{flushright}
\vspace{-2cm}
\small{
IFIC/22-25\\
WUB/22-00\\
DESY-22-051\\
CERN-TH-2022-015\\
HU-EP-22/04}
\vskip 0.7cm
\end{flushright}
Determination of $\alpha_s(m_Z)$ by the non-perturbative decoupling method\\[2ex]
\href{https://www-zeuthen.desy.de/alpha/}{\includegraphics[width=2.8cm]{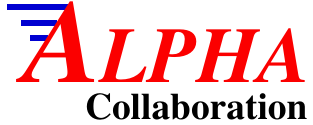}}}
\date{\today}

\author{\vspace{-1cm}Mattia~Dalla~Brida}
\address{\vspace{-4mm}Theoretical Physics Department, CERN, 1211 Geneva 23, Switzerland}
\author{Roman H{\"o}llwieser, Francesco Knechtli, Tomasz Korzec}
\address{\vspace{-5mm}Deptartment of Physics, University of Wuppertal, Gau{\ss}strasse 20, 42119 Germany}
\author{Alessandro~Nada}
\address{\vspace{-5mm}
Department of Physics, University of Turin,
Via Pietro Giuria 1, 10125 Turin, Italy}
\author{Alberto~Ramos} 
\address{\vspace{-5mm}Instituto de F\'isica Corpuscular (IFIC), CSIC-Universitat de Valencia, 46071, Valencia, Spain}
\author{Stefan~Sint}
\address{\vspace{-5mm}School of Mathematics and Hamilton Mathematics Institute, Trinity College Dublin, Dublin 2, Ireland}
\author{Rainer~Sommer}
\address{\vspace{-4mm}Deutsches Elektronen-Synchrotron DESY, Platanenallee~6, 15738~Zeuthen, Germany\\
Institut~f\"ur~Physik, Humboldt-Universit\"at~zu~Berlin, Newtonstr.~15, 12489~Berlin, Germany\vspace{-1.5cm}}

\begin{abstract}
  We present the details and first results of a new strategy for the determination of $\alpha_s(m_Z)$~\cite{DallaBrida:2019mqg}. 
  By simultaneously decoupling 3 fictitious heavy quarks we establish a relation between the  
  $\Lambda$-parameters of three-flavor QCD and pure gauge theory.
  Very precise recent results in the pure gauge theory~\cite{DallaBrida:2019wur,Nada:2020jay} can thus be leveraged 
  to obtain the three-flavour $\Lambda$-parameter in units
  of a common decoupling scale. Connecting this scale to hadronic physics in 3-flavour QCD leads to  
  our result in physical units, $\Lambda^{(3)}_{\overline{\rm MS}} = 336(12)\, \MeV$, which translates to $\alpha_s(m_Z) = 0.11823(84)$. {This is compatible
  with both the FLAG average~\cite{Aoki:2021kgd} and the previous
  ALPHA result~\cite{Bruno:2017gxd}, with a comparable, yet still  
  statistics dominated, error.} This constitutes a highly non-trivial check, as the decoupling strategy is conceptually very different 
  from the 3-flavour QCD step-scaling method, and so are their systematic errors. 
  These include the uncertainties of the combined decoupling and continuum limits, 
  which we discuss in some detail. We also quantify the correlation between both results, 
  due to some common elements, such as the scale determination in
  physical units and the definition of the energy scale where we apply
  decoupling.
\end{abstract}


\maketitle

\tableofcontents

\newpage

\section{Introduction}
\label{sec:intro}
Experiments in high energy physics have established the Standard Model
as a very good effective theory for elementary particle physics up to the TeV scale.
Consequently, the discovery of new physics will require excellent quantitative control 
over Standard Model (SM) predictions including QCD effects~\cite{Atlasresults}.
In particular, for the strong coupling, $\alpha_s(m_Z)$, as one of the 
fundamental Standard Model parameters, a sub percent uncertainty will be required, which is
significantly less than the current error for the PDG (non-lattice) average 
$\alpha_s(m_Z)=0.1176(11)$~\cite{ParticleDataGroup:2020ssz}.

At present, the most accurate results for $\alpha_s$ are obtained from lattice QCD, as
illustrated by the FLAG 2019 average  $\alpha_s(m_Z)=0.1182(8)$~\cite{Aoki:2019cca},
which was recently updated to $0.1184(8)$ for FLAG 2021~\cite{Aoki:2021kgd}. 
While lattice QCD does not require any model assumptions on hadronization, the determination
of $\alpha_s(m_Z)$ requires the solution of a multiscale or ``window'' 
problem~(for an introduction cf.~\cite{DelDebbio:2021ryq,DallaBrida:2020pag}). 
Therefore, most lattice studies attempt to extract the coupling at relatively low energy scales 
where perturbative truncation effects are hard to control. In particular, there is now
some evidence that error estimates obtained \emph{within perturbation theory} 
can be rather misleading unless large energy scales are reached 
non-perturbatively~\cite{Brida:2016flw,DallaBrida:2018rfy}. As a result many
lattice determinations of $\alpha_s$ are now limited by systematic errors.~cf.~\cite{DelDebbio:2021ryq,Aoki:2021kgd}.

A solution to this multiscale problem has been known for 30 years in the form of the 
recursive step-scaling method~\cite{Luscher:1991wu}. The method has since been applied 
to the running of the coupling in QCD with $\Nf=0$~\cite{Luscher:1993gh,DallaBrida:2019wur,Nada:2020jay,Bribian:2021cmg},
$\Nf=2$~\cite{DellaMorte:2004bc,Fritzsch:2012wq}, $\Nf=3$~\cite{Aoki:2009tf,Bruno:2017gxd} 
and $\Nf=4$~\cite{Tekin:2010mm,Perez-Rubio:2010zyf} quark flavours (for a review cf.~\cite{Sommer:2015kza}),
and in various candidate models of physics beyond the Standard Model~(for reviews, cf.~\cite{Nogradi:2016qek,DeGrand:2015zxa}).
Its most recent application in 3-flavour QCD has allowed the ALPHA collaboration to non-perturbatively trace the
scale evolution of the coupling in 3-flavour QCD between $0.2$ and $128$~GeV~\cite{Bruno:2017gxd}.
The corresponding result obtained for $\alpha_s(m_Z) = 0.11852(84)$ in 5-flavour QCD defines a benchmark against which to measure progress.
Knowing the scale dependence of the coupling is a pre-requisite for the step-scaling solution of other renormalization problems.
This is illustrated by a recent step-scaling study of the running quark mass in 3-flavour QCD~\cite{Campos:2018ahf,Plasencia:2021eon}
which will provide essential input for this paper.

Current lattice QCD simulations include the light up, down and strange quarks ($\Nf=2+1$), and
sometimes also the much heavier charm quark as active degrees of freedom ($\Nf=2+1+1$). 
The Applequist-Carazzone decoupling theorem~\cite{Appelquist:1974tg} ensures that the effects of heavy sea quarks on low energy observables
can be absorbed in parameter renormalizations up to effects that are power suppressed in the heavy quark masses \cite{Weinberg:1980wa}. 
In ref.~\cite{Athenodorou:2018wpk} the perturbative treatment of decoupling, 
known to 4-loops in the $\MSbar$ scheme~\cite{Bernreuther:1981sg,Grozin:2011nk,Chetyrkin:2005ia,Schroder:2005hy,Kniehl:2006bg,Gerlach:2018hen},
was shown to provide an excellent quantitative description of decoupling, even at scales as low as the charm quark mass. 
Hence, perturbative  matching of the $\Nf=3$ coupling across the charm and bottom quark thresholds 
yields a reliable estimate of the $\Nf=5$ coupling $\alpha_s(m_Z)$ in terms of the 
3-flavour $\Lambda$-parameter\footnote{One also needs to input values for the charm, 
bottom and $Z$-boson masses, taken e.g.~from~\cite{ParticleDataGroup:2020ssz}; corresponding uncertainties are negligible.}. 
The corresponding error is small and will remain sub-dominant for the foreseeable future.

The high accuracy of perturbative decoupling means that the inclusion of the charm quark 
is not required for a lattice determination of $\alpha_s(m_Z)$. There is much more
to gain from focusing on a reliable and precise determination of  $\Lambda_\msbar^{(3)}$. The currently 
best lattice result by the ALPHA collaboration, $341(12)$~MeV, has an error of $3.5\%$~\cite{Bruno:2017gxd}.
For comparison, the recent high precision study in the pure gauge theory~\cite{DallaBrida:2019wur} quotes $\Lambda_\msbar^{(0)}\sqrt{8t_0}=0.6227(98)$ i.e.~an
error of $1.6\%$.  Given that the error due the physical scale setting is subdominant,
it is thus conceivable that a substantial error reduction for $\Lambda_\msbar^{(3)}$ can still be achieved
by pushing the 3-flavour step-scaling method to higher precision.
While such a project would be very interesting, we emphasize that %
it is just as important to corroborate results with a different method.

The decoupling project, introduced in \cite{DallaBrida:2019mqg} and reviewed in \cite{DallaBrida:2020pag}, 
aims to deliver on both counts. It uses decoupling as a tool to connect $\Nf=3$ QCD to $\Nf=0$ QCD, 
and thus leverage the higher precision that can be achieved with step-scaling methods in 
the $\Nf=0$ theory~\cite{DallaBrida:2019wur,Nada:2020jay}.
This connection is achieved by varying the RGI mass $M$ of a fictitious triplet
of mass-degenerate quarks compared to a hadronic scale
$\mudec\approx 800$~MeV. We call this low energy scale the
decoupling scale.
Reaching values for $M$ of up to O($10$)~GeV and using 
perturbative 4-loop decoupling then establishes a relation
for $\Lambda_\msbar^{(\Nf)}/\mudec$ between both theories, with corrections that are either perturbative in 
the $\MSbar$-coupling at the scale of the heavy quark mass, $M$, or power suppressed in $1/M$.
Obviously, the heavy quark mass defines another scale and thus creates a potentially difficult multi-scale
problem. To alleviate this problem, the choice of $\mudec$ somewhat above $\Lambda_\text{QCD}$ is convenient, since
the matching to a hadronic scale can be safely performed in a separate computation.  
The use of a finite volume renormalization scheme for the coupling at scale $\mudec=1/L$ then reduces the decoupling project
to a problem involving two physical scales, $\mudec$ and $M$, where the challenge remains to reach large $z=M/\mudec \gg 1$ while
keeping the lattice spacing small enough so that $M \ll 1/a$.
In this paper we discuss the details of the decoupling strategy and the numerical simulations we performed.
When combined with earlier scale setting results~\cite{DallaBrida:2016kgh} and precision $\Nf=0$ studies~~\cite{DallaBrida:2019wur,Nada:2020jay}, 
the very accurate result, $\Lambda_\msbar^{(3)}=336(12)$~MeV, is obtained with uncertainties 
still dominated by statistical errors.  Relying on the usual perturbative matching to 5-flavour QCD this translates 
to our result $\alpha_s(m_Z)= 0.11823(84)$.

The paper is organized as follows. In Section 2 we give a step by step overview of the decoupling strategy, 
in a language aimed also at non-lattice experts. Section 3 proposes a closer look at the continuum and decoupling
limits. In particular, the leading corrections are derived in order to guide the analysis of the numerical data.
Section~4 starts with the chosen set-up of non-perturbatively O($a$) improved lattice QCD with Wilson quarks,
continues with a summary of the simulations with massive quarks and
then presents the continuum and heavy mass extrapolations of the data which lead to the $\Lambda$-parameter.
In Section~5 we obtain the corresponding result for $\alpha_s(m_Z)$ and we conclude with an outlook (Section~6).
{A number of appendices have been included. Appendix \ref{app:boundary} explains how we estimated heavy 
mass effects of order $1/M$ originating from the space-time boundaries. Appendix \ref{sec:summary-pure-gauge} summarizes
the required $\Nf=0$ results, obtained either by dedicated $\Nf=0$ simulations or taken from the literature.
Appendix \ref{sec:simulations} contains details about $\Nf=3$ simulations, both with massive and massless quarks. 
The latter are required to ensure O($a$) improvement of the renormalized quark masses. Appendix \ref{app:betashifts} 
discusses the derivation and numerical implementation of formulas (\ref{eq:deriv0}) and (\ref{eq:deriv1}), which allow us 
to perform shifts to the data and correct for small mistunings to the relevant lines of constant physics. 
The ingredients for quark mass renormalization and O($a$) improvement, as well as some consistency 
checks, are then given in Appendix~\ref{app:renormass}. The impact of errors in the quark mass determination is 
discussed in Appendix \ref{app:errz}. Finally, Appendix \ref{sec:lcp4048}
explains how bare parameters were chosen on the larger lattices to ensure constant physical conditions.}
 
\section{The decoupling strategy}
\label{sec:strategy}
The decoupling strategy has been introduced and explained in ref.~\cite{DallaBrida:2019mqg}.
We assume the reader to be familiar with this reference and try to keep the overlap minimal. 
Some aspects of the chosen strategy may not seem obvious at first sight, 
as they are conditioned by previous projects of the ALPHA 
collaboration~\cite{Fritzsch:2018yag,Bruno:2017gxd, DallaBrida:2016kgh, Campos:2018ahf, DallaBrida:2019wur,Nada:2020jay}. 
Besides the lattice action, this concerns the choice of boundary conditions and renormalized couplings.
Further technical issues arise from the necessity of O($a$) improvement and 
the need to control boundary effects both at O($a$) and 
in the heavy mass expansion. We will address these points in the following sections. 
Here we use a continuum language to discuss how the decoupling strategy is set up in principle.

\subsection{Renormalization group invariant parameters}
\label{subsec:RGI}

To set the stage we consider continuum QCD with $\Nf$ mass-degenerate quarks. 
The theory only has two bare parameters, the coupling and the quark mass. 
If these are renormalized in a mass-independent scheme $s$,
their scale dependence gives rise to the definition of the $\beta$-function
\begin{equation}
 \frac{\partial \bar{g}_s(\mu)}{\partial\ln\mu} = \beta_s(\bar{g}_s) \simas{\bar g_s \to 0} - \bar{g}_s^3\sum_{n\geq0} b_n \bar{g}_s^{2n}
\end{equation}
and the quark mass anomalous dimension,
\begin{equation}
   \frac{\partial\ln\bar{m}_s(\mu)}{\partial\ln\mu} = \tau_s(\bar{g}_s)  \simas{\bar g_s \to 0}   -\bar{g}_s^2\sum_{n\geq0} d_n \bar{g}_s^{2n}\,.
\end{equation}
The renormalization scheme dependence begins with $b_2$ and $d_1$, and the universal first coefficients
are, for 3 colours,
\begin{equation}
  b_0 = \left(11-\frac23\Nf\right)\times(4\pi)^{-2},\quad b_1 = \left(102-\frac{38}{3}\Nf\right)\times (4\pi)^{-4} , \quad d_0 = 8 \times(4\pi)^{-2}\,.
\end{equation}
Given a non-perturbative definition of the running parameters in a scheme $s$ and thus non-perturbative results
for the RG functions $\beta_s$ and $\tau_s$, one may define the renormalization group invariant (RGI) parameters, 
\begin{eqnarray}
\label{eq:LRGI1}
  \Lambda_s &=& \mu \varphi_s^{}(\bar g_s (\mu))\,, \\ 
  \label{eq:varphi}
    \varphi_s^{}(\gbar_s) &=& ( b_0 \gbar_s^2 )^{-b_1/(2b_0^2)} 
        \rme^{-1/(2b_0 \gbar_s^2)} \times \exp\left\{-\int\limits_0^{\gbar_s} \rmd x\ 
        \left[\frac{1}{\beta_s(x)} 
             +\frac{1}{b_0x^3} - \frac{b_1}{b_0^2x} \right] \right\} \, , \nonumber\\
   \label{eq:MRGI}
   M &=& \overline{m}_s(\mu) \left[2b_0\bar{g}_s^2(\mu) \right]^{-\frac{d_0}{2b_0}}
        \exp\left\{-\int\limits_0^{\bar{g}_s(\mu)} \left[\frac{\tau_s(x)}{\beta_s(x)}-\frac{d_0}{b_0x} \right]\, dx \right\}\, ,
\end{eqnarray}
where the RGI quark mass $M$ is scheme independent, while $\Lambda_s$ depends on the renormalization scheme $s$, 
the standard reference being the $\MSbar$ scheme of dimensional regularization\footnote{The scheme dependence is however
{\em exactly} computable from the perturbative 1-loop relation between the respective couplings.}.
The running coupling, the quark mass, and the RGI parameters are defined for QCD with a fixed flavour number $\Nf$. 
We occasionally indicate this dependence by a superscript; when omitted we refer to QCD with non zero $\Nf$ 
implying $\Nf=3$ in our numerical work. 
Note that the ratio $\Lambda_s/\mu$ is, for large enough $\mu$, in
one-to-one correspondence with the coupling $\bar{g}_s^2(\mu)$ in
scheme $s$. Hence, the running parameters in scheme $s$ at large $\mu$ 
can be traded for $\Lambda_{\msbar}/\mu$ and the scheme independent RGI quark mass $M$.

\subsection{Decoupling relations}

So far we have assumed that the coupling and quark mass are
renormalized in a mass-independent scheme.  
In practice, this is achieved by imposing the renormalization condition at vanishing 
quark mass~\cite{Weinberg:1973xwm}.
A renormalized coupling at finite quark mass defines a function
$\bar{g}_s^{}(\mu,M)$ such that, for $M\ll \mu$,  it coincides with the coupling in a massless scheme, 
$\bar{g}_s^{}(\mu,0) = \bar{g}_s(\mu)$, whereas, for $M \gg \mu$ its scale dependence is well described
by an effective theory with the $\Nf$ heavy quarks removed. In the absence of other light quarks, 
this simply is the pure gauge theory or ``quenched ($\Nf=0$) QCD''. We can thus write
\begin{equation}
  \bar{g}_s^{}(\mu,M) = \bar{g}_s^{(0)}(\mu) +\rmO(\mu^2 /M^2) \,,
\end{equation}
where on the r.h.s. the scale $\mu$ in units of the pure gauge theory $\Lambda$-parameter has an implicit $M$-dependence. 
The mass-dependent coupling is thus seen to interpolate 
the couplings in QCD with $\Nf$ and zero flavours. This in turn implies a relation
between the respective mass independent couplings.
In perturbation theory and in the $\MSbar$ scheme  the ensuing relations have been 
computed up to 4-loop order~\cite{Bernreuther:1981sg,Grozin:2011nk,Chetyrkin:2005ia,Schroder:2005hy,Kniehl:2006bg,Gerlach:2018hen}:
\begin{equation}
   [\bar{g}^{(0)}_\msbar(\mu)]^2 = C(\bar{g}^{}_{\msbar}(m_\star)) \bar{g}^2_{\msbar}(m_\star) 
\end{equation}
where the scale choice $\mu = m_\star=\overline{m}^{}_{\msbar}(m_\star)$
eliminates the 1-loop coefficient in
\begin{equation}
C(x) = 1 + c_2x^4 + c_3 x^6 + c_4x^8 + \ldots\,,
\end{equation}
and, for $\Nf=3$, one obtains~\cite{Gerlach:2018hen}
\begin{equation}
  c_2 = 2.940776\times 10^{-4},\qquad
  c_3 = 4.435355\times 10^{-5},\qquad
  c_4 = 5.713208\times 10^{-6} \,.
  \label{eq:Cfunc}
\end{equation}
Beyond perturbation theory, the limit of infinite $M/\mu$ requires the extrapolation
of numerical data and one would like to understand how it is approached.
To this end, the language of effective field theory is most helpful, cf.~\cite{Athenodorou:2018wpk}.
Assuming that the heavy quark limit is described by a local effective theory,
one obtains a systematic expansion in inverse powers of the quark mass $M$.
In particular, with all fermions decoupled simultaneously the effective theory
takes the form of the pure gauge theory where the inverse mass corrections
are proportional to insertions of higher dimensional local gluonic operators.
One naturally wonders whether all powers of $1/M$ may appear in this expansion. In the absence of
space-time boundaries, and for gluonic observables defining the running couplings, 
the locality of the effective decoupling theory, Euclidean O(4) symmetry and 
gauge invariance rule out any odd-dimensional terms so that the expansion is effectively in $1/M^2$.

The situation changes if space-time is a manifold with boundaries, as this allows for
additional local boundary terms at order $1/M$ in the effective Lagrangian. 
This is relevant in our case, where we use the standard Schr\"odinger functional
set-up on a space-time hyper-cylinder of 4-volume $L^3\times T$, with Dirichlet boundary conditions
imposed on some of the fermionic and gauge field components at Euclidean times $x_0=0$ and $x_0=T$~\cite{Luscher:1992an,Sint:1993un}.
In order to minimize the impact of such boundary contributions, we
chose a geometry such as to have large distances from the boundary to
the observable defining the coupling in the mass-dependent scheme.  
Using  the decoupling effective field theory, with a single local term at the boundaries $x_0=0$ and $x_0=T$, 
we are able to compute the $1/M$ contributions to the coupling. They are given in terms of a 1-loop matching 
coefficient (see \sect{app:omegab}) and a
non-perturbative matrix element. The latter is evaluated 
by simulations in the decoupled theory with $\nf=0$ and
extrapolated to the continuum
(see \sect{app:boundarycontribution}). We can thus confirm that the
boundary effects are negligible for $T=2L$.

\subsection{Master formula and strategy breakdown}

Following Ref.~\cite{DallaBrida:2019mqg}, the decoupling strategy can be cast in the form
\begin{eqnarray}
  \frac{\Lam_{\msbar}}{\mudec} = 
  \frac{\Laml}{\Lambda_s^\eff}\times\lim_{M/\mudec \rightarrow \infty}\left[ 
  \dfrac{\varphi_s^\eff\left( \bar g_s(\mudec,M)\right)}
        {P\left(\frac{M}\mudec \left/\frac{\Lam_{\msbar}}{\mudec}\right.\right)}\right]\,. \nonumber \\[-2ex]
 \label{eq:basic}
\end{eqnarray} 
Note that the desired ratio $\Lam_{\msbar}/\mudec$ on the l.h.s. in $\Nf=3$ QCD also appears on the r.h.s. in the argument
of the function $P$, i.e~the equation is implicit. In order to solve it one needs to be able to both evaluate the pure gauge theory
function $\varphi_s^\eff$ and the function $P$. The latter corresponds to $P_{0,\Nf}$ in the notation of \cite{Athenodorou:2018wpk},
where it was shown that non-perturbatively $P$ has an ambiguity of order $(\Lambda/M)^2$,
which arises once the reference to a specific matching condition is removed. A major result of \cite{Athenodorou:2018wpk}
is the observation that the perturbative evaluation of $P$ using the $\msbar$-scheme is numerically very
accurate already for quark masses in the charm region. With the heavy quark mass setting the scale for the 
$\msbar$-coupling, the accuracy further improves towards the decoupling limit.
In practice, Eq.~(\ref{eq:basic}) will be used for a range of finite
but reasonably large values $M/\mudec$.  
When using a perturbative approximation for $P$ in the $\overline{\rm
  MS} $ scheme, deviations from the limit are expected to be on one
hand proportional to
$1/M^2$ and on the other hand logarithmic corrections of
$\text{O}(\alpha_{\overline{\rm MS} }^4(M))$. This assumes that linear terms in 
$1/M$ are either completely  
absent by symmetry or that they can be controlled or subtracted explicitly.

While decoupling can be studied in the infinite volume regime~\cite{Athenodorou:2018wpk},
for a lattice QCD approach it is advantageous to separate the determination of the hadronic scales 
from the study of decoupling, by using a finite volume renormalization scheme~\cite{DallaBrida:2019mqg}. 
We use the GF scheme with SF boundary conditions, 
first introduced in \cite{Fritzsch:2013je}, with details given
in~\cite{DallaBrida:2016kgh}. 
In a continuum language it is given by
\begin{equation}
  \bar{g}^2_\text{GF}(\mu) = {\cal N}^{-1} \left.\sum_{k,l=1}^3
    \dfrac{t^2\langle {\rm tr}\left\{G_{kl}(t,x)G_{kl}(t,x)\right\}\delta_{Q,0} 
    \rangle}{\langle \delta_{Q,0}\rangle}\right\vert_{\mu=1/L,T=L,\, M=0}^{x_0=T/2,\,c = \sqrt{8t}/L}
 \label{eq:GF}
\end{equation}
where $G_{\mu\nu}(t,x)$ denotes the field tensor for the gauge field
at flow time $t$, and ${\cal N}$ is a known normalization 
factor which ensures $\bar{g}^2_\text{GF}(\mu) = g_0^2 + \rmO(g_0^4)$, with $g_0$
the bare coupling. The projection onto the topological charge $Q=0$ sector
is part of the scheme definition and merely introduced in order 
to avoid technical difficulties with the numerical simulation algorithms. 
The remaining parameter $c$ fixes the ratio between the
scales set by the flow time and the finite volume.

We may now break down the decoupling strategy into several steps:
\begin{enumerate}
\item Decoupling scale $\mudec$:  Given the coupling in the massless fundamental theory, we fix 
$\mudec$ by setting
\begin{equation}
   \bar{g}^2_\text{GF}(\mudec) = 3.949\,.
\label{eq:LCP1}
   \end{equation}
From previous work~\cite{DallaBrida:2016kgh,Bruno:2017gxd} one finds $\mudec = 789(15)\,\MeV$, which is a typical QCD scale.
Eq.~(\ref{eq:LCP1}) defines a so-called line of constant physics (LCP); following it towards the 
continuum, lattice spacing $a\to 0$, means 
that the limit is approached at fixed $\mudec/\Lam_{\msbar}$. Evaluating the LCP for a given lattice size $L/a=1/(a\mudec)$ 
defines a corresponding value for the bare coupling, $g_0^2\equiv 6/\beta$, and vice versa. We have implicitly
assumed here that $M$ vanishes. With Wilson fermions this requires a further tuning condition on the bare mass parameter. Details are discussed in \sect{sec:line-const-phys}.

\item Definition of $z=M/\mudec$:
A further set of constant physics conditions is obtained by fixing the RGI quark mass in units of $\mudec$.
Choosing a set of values in the range $\in [2,12]$, one needs to work out, for given lattice spacing 
(as obtained from Eq.~(\ref{eq:LCP1})) the corresponding bare mass parameters $am_0$. 
The details of this procedure will be discussed in Section~\ref{sec:massive-line-const-phys}.

\item Determination of $\bar{g}^{(0)}_\text{GFT}(\mudec)$: The value of a renormalized coupling in the $\Nf=0$ theory,
at a known scale $\mudec$ is obtained by evaluating the same coupling in the fundamental theory at a
heavy mass $M$ and assuming decoupling. The main problem with a mass dependent GF-coupling are boundary $1/M$ terms, 
which render decoupling slower than necessary.
In order to minimize these effects we use a variant of the coupling with $T=2L$,
\begin{equation}
   \bar{g}^2_\text{GFT}(\mu,M) = \left.\bar{g}^2_\text{GF}(\mu)\right|_{T\to 2L, M\to z\mu}\,.
   \label{eq:GFT}
\end{equation}
Compared to the GF coupling (\ref{eq:GF}) this doubles
the distance of the magnetic energy density to the boundaries, thereby reducing the
coefficient of the $1/M$ boundary contribution substantially.
Calling this scheme GFT, the main computational effort was required for
the evaluation of $\bar{g}^2_\text{GFT}(\mudec, M)$, at the lattice spacings and
bare quark masses which follow from the chosen lines of constant physics.

\item Determination of $\bar{g}^{(0)}_\text{GF}(\mudec)$: Obtaining this input value for the precisely known
$\varphi_\text{s}^\eff$ (with the scheme $s=\text{GF}$) in Eq.~(\ref{eq:basic}) requires the establishment of
a non-perturbative relation between the GF and the GFT schemes in the $\Nf=0$ theory at scale $\mudec$.
This is achieved by evaluating the GF coupling along a LCP defined by a fixed value of the GFT coupling,
and continuum extrapolating.

\item Determination of $\Lam_{\msbar}^{(0)}/\mudec$: The recent step-scaling study of the GF coupling in \cite{DallaBrida:2019wur}
allows us to evaluate $\varphi_\text{GF}^{}(\bar g^{(0)}_\text{GF}(\mudec)) =  \Lambda^{(0)}_\text{GF}/\mudec$
which completes the numerator in the square brackets  of Eq.~(\ref{eq:basic}).
The conversion to the $\MSbar$ $\Lambda$-parameter then simply requires the one-loop matching 
between the GF and $\MSbar$-couplings in the pure gauge theory~\cite{DallaBrida:2017tru}.

\item
The function $P$ gives the ratios of $\Lambda$-parameters between the fundamental and effective theories
and can be reliably evaluated in {\em massless} continuum perturbation theory
\begin{equation}
   P = \varphi^{(0)}_{\msbar}(g_\star \sqrt{C(g_\star)})/\varphi^{}_{\msbar}(g_\star)\,,
\end{equation}
with $C(g)$ known to 4-loop order, cf.~Eq.~(\ref{eq:Cfunc}) and the notation $g_\star=\bar g_\msbar(m_\star)$.
In particular, the quark mass $M$ only enters to set the scale.
For given $z=M/\mu_\mathrm{dec}$, the l.h.s. and the function $P$ in  Eq.~(\ref{eq:basic}) only depend on $\Lam_{\msbar}/\mudec$ and with
$\varphi_{\text{GF}}$ known it remains to numerically solve for $\Lam_{\msbar}/\mudec$. 
This is to be repeated for all available $z$-values
and the result for $\Lam_{\msbar}/\mudec$ is then obtained by extrapolation 
to the decoupling limit.
\end{enumerate}
Concluding this overview, we see that, besides the evaluation of the GFT coupling
in a mass-dependent scheme, the main ingredients are precision results in the pure gauge theory 
for the running GF coupling and  the matching between GF and GFT schemes. Together 
with available 5-loop perturbative results for the function $P$, this allows us 
to infer the $\Nf=3$ $\Lambda$-parameter in units of $\mudec$ and thus in MeV, given
the relation of $\mudec$ to a hadronic scale from~\cite{DallaBrida:2016kgh}.

\section{The continuum and decoupling limits: a closer look}
\label{sec:limits}

In this section we discuss the approach to the continuum and decoupling
limits in some more detail, in order to provide the theoretical
underpinning for the analysis of the lattice data.
While the limits are conceptually independent, in practice they are
best dealt with together, in terms of effective continuum field theories.
The methods of refs.~\cite{Husung:2019ytz,Husung:2021geh,Husung:2021mfl,Husung:2021tml,Husung:2022kvi} 
allow us in principle to go beyond power law behaviour and use renormalization group improved
perturbation theory to obtain the correct leading asymptotics. This holds
true for both limits, with the small parameter being either the lattice 
spacing or the inverse quark mass.
While the information for the bulk effects is still incomplete, the discussion serves
to motivate the fit ans\"atze which will be used in the data analysis, cf.~Section~\ref{sec:lattice}.
For the boundary $1/m$ effects we are able to estimate the full 
contribution in Section~\ref{subsec:boundary} without a fit to the data.
We will focus on the bulk effects first and address the influence of the boundaries 
in the Euclidean time direction in the end.

\subsection{Symanzik's effective theory for lattice QCD} 
\label{subsec:Symanzik}

Following Symanzik~\cite{Symanzik:1983dc,Symanzik:1983gh,Luscher:1984xn,Sheikholeslami:1985ij}, 
the approach of a connected lattice correlation function
to the continuum limit can be described in terms of an effective continuum theory,
with action
\begin{equation}
  S_\text{eff} = S_0 + a S_1 + a^2 S_2 + \ldots\,.
\end{equation}
Here, $S_0$ is the continuum action and $S_k$ are space-time integrals
over linear combinations of local composite fields of mass dimension $4+k$, $k=1,2,\ldots$,
which respect all the symmetries of the lattice action. We 
will omit $S_1$ in the following, assuming a non-perturbatively O($a$) improved 
lattice set-up. Residual effects due to $S_1$ are dealt with separately 
(cf.~Sect.~\ref{subsec:boundary} and Appendix \ref{app:renormass}).

Local fields $O$ defining the observables 
are represented by corresponding effective fields and expanded similarly,
\begin{equation}
  O_\text{eff} = O_0 + a O_1 + a^2 O_2 + \ldots\,.
\end{equation}
Gluonic gradient flow observables $O_\text{gf}$  can be formulated in terms of a local 4+1 dimensional field 
theory~\cite{Luscher:2011bx, Ramos:2015baa} with the flow time $t$ as
the extra coordinate. This allows us to  work entirely in terms of local observables $O$ and improve them 
to O($a^2$), such that $O_1$ and $O_2$ vanish in the effective field description.
We  also assume that the O($a^2$) effects originating from the 4+1 dimensional bulk action are removed by
an appropriate O($a^2$) modification of the flow equation~\cite{Ramos:2015baa}.
The Symanzik expansion for such observables then takes the form
\begin{equation}
   \langle O_\text{gf}\rangle_\text{lat} = \langle O_\text{gf} \rangle_\text{cont} 
   - a^2 \langle O_\text{gf} S_2\rangle_\text{cont} + \ldots,
  \label{eq:Sym-expansion}
\end{equation}
in terms of connected correlation functions. Although
$S_2$ contains an integral over space-time, no contact terms are generated for gradient
flow observables $O_\text{gf}$, as they are separated from $S_2 $ by the finite flow time $t$.
For the lattice set-up with non-perturbatively O($a$) improved, mass degenerate Wilson quarks, $S_2$ is given as
a linear combination of 18 local dimension-6 operators, 
\begin{equation}
   S_2 = \int d^4x\, \sum_{i=1}^{18} \omega_i \op{i}(x)\,,
 \label{eq:omegai}
\end{equation}
integrated over space-time. This constitutes an operator basis after the use of the equations of motion, the elimination
of total derivative terms and the use of relations among 4-quark operators due to Fierz transformations.
In the absence of gradient flow observables, the equations of motion simplify and allow for the elimination of 2 operators~\cite{Husung:2021tml}.

Note that Eq.~(\ref{eq:Sym-expansion}) makes the power dependence on the lattice
spacing explicit. An additional $a$-dependence arises through the coefficients $\omega_i$ of the operators in $S_2$, as these can be
understood as functions of a renormalized coupling at the cutoff scale, $\mu=1/a$. Close to the continuum limit 
asymptotic freedom implies that their leading asymptotic behaviour is {\em exactly} computable. 
This was first used by Balog, Niedermayer and Weisz~\cite{Balog:2009yj,Balog:2009np} 
in their analysis of the 2-dimensional O($n$) $\sigma$-model.
The technique has recently been extended and applied to gauge theories in various lattice regularisations, including
lattice QCD with quarks of both the Wilson and Ginsparg-Wilson type~\cite{Husung:2019ytz,Husung:2021mfl,Husung:2021geh,Husung:2021tml,Husung:2022kvi}. 
Technically, one needs to compute, to 1-loop order, the anomalous dimension matrix for the set 
of mass dimension 6 operators entering $S_2$. The operator basis mixes under renormalization,
$\op{\text{R},i} = \sum_{j=1}^{18} Z_{ij} \op{j}$,
and, following our conventions from Sect.~\ref{sec:strategy}, we define the corresponding anomalous dimension matrix 
\begin{equation}
   \gamma_{ij}^{\cal O} = \sum_{k=1}^{18} \left(\mu\dfrac{d}{d\mu} Z_{ik}\right) \left(Z^{-1}\right)_{kj}  
   = - g^2 \left[{(\gamma_0^{\cal O})}_{ij} + {(\gamma_1^{\cal O})}_{ij} g^2 + \rmO(g^4)\right] \,.
\end{equation}
A change of basis, ${\base}_i = \sum_j V_{ij}{\cal O}_j$, may then be performed in order to
diagonalize the one-loop anomalous dimension matrix, $\gamma_0^{\cal O}$, 
and determine its eigenvectors and eigenvalues. Denoting the transformed
anomalous dimension matrix by
\begin{equation}
  \gamma^{\base} =  V\gamma^{\cal O}V^{-1} = -g^2\left(\gamma^{\base}_0 + \gamma^{\base}_1 g^2+ \ldots\right), \qquad  
  \left(\gamma_0^{\base}\right)_{ij} = \delta_{ij}\gamma_{0,i}^{\base}\,, 
\end{equation}
renormalization group invariant (RGI) operators can be defined 
through\footnote{The absolute normalization of the RGI operator
conventionally includes a constant (but $\Nf$-dependent) factor $[8\pi b_0]^{-\hat\gamma}$
which we omit for the sake of readability and because 
the normalization of ${\base}^\text{RGI}$ will be irrelevant in the following.}
\begin{equation}
   {\base}_{i}^\text{RGI} = \lim_{\mu\rightarrow \infty}  
   \left[\alpha_{\MSbar}(\mu)\right]^{-\hat\gamma_i^\base}{\base}_{i}(\mu)\,,
   \qquad \hat\gamma^\base_i = \gamma_{0,i}^{\base}/(2b_0)\,.
  \label{eq:BRGI}
\end{equation}
At finite $\mu$ there are corrections of O($\alpha$) stemming from the two- and higher loop anomalous dimensions.
Note also the $\Nf$-dependence of $\hat\gamma^\base_i$, due to the normalization by $2b_0$.

In terms of the eigenbasis of operators, $\{{\base}_i\}$, the cutoff effects take the form
\begin{equation}
   \langle O_\text{gf}\rangle_\text{lat} = \langle O_\text{gf} \rangle_\text{cont} 
- a^2 \sum_i  b_i\left(\alpha_{\MSbar}(1/a)\right) [\alpha_{\MSbar}(1/a)]^{\hat{\gamma}^\base_i} 
 \int d^4 x\langle O_\text{gf}{\base}^\text{RGI}_i(x)\rangle+\ldots\,,
\end{equation}
where the coefficient functions
\begin{equation}
  b_i(\alpha) = \sum_{j=1}^{18} \omega_j(\alpha)\,(V^{-1})_{ji}  = \sum_{n\geq 0} \alpha^{n} b_i^{(n)},
\end{equation}
are given as a linear combination of the $\omega_i$, Eq.~(\ref{eq:omegai}), and are
thus perturbatively computable.
For tree-level O($a^2$) improved lattice actions, $b^{(0)}_i = 0$, 
and the higher coefficients can be successively eliminated by perturbative O($a^2$) improvement of the lattice action.
For lattice QCD with O($a$) improved Wilson quarks, one then expects that the leading cutoff effects in the bulk are of the form
\begin{eqnarray}
  \langle O_\text{gf}\rangle_\text{lat} &=& \langle O_\text{gf} \rangle_\text{cont} - a^2 \sum_{i=1}^{18} A_i [\alpha_{\MSbar}(1/a)]^{\hat{\Gamma}_i} 
    \left \{1 + \rmO(\alpha_{\MSbar}(1/a))\right\}  + \rmO(a^3)\,,  \\
   A_i &=& b_i^{(n_i^\mathrm{I})} \int d^4x\,\left\langle O_\text{gf}{\base}^\text{RGI}_i(x)\right\rangle_\text{cont}\,,
  \label{eq:Ai}
\end{eqnarray} 
where the neglected powers in $\alpha$ include both the expansion of $b_i(\alpha)$ and terms containing the (non-diagonal) 
higher order anomalous dimensions. The constants $A_i$ contain the insertions of the scale-independent RGI operators 
and $\hat{\Gamma}_i=\hat{\gamma}^\base_i +n_i^\mathrm{I}$ depends on the
degree of perturbative O($a^2$) improvement of the lattice action. For example, a tree-level (completely) improved action 
leads to $n_i^\mathrm{I}\geq1$ and in general we have $b_i=\hat{b}_i\alpha^{n_i^\mathrm{I}} (1+\rmO(\alpha))$.

For $\Nf=3$ lattice QCD, with O($a$) improved Wilson quarks,
Husung et al.~\cite{Husung:2021tml,Husung:2021mfl,Husung:2022kvi} found that the spectrum for the 
1-loop anomalous dimensions is bounded from below by $\hat\Gamma_i \ge -1/9$,
for the basis of 16 operators needed for observables not involving the gradient flow.
There are then 6 operators found with 1-loop anomalous dimensions $-1/9 \le \hat\Gamma_i < 8/9$. 
The remaining operators describe cutoff effects accompanied by powers of $\alpha$ equal or higher
than other neglected terms and may therefore be discarded. Explicit expressions 
for the eigen-operators of the 1-loop anomalous dimension matrix are, in general, rather complicated and
will not be required here. For the case of gradient flow observables this result is not complete,
as there are two further dimension 6 operators which must be included to obtain the full matrices $V$ 
and $\gamma^{\cal O}_0$~\cite{Husung:2021tml}. They have so far only been computed in the pure gauge theory \cite{Husung:2021tml}.

In view of the heavy mass expansion, there is a very interesting block structure
in $\gamma^{\cal O}_0$, for the subset of operators in $S_2$ which come with a positive power 
of the quark mass. For $\Nf=3$ with non-degenerate quarks there are eleven operators~\cite{Husung:2021tml,Husung:2021mfl,Husung:2022kvi},
which reduce to just three for degenerate quarks, namely
\begin{equation}
 \op{m,1} = \frac1{g_0^2}\sum_{\mu,\nu} m^2 \tr\left(F_{\mu\nu}F_{\mu\nu}\right)\,,\qquad
 \op{m,2} = m^3 \psibar \psi\,,\qquad
 \op{m,3} = \frac14 \sum_{\mu,\nu} m\psibar\, i\sigma_{\mu\nu} F_{\mu\nu} \psi\,.
 \end{equation}
 Note that this subset will remain the same for gradient flow observables, as the additional
 operators do not come with mass factors. Moreover, the tridiagonal block structure of $\gamma^{\cal O}_0$
 means that their anomalous dimensions will not be affected by enlarging the basis 
 and their renormalization can be consistently carried out ignoring the remainder of the basis~\cite{Husung:2021tml,Husung:2021mfl,Husung:2022kvi}.
 This is fortunate, as it means that the structure of the leading 
 $a^2m^2$ lattice effects can be inferred with current knowledge.
 We denote the corresponding basis of eigen-operators for the 1-loop anomalous dimension matrix by 
$\{{\base}_{m,i}\}_{i=1,2,3}$,
and their 1-loop anomalous dimensions are then given by~\cite{Husung:2021tml,Husung:2021mfl,Husung:2022kvi},
\begin{equation}
  \hat\gamma^\base_{m,1} = -1/9\,,\qquad \hat\gamma^\base_{m,2} = 14/27\,, \qquad   \hat\gamma^\base_{m,3} = 8/9\,.
\end{equation}
Furthermore, from~\cite{Husung:2021tml,Husung:2021mfl,Husung:2022kvi} one infers that, for our lattice action (cf.~Appendix~\ref{sec:simulations}), 
we have ${b}_{m,i} = \hat{b}_{m,i} + \rmO(\alpha)$ so that $\hat{\Gamma}_{m,i}=\hat{\gamma}^\base_{m,i}$ for $i=1,2,3$. 
We note that one only needs to retain the first two operators as the 
difference $\hat\gamma^\base_{m,3}-\hat\gamma^\base_{m,1}=1$ translates to a relative factor of $\alpha$,
i.e.~${\base}_{m,3}$ contributes at the same order as other neglected contributions.

\subsection{The decoupling expansion}

The Symanzik expansion renders the $a$-dependence explicit, both for the powers of $a$ and
the leading logarithmic terms given as fractional powers of $\alpha_{\MSbar}(1/a)$.
The connected correlation functions which appear in this expansion are thus defined in
the continuum limit, with respect to the continuum QCD action. In a second step,
we now determine how the continuum correlation functions,
$ \langle O_\text{gf}\rangle_\text{cont}$, $\langle O_\text{gf} S_2\rangle_\text{cont},$
behave as the quark mass $m$ is taken large. 
The effective decoupling theory bears formal similarities with Symanzik's effective theory, in particular
it renders both the powers in $1/m$ and the logarithmic corrections explicit.  
The effective decoupling action can be expanded,
\begin{equation}
  \label{eq:Sdec}
  S_\text{dec} = S_{0,\text{dec}} + \frac{1}{m} S_{1,\text{dec}}+\frac{1}{m^2} S_{2,\text{dec}} + \rmO(1/m^3)\,,
\end{equation}
with
\begin{eqnarray}
  S_{0,\text{dec}} &=& -\frac12 \int d^4x \,{\cal D}_0(x)\,,\qquad  {\cal D}_{0} = \frac{1}{g_0^2} \tr(F_{\mu\nu} F_{\mu\nu})\,, \\
  S_{2,\text{dec}} &=&  \int d^4x \left(\dS_1 {\cal D}_{1}(x) + \dS_2 {\cal D}_{2}(x)\right).   
  \label{eq:S2dec}
\end{eqnarray}
Due to the simultaneous decoupling of all quarks the leading term, $S_{0,\text{dec}}$, is given by the pure gauge action. 
$S_{k,\text{dec}}$ are given space-time integrals of gauge invariant local operators of mass dimension $4+k$, 
polynomial in the gauge field and its derivatives.
Gauge and $O(4)$ symmetries do not allow for odd values of $k$, so that the first order term must vanish.
The dimension-6 pure gauge operators in Eq.~(\ref{eq:S2dec}) take the form,
\begin{equation}     
   {\cal D}_{1} = \frac1{g_0^2}\sum_{\mu,\nu,\rho}\tr\left(D_\mu F_{\mu\nu}D_\rho  F_{\rho\nu}\right) \,,\qquad
   {\cal D}_{2} = \frac1{g_0^2}\sum_{\mu,\nu,\rho}\tr\left(D_\mu F_{\rho\nu} D_\mu F_{\rho\nu}\right) -\frac{23}{7}{\cal D}_{1}\,, 
\end{equation}
where we have directly chosen the eigenbasis of the one-loop anomalous dimension matrix, with eigenvalues $\hat{\gamma}^{\cal D}_{0,1,2} = -1,0,7/11$,
respectively. The coefficients  $\dS_{1,2}$ are matching coefficients between QCD with $\Nf=3$ heavy quarks and the $\Nf=0$ effective theory 
and can be perturbatively expanded in $\alpha_\msbar$, taken at the decoupling scale. 
In perturbation theory, this scale is most naturally defined as the running quark mass $\overline{m}^{}_{\MSbar}\left(\mu\right)$ 
at its own scale,
\begin{equation}
  m_\star = \overline{m}^{}_{\MSbar}\left(m_\star\right),
\end{equation}
which also defines the (inverse) expansion parameter of the effective decoupling theory.

Besides the effective action, observables $O$ have an effective large mass description, too. 
For the case of linear combinations of the fields $\base_i$, $O= \sum_i c_i{\base}_i$, it starts with a term of O($m^2$).
We thus expect the form
\begin{equation}
 [O]_\text{dec} =  m^2\sum_{k\geq0} \frac{1}{m^{2k}} {O_\text{$2k$,dec}} \,,
\end{equation}
where the fields ${O_\text{$2k$,dec}}$ are linear combinations of gauge invariant local composite
fields, polynomial in the gauge field and its derivatives, of mass dimension $d_O+2(k-1)$, where $d_O$ is
the dimension of the observable. 
For gradient flow observables $O_\text{gf}$ we will assume that the effective observable description reduces to the term with $k=1$.

We will now look at the combined Symanzik and decoupling expansion in $a$ and $1/m$ and discuss
in turn the corrections terms of order O($1/m^2$), O($a^2m^2$) and O($a^2)$.
We emphasize that the decoupling expansion is applied to the Symanzik effective theory. Hence, the combined expansion is valid for 
\begin{equation} 
   q \ll m \ll 1/a \,,
\end{equation}
for all scales $q$ present in the observable considered. In our application these are $q\in\{1/\sqrt{8t},\Lambda_\mathrm{QCD}\}$.

\subsubsection{Corrections of O($1/m^2$)}
\label{subsec:decouplinglogcorrections}

To order $1/m^2$ in the heavy mass expansion, we formally have,
\begin{equation}
  \langle O_\text{gf} \rangle_\text{cont} = \langle O_\text{gf}\rangle_\text{dec}  
  - \frac{1}{m^2} \langle O_\text{gf} S_{2,\text{dec}}\rangle_\text{dec} + \ldots\,,
\end{equation}
which evaluates to
\begin{equation}
   \label{eq:O1oM2}
  \langle O_\text{gf} \rangle_\text{cont} = \langle O_\text{gf}\rangle_{\text{dec}} 
  -\frac{1}{m_\star^2}\sum_{i=1}^2 \dS_{i} [\alpha_{\msbar}^{(0)}(m_\star)]^{\hat{\gamma}^{\cal D}_i} 
       \int d^4x\langle O_\text{gf}{\cal D}_i^\text{RGI}(x)\rangle_{\text{dec}} + \ldots\,,
\end{equation}
where we have converted to the RGI operators in the $\Nf=0$ theory. Without performing an
explicit matching calculation, the leading order, $\leads_i$, in $\dS_i = \hat{d}^{\rm S}_i \alpha^{\leads_i}+\rmO(\alpha^{\leads_i+1})$, $i=1,2$,
is not known and we will have to use assumptions for $\leads_i$. Also converting to 
the RGI quark mass, $M$,
\begin{equation}
   m_\star = [8\pi b_0 \alpha_\msbar(m_\star)]^{\hat\gamma_m} M\,,  \qquad
   \hat\gamma_m = 4/9 \quad (\Nf=3)\,,
\end{equation}
then leads to the asymptotic large mass behaviour in the continuum limit of the form
\begin{equation}
  \langle O_\text{gf} \rangle_\text{cont} = \langle O_\text{gf}\rangle_{\text{dec}} 
  -\frac{1}{M^2}\sum_{i=1}^2 D_i [\alpha_{\msbar}(m_\star)]^{\leads_i-2\hat\gamma_m +\hat{\gamma}^{\cal D}_i}
  +\ldots\,,
\end{equation} 
where we have used that the couplings of the $\Nf=3$ and $0$ theory
coincide at the decoupling scale, i.e.~$\alpha_\msbar(m_\star)\equiv\alpha^{(3)}_\msbar(m_\star)=\alpha^{(0)}_\msbar(m_\star)$,
up to terms of O($\alpha^2$), which are neglected here.
The constants $D_i$ parametrize the matrix elements in the decoupled theory 
and the exponents of $\alpha$ are further specified as
\begin{equation}
   \leads_i-2\hat\gamma_m +\hat{\gamma}_i^{\cal D} = 
   \begin{cases} \leads_1 - 8/9 &  (i=1)\,, \cr \leads_2 -25/99  &  (i=2)\,.\end{cases}
\end{equation}
Assuming, e.g.~$\leads_1=\leads_2=1$ (at least one fermion loop has to be present in QCD), 
then fixes a possible ansatz for the heavy mass extrapolation
of continuum extrapolated data for the gradient flow observable, with leading correction 
terms  $\propto[\alpha_\msbar(m_\star)]^{1/9}/M^2$ and $\propto[\alpha_\msbar(m_\star)]^{74/99}/M^2$, for $i=1,2$,
respectively.

\subsubsection{Corrections of O($a^2m^2$) and O($a^2$)}
\label{sec:corr-oa2m2-oa2}
We now turn to the large mass expansion of  $\langle O_\text{gf}{\base}_i^\text{RGI}\rangle_\text{cont}$ which
appears at O($a^2$) in the Symanzik expansion. We first transform the RGI operators to the relevant
scale $\mu=m_\star$, by applying Eq.~(\ref{eq:BRGI})
 \begin{equation}
   {\base}_{i}^\text{RGI} = \left[\alpha_{\MSbar}(m_\star)\right]^{-\hat\gamma^\base_i}{\base}_{i}(m_\star) 
   \left[1 + \rmO(\alpha(m_\star))\right],
 \end{equation}
and inserting into the Symanzik expansion coefficient,
\begin{equation}
   \langle O_\text{gf} S_2 \rangle_\text{cont} = 
   \sum_{i=1}^{18}b_i\left(\alpha_\msbar(1/a)\right) R_\alpha^{\hat\gamma^\base_i} 
   \int d^4x\, \langle O_\text{gf}{\base}_i(m_\star;x)\rangle_\text{cont} \,,
   \label{eq:OgfS2evolved}
\end{equation}
where we have neglected terms of relative $\rmO(\alpha)$ and introduced the notation,
\begin{equation}
  R_\alpha= \frac{\alpha^{(3)}_\msbar(1/a)}{\alpha^{(3)}_\msbar(m_\star)}\,.
\end{equation}
In this approximation, we expect that less than half of the 18 operators contribute
terms that are parametrically leading in $\alpha$. However, a precise statement
can only be made once the full one-loop anomalous dimension matrix and the coefficients $b_i$ are known.

With these preliminaries we use the effective decoupling description for the operators ${\base}_i$,
\begin{equation}
 [{\base}_{i}]_\text{dec} = m^2 \dB_{i,0} {\cal D}_0 +  \dB_{i,1} {\cal D}_1 + \dB_{i,2} {\cal D}_2 + \rmO(1/m^2)
\end{equation}
with matching coefficients $\dB_{i,j}$. 
Inserting the expansion of both the decoupling action (\ref{eq:Sdec}) and these fields we obtain,
\begin{eqnarray}
  \langle O_\text{gf}{\base}_i(x) \rangle_\text{cont} &=& 
   m^2 \dB_{i,0} \langle O_\text{gf}{\cal D}_0(x) \rangle_\text{dec} \nonumber\\
&&- \dB_{i,0} \sum_{j=1}^2 \dS_j\int d^4 y \langle O_\text{gf}{\cal D}_0(x) {\cal D}_j(y) \rangle_\text{dec}\nonumber\\
&&  + \sum_{j=1}^2 \dB_{i,j} \langle O_\text{gf}{\cal D}_j(x) \rangle_\text{dec}\,,
\end{eqnarray}
up to terms of order $1/m^2$.
In the next step we pass back to RGI operators, now in the decoupled, $\Nf=0$ theory.
With the anomalous dimension of ${\cal D}_0$ given by $\hat\gamma^{\cal D}_{0}=-1$~\cite{Husung:2021tml}, 
and after conversion to the RGI quark mass with $\hat\gamma_m=4/9$, we find
\begin{eqnarray}
  \langle O_\text{gf}{\base}_i(m_\star;x) \rangle_\text{cont} &=& 
   M^2 \dB_{i,0}[\alpha^{}_\msbar(m_\star)]^{-1+8/9}   \langle O_\text{gf}{\cal D}^\text{RGI}_0(x) \rangle_\text{dec} \nonumber\\
&&\mbox{} -\dB_{i,0}\sum_{j=1}^2 \dS_j [\alpha^{}_{\msbar}(m_\star)]^{-1+\hat\gamma^{\cal D}_j}\int d^4 y\,
\langle O_\text{gf}{\cal D}^\text{RGI}_0(x) {\cal D}^\text{RGI}_j(y) \rangle_\text{dec}\nonumber\\
&&\mbox{}  + \sum_{j=1}^2 \dB_{i,j} [\alpha^{}_{\msbar}(m_\star)]^{\hat\gamma^{\cal D}_j}    
\langle O_\text{gf}{\cal D}^\text{RGI}_j(x) \rangle_\text{dec}  +\rmO(1/m^2_\star)\,,
\label{eq:leadasquare}
\end{eqnarray}
where we have used once again that the couplings coincide at the decoupling scale, i.e.~
$\alpha_\msbar(m_\star)\equiv \alpha^{(3)}_\msbar(m_\star) = \alpha^{(0)}_\msbar(m_\star)$,
up to terms of O($\alpha^2$), which are negligible in this context. 
Inserting this expansion into Eq.~(\ref{eq:OgfS2evolved}) one notices that each term is weighted by $b_i \times R_\alpha^{\hat\gamma^\base_i}$. 
The matching coefficients $\dB_{i,j}$ for the observable and $\dS_j$ for the action have expansions
in $\alpha$, but their leading orders are not known. However, as we are interested in the leading $M^2$ behaviour,
we focus on the massive operators $\base_{m,i}$, for $i=1,2$ (cf.~Sect.~\ref{subsec:Symanzik}). Both operators contain the gluonic component 
${\cal O}_{m,1}$, so that one expects the expansion of their matching coefficients to start at tree level, 
i.e.~$\dB_{(m,i),0}=\hat{d}^\base_{i,0} +\rmO(\alpha)$.  
Combining this with the Symanzik expansion we obtain the form of the leading $a^2M^2$ lattice effects,
\begin{equation}
 a^2 \langle O_\text{gf} S_2\rangle_\text{cont} = a^2M^2 D_0\hat{b}_{m,1}\hat{d}^\base_{1,0} [\alpha_\msbar(1/a)]^{-1/9}
 \left[ 1 + \frac{\hat{b}_{m,2}\hat{d}^\base_{2,0}}{\hat{b}_{m,1}\hat{d}^\base_{1,0}} R_\alpha^{17/27} + \rmO(\alpha)\right]+\ldots
 \label{e:symanzikfinal1}
\end{equation}
where $D_0$ denotes the matrix element of ${\cal D}^\text{RGI}_0$. Note that $\alpha_\msbar(m_\star)$ accidentally cancels out in the leading term.

Proceeding to the subleading $a^2$-effects, there are two types of contributions in Eq.~(\ref{eq:leadasquare}). The
first arises from the cancellation of the $m^2$ leading term with the subleading $1/m^2$ contribution from the effective decoupling
action $S_\text{2,dec}$, Eq.~(\ref{eq:S2dec}). Counting powers of $\alpha$, we expect
that only the massive operators $\base_{m,i}$ have a tree level matching coefficient to ${\cal D}_0$, rendering 
all non-massive operators negligible. For the matching coefficients in $S_\text{2,dec}$ we assume 
$\dS_{1,2} = \hat d^\mathrm{S}_{1,2}\alpha$, so that we obtain the form of the first subleading $a^2$-effect,
\begin{equation}
\begin{split}
 &  a^2 \langle O_\text{gf} S_2\rangle_\text{cont} = \text{$a^2M^2$-terms} \quad    \\
 &- a^2\hat{b}_{m,1}\hat{d}^\base_{1,0}  [\alpha_\msbar(1/a)]^{-1/9}
\left[ 1 + \frac{\hat{b}_{m,2}\hat{d}^\base_{2,0}}{\hat{b}_{m,1}\hat{d}^\base_{1,0}} R_\alpha^{17/27}\right]
  \left(\hat{d}^\mathrm{S}_1 D_{01} + [\alpha_\msbar(m_\star)]^{7/11} \hat{d}^\mathrm{S}_2 D_{02}\right)+\ldots\,.
\end{split}
 \label{e:symanzikfinal2}
\end{equation}
Here, we have used $\hat\gamma_{1,2}^{\cal D}=0,7/11$ and $D_{0i}$ denotes the matrix elements of ${\cal D}^\text{RGI}_0{\cal D}^\text{RGI}_i$, for $i=1,2$.
The leading term is proportional to  $a^2\times \alpha_\msbar(1/a)^{-1/9}$ and $\alpha_\msbar(m_\star)$ thus cancels yet again. 

For the second subleading $a^2$-term, the main question is which operators $\base_i$ match to ${\cal D}_{1,2}$
with a non-zero tree-level coefficient $\dB_{i,j}=\hat{d}^{\base}_{i,j} + \rmO(\alpha)$. This is certainly the case 
for those operators in $S_2$ which contain the gluonic dimension-6 operators of the same form as ${\cal D}_{1,2}$.
Including only such operators $\base_i$ in $S_2$ we then expect
\begin{equation}
\begin{split}
   a^2 \langle O_\text{gf} S_2\rangle_\text{cont} = &\text{$a^2M^2$-terms $+$ $(a^2M^2)/M^2$ terms} \\
  &+ a^2\sum_{i} \hat{b}_{i} R_\alpha^{\hat{\Gamma}_i} \left(\hat{d}^{\base}_{i,1} D_{1} 
 + \alpha_\msbar(m_\star)^{7/11} \hat{d}^{\base}_{i,2} D_{2}\right)+\ldots\,,
 \end{split}
 \label{eq:asquare_sublead}
\end{equation}
where $D_i$ denotes the matrix element of ${\cal D}^\text{RGI}_i$, $i=1,2$.
The possible powers $\hat\Gamma_i$ could be obtained from a complete basis of operators for gradient flow observables.
Until this becomes available we assume that the $a^2$-effects in Eq.~(\ref{eq:asquare_sublead}) are subleading,
i.e.~$\hat\Gamma_i > \hat\Gamma_1 =-1/9$, with $\hat\Gamma_1$ corresponding to the massive operator $\base_{m,1}$. 

\subsection{Boundary effects} 
\label{subsec:boundary}

So far we have not considered the effect of boundaries, where chiral symmetry 
can be broken by the boundary conditions. This is the case for standard
SF~\cite{Sint:1993un}, open~\cite{Luscher:2011kk}, and open-SF~\cite{Luscher:2014kea}
boundary conditions. Locality means that these effects can be discussed separately. 
In particular, boundary O($a^k$) and O($1/m^k$) effects can be respectively described
in the Symanzik and decoupling effective theory, in terms of local gauge-invariant fields
of dimension $3+k$ localized at the boundaries~\cite{Symanzik:1981wd}. In fact, the
counterterm fields that appear at O($a$) in the Symanzik expansion and at O($1/m$) in 
the large-mass expansion are the same. A complete set of fields can be found in
ref.~\cite{Luscher:1996sc}, where a detailed discussion of the O($a$) contributions 
to the Symanzik effective action in the presence of SF boundary conditions is presented. 
Below we shall focus on the decoupling expansion in the presence of SF boundary conditions.
For a discussion on the boundary O($a$) effects affecting the observables of
interest, instead, we refer the reader to refs.~\cite{DallaBrida:2016kgh,DallaBrida:2019wur}.
In these references, a detailed analysis for the case of the GF-couplings in
the $\Nf=0$ and $3$ theory is presented. Here we note that for the case of the GFT-couplings,
due to the larger separation between the flow energy density defining the couplings and the SF boundaries
(cf.~eq.~(\ref{eq:GFT})), these effects are expected to be significantly smaller than the 
estimates obtained in refs.~\cite{DallaBrida:2016kgh,DallaBrida:2019wur}. In practice, this renders these 
effects irrelevant in the context of the analysis presented in Sect.~\ref{subsec:context} and we neglect them.

As in the previous subsections, we are interested in the situation where the resulting 
effective theory for large quark masses is the pure gauge theory, i.e.~all quarks simultaneously 
decouple. In this case, we have (cf.~eq.~(\ref{eq:Sdec})), 
\begin{equation}
	\label{eq:SdecSF}
	S_\text{dec} = S_{0,\rm dec} + \frac{1}{m} S_{1,\rm dec} + \frac{1}{m^2} S_{2,\rm dec} + \rmO(1/m^3)\,,
\end{equation}
where
\begin{equation}
	S_{1,\rm dec}= \sum_{i=1}^{2}\int d^3x\,\omega_{i,{\rm b}}\,
	\big[\mathcal{O}_{i,{\rm b}}(0,\boldsymbol{x})+  \mathcal{O}_{i,{\rm b}}(T,\boldsymbol{x})\big]\,,
\end{equation}
with
\begin{equation}
	\mathcal{O}_{1,{\rm b}}= -{1\over g_0^2}\sum_{k=1}^3\tr(F_{0k}F_{0k})\,,
	\qquad
	\mathcal{O}_{2,{\rm b}}= -{1\over 2g_0^2}\sum_{k,l=1}^3\tr(F_{kl}F_{kl})\,.
\end{equation}
The SF boundary conditions commonly considered in applications are defined in 
terms of spatially constant Abelian fields~\cite{Luscher:1992an,Luscher:1993gh}. 
These include in particular the case of vanishing boundary conditions for the 
gauge field. For this class of fields, the only operator that 
contributes to the effective action is $\mathcal{O}_{1,{\rm b}}$, since $F_{kl}(x)=0$ 
at $x_0=0,T$.%
\footnote{In the case of open boundary conditions $F_{0k}(x)=0$ at $x_0=0,T$, and
only the operator $\mathcal{O}_{2,{\rm b}}$ is relevant. For open-SF boundary
conditions, instead, $\mathcal{O}_{2,{\rm b}}$ contributes at $x_0=0$, while 
$\mathcal{O}_{1,{\rm b}}$ at $x_0=T$, if spatially constant Abelian boundary fields
are considered.}
Thus, in this situation, we can take,
\begin{equation}
	S_{1,\rm dec}= \int d^3x\,\omega_{{\rm b}}
	\big[\mathcal{O}_{{\rm b}}(0,\boldsymbol{x})+  \mathcal{O}_{{\rm b}}(T,\boldsymbol{x})\big]\,,
\end{equation}
where we simplified the notation by setting $\omega_{\rm b}\equiv\omega_{1,\rm b}$ 
and $\mathcal{O}_{\rm b}\equiv\mathcal{O}_{1,\rm b}$. 

The knowledge of the matching coefficient $\omega_{\rm b}$ between QCD with 
$\Nf$ heavy quarks and the pure gauge theory, would allow us to compute the 
O($1/m$) corrections to any observable stemming from the effective action. 
In the case of gradient flow quantities, $O_\text{gf}$, these are the only 
O($1/m$) effects. As a result, we have that,
\begin{equation}
	\label{eq:LeadingO1MCorr}
	\langle O_\text{gf}\rangle_{\rm cont} = \langle O_\text{gf} \rangle_{\rm dec} -
	{1\over m_\star}\omega_{\rm b}
	\int d^3x\,
	\big\langle O_\text{gf}\big[\mathcal{O}_{{\rm b}}(0,\boldsymbol{x})+  \mathcal{O}_{{\rm b}}(T,\boldsymbol{x})\big]\big\rangle_{\rm dec}
	+{\rm O}(1/m_\star^2)\,.
\end{equation}
Comparing eqs.~(\ref{eq:LeadingO1MCorr}) and (\ref{eq:O1oM2}), the attentive reader might have 
noticed the absence of a factor $[\alpha^{(0)}_{\MSbar}(m_\star)]^{\hat{\gamma}_{\rm b}}$, with $\hat{\gamma}_{\rm b}$ 
the anomalous dimension of the relevant boundary field. This can be understood by noticing that for $x_0=0,T$, 
$\mathcal{O}_{\rm b}$ coincides  with the Hamiltonian density operator of the pure gauge theory,
\begin{equation}
	\mathcal{H}=-{1\over g_0^2}\bigg[\sum_{k=1}^3\tr(F_{0k}F_{0k})-{1\over2}\sum_{k,l=1}^3\tr(F_{kl}F_{kl})\bigg]\,.
\end{equation}
As is well-known, in continuum regularizations where the Euclidean space-time 
symmetries are preserved, the latter is protected against renormalization 
and its insertion in correlation functions is $x_0$-independent. 
The result in eq.~(\ref{eq:LeadingO1MCorr}) is thus expected 
to be valid to all orders in the perturbative expansion. On the lattice, 
where the continuum space-time symmetries are reduced to the symmetries of 
the hypercube, $\mathcal{O}_{\rm b}$ still has vanishing anomalous dimension,
but it requires a scale-independent multiplicative renormalization~\cite{Caracciolo:1989pt}.

Since the matching coefficient $\omega_{\rm b}$ is independent of the specific
correlator considered, we may impose the validity of eq.~(\ref{eq:LeadingO1MCorr})
for some convenient observable (neglecting O($1/m^2$) terms) in order to determine 
$\omega_{\rm b}$ (up to O($1/m$) ambiguities). It can then be used to compute the 
corresponding O($1/m$) corrections to any other quantity of interest. 
While in principle eq.~(\ref{eq:LeadingO1MCorr}) could be imposed non-perturbatively, 
in practice this is expected to be very challenging, since the $1/m$ contributions have 
to be separated numerically from other powers. For large enough masses $m_\star$, however, 
we may rely on a perturbative determination of $\omega_{{\rm b}}$, since $\alpha_{\MSbar}(m_\star)$ 
is then small. A convenient observable that can be used to determine $\omega_{\rm b}$ is the finite-volume 
SF coupling,  $\bar{g}^2_{\rm SF}(\mu)$~\cite{Luscher:1992an,Luscher:1993gh,Sint:1995ch}. 
Although it is not a gradient flow quantity, it receives $1/m$ corrections only from terms 
in the action. This is so because it is defined through the variation of the (logarithm of the) 
partition function with respect to the boundary conditions for the gauge fields, and not by 
the correlation function of some field. Thus, eq.~(\ref{eq:LeadingO1MCorr}) still holds in this
form. Solving this equation in perturbation theory, where on the l.h.s.~the coupling is computed 
in $\Nf$-flavour QCD with $\Nf$ heavy quarks, while on the r.h.s.~the correlators are computed 
in the pure gauge theory, we can extract 
\begin{equation}
	\omega_{\rm b}(\alpha_\star) = 
	\omega_{{\rm b}}^{(1)} \alpha_\star +\omega_{{\rm b}}^{(2)} \alpha_\star^2+ \ldots\,,
	\qquad
	\alpha_\star\equiv\alpha_{\MSbar}(m_\star)\,,
\end{equation}
by studying the limit $m_\star\to\infty$. We refer the interested reader to Appendix 
\ref{app:omegab}  for the details. Here we simply quote the result,
\begin{equation}
	\omega_{{\rm b}}^{(1)}=-0.0541(5) N_{\rm f} \,.
\end{equation}
A couple of remarks are in order at this point. While the strategy based on
eq.~(\ref{eq:LeadingO1MCorr}) is a general way to compute (and therefore eliminate) 
the O($1/m$) effects due to the SF boundary conditions, other strategies are 
in principle possible. For an even number of flavours $N_{\rm f}$, considering 
a twisted mass $\mu_{\rm tw}$ rather than a standard mass for the heavy quarks, 
would imply having leading O($1/\mu_{\rm tw}^2$) corrections to 
observables~\cite{Sint:2007ug}. Entirely equivalent in the continuum is the choice 
of having a standard mass for the quarks,  but with chirally rotated SF boundary
conditions~\cite{Sint:2010eh}. Regular periodic boundary conditions are in principle
possible for any value of $N_{\rm f}$ with leading corrections of O($1/m^2$),
however, perturbation theory becomes unduly complicated~\cite{Gonzalez-Arroyo:1981ckv}.
In QCD with $\Nf=3$ flavours $1/m$ effects could also be avoided by choosing 
twisted-periodic boundary conditions~\cite{tHooft:1979rtg}.
 
\section{3-flavour QCD: set-up, simulations and results} 
\label{sec:lattice}
\newcommand{\gbargf}{\bar g_{\rm GF}}

We simulate three flavors of non-perturbatively $\rmO(a)$-improved Wilson fermions with
the tree-level Symanzik $\rmO(a^2)$-improved gauge action~ \cite{Bulava:2013cta}. 
The same discretization  was employed in our previous work~\cite{DallaBrida:2016kgh}.
The parameter $\beta=6/g_0^2$ in the gauge action and a parameter
$\kappa=1/(2am_0+8)$ in the mass-degenerate fermion action need to be fixed 
for each lattice size, $L/a = 1/(a\mudec)$ and for each physical heavy quark
mass $M$. Our line of constant physics (LCP) is identified in terms of the value of
the massless coupling $\gbargf^2(\mudec) = 3.949$ and the values of
$z=ML$. Our error analysis takes all (auto-) correlations into account using the publicly available
implementations of the $\Gamma$-method~\cite{Wolff:2003sm, Ramos:2018vgu} and a second independent analysis.
A preliminary analysis of our results was presented in~\cite{Brida:2021xwa}.

\subsection{Line of constant physics at $M=0$ }
\label{sec:line-const-phys}

The first four columns of table~\ref{tab:lcp} show results of simulations tuned such that the PCAC mass 
vanishes and $\gbargf^2\approx3.949$.  In order to precisely tune to our LCP, we apply a small shift,
\begin{equation}
  g^2_{0,\mathrm{LCP}} = g^2_{0,\mathrm{sim}} +\frac{3.949 - \gGF^2}{S} \label{eq:lopt}\,,\quad  S=\left. \frac{\partial \gGF^2}{\partial \tilde g_0^2}\right|_{L/a} \,,
\end{equation}
where $g^2_{0,\rm sim}$ are the simulated bare couplings (cf.~second column of table~\ref{tab:lcp}) and the slope
\begin{eqnarray}
  S    =
  \left.\frac{\partial \gGF^2}{\partial \log(a)}\right|_{ L/a} \,\frac{\rmd \log(a)}{\rmd g_0^2} = 
  \left.\frac{\partial \gGF^2}{\partial \log(L)}\right|_{ L/a} \,\frac{\rmd \log(a)}{\rmd g_0^2}\,  =  
  \frac{\gGF\beta^{(3)}_\mathrm{GF}(\gGF)}{ g_0 \beta_0^{(3)}(g_0) }\,.
	\label{eq:deriv0}
\end{eqnarray} 
All quantities here are defined at vanishing quark mass, but
we note that the shift in $\kappa$ to maintain the $m=0$ condition at
$\beta_\mathrm{LCP}$ is negligible.
We also convert the uncertainty in $\bar g^2_{\rm GF}$ into an uncertainty in the LCP $\beta$-value using the slope $S$.
The last column of table~\ref{tab:lcp} lists the resulting values
of $\beta_{\rm LCP}=6/g^2_{0,\mathrm{LCP}}$. 
Note that the decoupling scale $\mu_{\rm dec}$ is implicitly defined
by our LCP, i.e. $\bar g^2_{\rm GF}(\mu_{\rm dec}) = 3.949$.
Our estimates used for the three-flavor renormalized ($\beta^{(3)}_\mathrm{GF}$) and bare ($\beta^{(3)}_0$) beta-functions 
are described in Appendix \ref{app:betashifts}.
\begin{table}
  \centering
\begin{tabular}{r|lll|l}
\toprule
\(L/a\) & \(\beta\) & \(\kappa\) & \(\bar  g^2_{\rm GF}\) & $\beta_{\rm LCP}$\\
\midrule
  12 & 4.3030 & 0.1359947(18) &  3.9461(41)  &     4.3019(16)\\
16 & 4.4662 & 0.1355985(9) &  3.9475(61)     &     4.4656(23)\\
20 & 4.6017 & 0.1352848(2) &  3.9493(63)     &     4.6018(24)\\
24 & 4.7165 & 0.1350181(20) & 3.9492(64)     &     4.7166(25)\\
  32 & 4.9000 & 0.1345991(8) & 3.949(11)     &     4.9000(42)\\
  \midrule
  40 & -- & -- & --  &       5.0497(41)\\
  48 & -- & -- & --  &       5.1741(54)\\
  \bottomrule                             
\end{tabular}

   \caption{Massless line of constant physics. 
    The simulations described in the first four columns are taken from \cite{Fritzsch:2018yag}. They are used to
    fix $\beta_{\rm LCP}$ such that the renormalized
    massless coupling $\bar g^2_{\rm GF}(\mu) = 3.949$. 
    The last row ($L/a=48$) is obtained indirectly from our knowledge of the
    non-perturbative running of the coupling, while the previous one  ($L/a=40$) is an interpolation of the other data, 
    see Appendix~\ref{sec:lcp4048} for more details.}
  \label{tab:lcp}
\end{table}

\subsection{Massive simulations}
\label{sec:massive-line-const-phys}

With the value of the massless coupling fixed by our LCP, we 
proceed to simulate massive quarks with (again) homogeneous SF boundary
conditions but with $T=2L$ and various $z$. 

Namely, for a given $L/a$, the massive simulations have to be performed at fixed 
lattice spacing (defined in a massless scheme and with O($a$) improvement~\cite{Luscher:1996sc}) 
and for a set of renormalized quark masses, common to all $L/a$. 
Therefore, for each $L/a$ we fix the simulation parameters $\beta, \kappa$ for a
prescribed value of $z = M /\mu_{\rm dec}=ML$. 
This last quantity is given by
\begin{equation}
  z = \frac{L}{a} \frac{M}{\overline{m}(\mudec)}Z_{\rm m}(\tilde g_0^2,a\mudec) \left[ 1 + b_{\rm m}(\tilde g_0) am_{\rm q}\right]\,am_{\rm q}\,,
\label{eq:zlm}
\end{equation}
where the running factor ${M}/{\overline{m}(\mudec)} = 1.474(11)$ (with
$\overline{m}(\mudec)$ defined in the SF scheme) can be determined
from  results available in the literature~\cite{Campos:2018ahf} (see
Appendix~\ref{app:renormass}). The renormalization 
constant $Z_{\rm m}(\tilde g_0^2)$ and improvement coefficient $b_{\rm
  m}(\tilde g_0)$, instead, are determined in Appendix~\ref{app:renormass}. Once $z$ is fixed, 
eq.~(\ref{eq:zlm}) is just a quadratic equation in $am_{ \rm q}$. 
For our O$(a)$-improved Wilson fermions fixed lattice spacing corresponds to fixed \emph{improved} bare coupling
$\tilde g_0^2$. The simulation parameter $\beta$ of the massive simulation is thus
obtained from
\begin{equation}
\label{e:beta_of_bg}
  \beta = \frac{6(1+b_{\rm g}\, am_{\rm q})}{\tilde g_0^2}\,,
\end{equation}
where the values of $\tilde g_0^2 =6/\beta_{\rm LCP}$ are taken from table~\ref{tab:lcp}, since at zero mass the improved and unimproved
couplings coincide. For $b_{\rm g}$ as well as for all other improvement coefficients that are known only perturbatively, we use 1-loop 
values and treat the difference between tree-level and 1-loop as uncertainties, see below and Appendix~\ref{app:renormass}. 
The  largest effect arises from $\bg$.

The other simulation parameter, $\kappa$, is obtained from
the critical mass, $am_{\rm c}(g_0^2)$. 
Since table~\ref{tab:lcp} provides the values of
$\kappa_{\rm c} = 1/(2am_{\rm c}(\tilde g_0^2)+8)$ we perform a small shift
\begin{equation}
      am_{\rm c}(g_0^2) = am_{\rm c}(\tilde g_0^2) + \left(g_0^2-\tilde g_0^2\right)\frac{\partial }{\partial
        \tilde g_0^2} \,(am_{\rm c})\,,
\end{equation}
where the needed derivative can be obtained either from the 
literature~\cite{DallaBrida:2016kgh}, or from the simulations used to
extract $Z_{\rm m}, b_{\rm m}$ (see Appendix~\ref{app:renormass}). 
Both determinations of the derivative agree at the percent level. 
We thus obtain $\beta,\kappa$ needed to simulate
at fixed values of $z$. 
The uncertainty in $z$, propagated from our determinations of $Z_{\rm m}$,
$b_{\rm m}$, $\kappa$, are propagated into an error in the coupling according
to the discussion in Appendix~\ref{app:errz}. 
The error in $z$ contributes to a small part to the uncertainty of $\bar
g^2_{\rm GFT}$.

Our simulations were performed when only an incomplete data-set for
the determination of the LCP was available. 
This can be observed by comparing our simulation parameters at $z=0$
in table~\ref{tab:massless} with the final values of the LCP available
in table~\ref{tab:lcp}. We correct for this small mismatch by 
a linear shift in $\tilde g_0^2$ using
\begin{eqnarray}
  \left. \frac{\partial \gGFT^2}{\partial \tilde g_0^2}  \right|_{z, L/a}  &=&
  \left.\frac{\partial \gGFT^2}{\partial \log(a)}\right|_{z, L/a} \,\frac{\rmd \log(a)}{\rmd \tilde g_0^2} = 
  \left.\frac{\partial \gGFT^2}{\partial \log(L)}\right|_{z, L/a} \,\frac{\rmd \log(a)}{\rmd \tilde g_0^2}\, 
  \nonumber \\
  &=& 
  \frac{\gGFT\beta^{(0)}_\mathrm{GFT}(\gGFT)(1-\etaM(\gstar))}{\tilde g_0 \beta_0^{(3)}(\tilde g_0) }[1+R_\mathrm{z}+R_\mathrm{a}]\,.
	\label{eq:deriv1}
\end{eqnarray} 
Here, the numerator uses decoupling and thus the pure gauge 
theory $\beta$-function of the GFT coupling appears together 
with the factor $(1-\etaM(\gstar)) \approx b_0^{(3)}/b_0^{(0)}=9/11$~\cite{Athenodorou:2018wpk}. 
The denominator is the $\nf=3$ bare $\beta$-function at $z=0$. 
The terms $R_\mathrm{z},\;R_\mathrm{a}$ are corrections for 
O($1/z^2$) and O($a^2$) terms, respectively. The derivation of this equation 
as well as its numerical approximation are explained in Appendix \ref{app:betashifts}. 
In all cases the resulting shifts in $\gbar^2_{\rm GFT}$ are small. In general they are 
well below our statistical uncertainties; only at $L/a=40$ the shifts amount to more
than one standard deviation. 

\subsection{Choice of $c$} \label{sec:note-diff-valu}
Within the same simulation, the massive coupling $\bar g^2_{\rm GFT}(\mu_{\rm dec},M)$ can be
obtained at different values of $c = \mu_{\rm  dec}\sqrt{8t}$, which defines the given \emph{gradient 
flow scheme} (cf.~eqs.~(\ref{eq:GF}),(\ref{eq:GFT})). For better clarity, in the
following discussion we shall thus employ the notation $\bar g^2_{\rm GFT,c}(\mu_{\rm dec},M)$.  
Typically, in finite size scaling studies (in massless renormalization schemes), the choice of $c$ 
represents a compromise between scaling violations (larger at small values of
$c$) and statistical uncertainties (larger at large values of $c$),
with $c$ in the range $0.3-0.5$ representing a good choice~\cite{Fritzsch:2013je}. 
Here, however, we do not intend to use the massive coupling to compute a step-scaling function, 
but rather as an observable to which decoupling can be applied. Its value is the same as in the pure
gauge theory up to power corrections in the inverse heavy-quark mass. In particular, the leading 
corrections are expected to be of $\rmO(\mu_\mathrm{max}^2/M^2)$, where $\mu_\mathrm{max}$ is 
the largest mass scale present. For $\bar g^2_{\rm GFT,c}$ we have: $\mu_\mathrm{max}=1/\sqrt{8t} = \mudec/c$. 
This implies that at a fixed scale $\mu_{\rm dec}$, a scheme with larger $c$ is expected to have smaller corrections 
to the infinite mass limit. In addition, contrary to  standard finite-size scaling studies, in the present context 
we do not expect a larger value of $c$ to reduce the discretization errors in our data. Discretization errors are in fact  
dominated by ${\rm O}((aM)^2)$ terms.

We have performed the analysis for $c=\sqrt{8t}/L=0.30,0.33,0.36,0.39,0.42$. 
In general, we will focus the discussion on the cases $c=0.30$ and
$c=0.36$, although the conclusions are similar for other values (see
table~\ref{tab:cont_res}). The case $c=0.3$ represents the most precise dataset. 
It is therefore an ideal choice to explore different mass cuts and study the
systematics involved in the continuum extrapolations. On the other hand, $c=0.36$ is an intermediate value from which 
we will extract our central results.

\subsection{Continuum extrapolations}
\label{subsec:context}

We turn to the continuum extrapolations of the massive couplings $\bar g^2_{\rm GFT,c}(\mu_{\rm dec},M,a\mu_{\rm dec})$ 
for different $z = M/\mu_{\rm dec}$ and $c$. 
In Section~\ref{sec:limits} we gave a detailed description of the scaling violations in the framework of the Symanzik
effective theory. In particular, we explained the absence of corrections $\sim a^2M\mu_{\rm dec}$. 
Still, continuum extrapolations are difficult due to the complicated
functional form, eq.~(\ref{eq:asquare_sublead}), of the $\rmO(a^2)$ corrections. 
Even when the leading anomalous dimensions are known (see
Section~\ref{sec:corr-oa2m2-oa2}) there are 
higher order corrections in $a$ and in $\alpha(1/a)$. Hence, in practice, 
the extrapolations have to be approached from an empirical point of view. 
The effect of the different logarithmic corrections can be explored by
varying the values of their exponents, see below.
Our values for $a\mu_{\rm dec}$ are very small, but having large 
masses we expect to have significant cutoff effects of $\rmO((aM)^2)$. 
Different cuts in $aM$ will thus be used to 
test the different assumptions regarding logarithmic corrections and
higher order terms.

\begin{figure}[ht]
  \centering
  \begin{subfigure}{0.475\textwidth}
    \includegraphics[width=\textwidth]{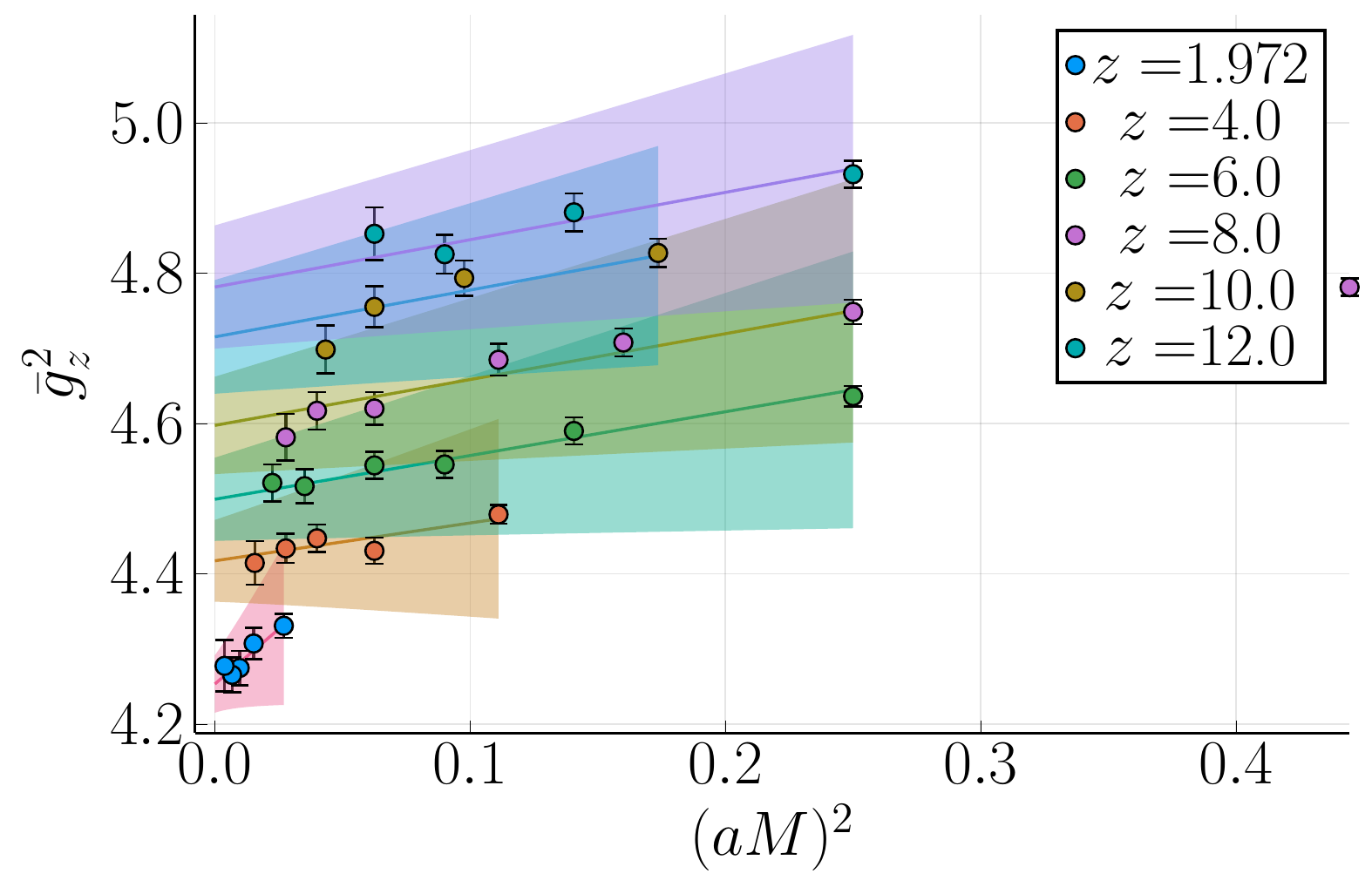}
    \caption{$(aM)^2<0.25$.}
  \end{subfigure}
  \begin{subfigure}{0.475\textwidth}
    \includegraphics[width=\textwidth]{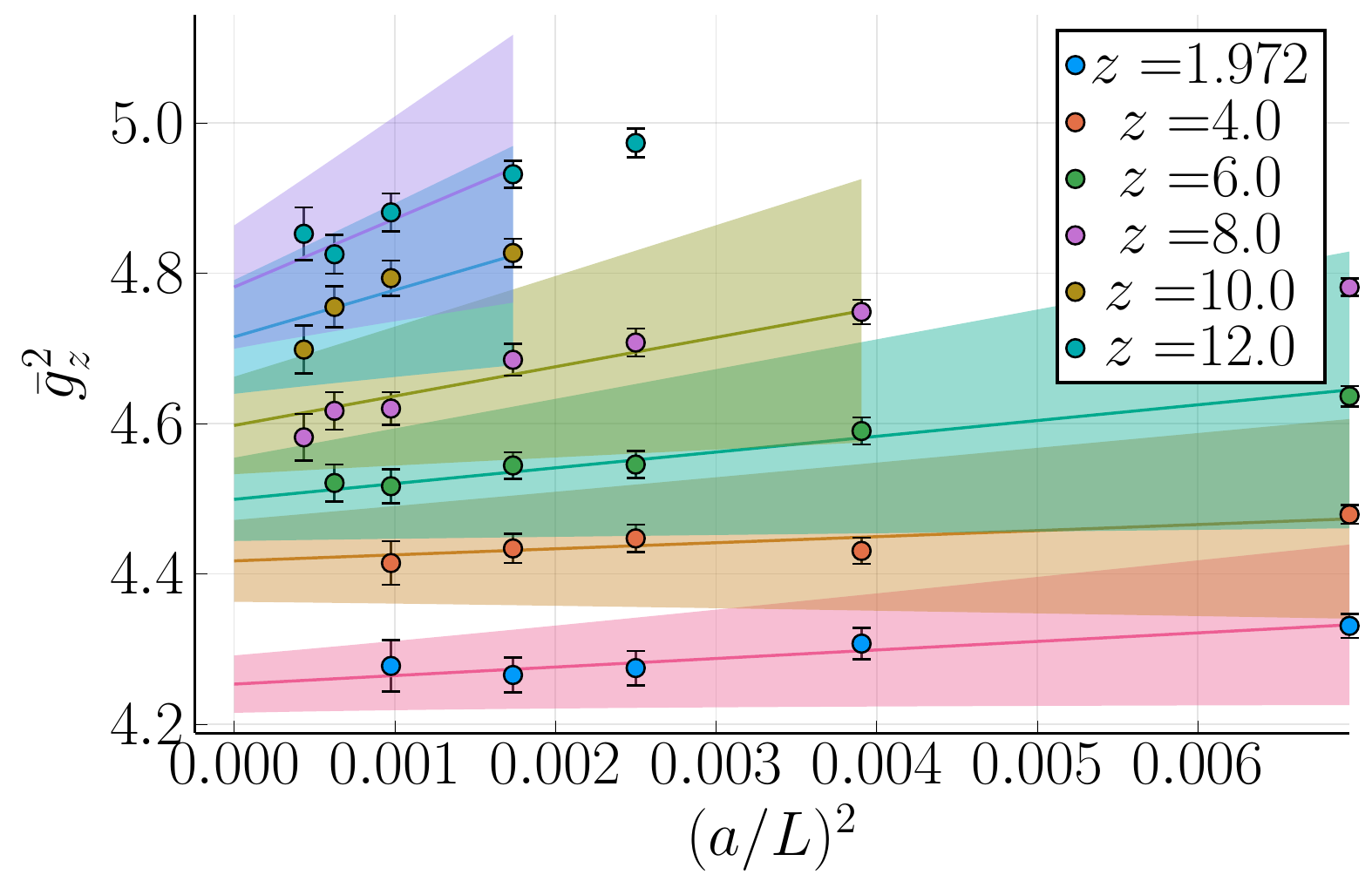}    
      \caption{$(aM)^2<0.25$.}
  \end{subfigure}

  \begin{subfigure}{0.475\textwidth}
  \includegraphics[width=\textwidth]{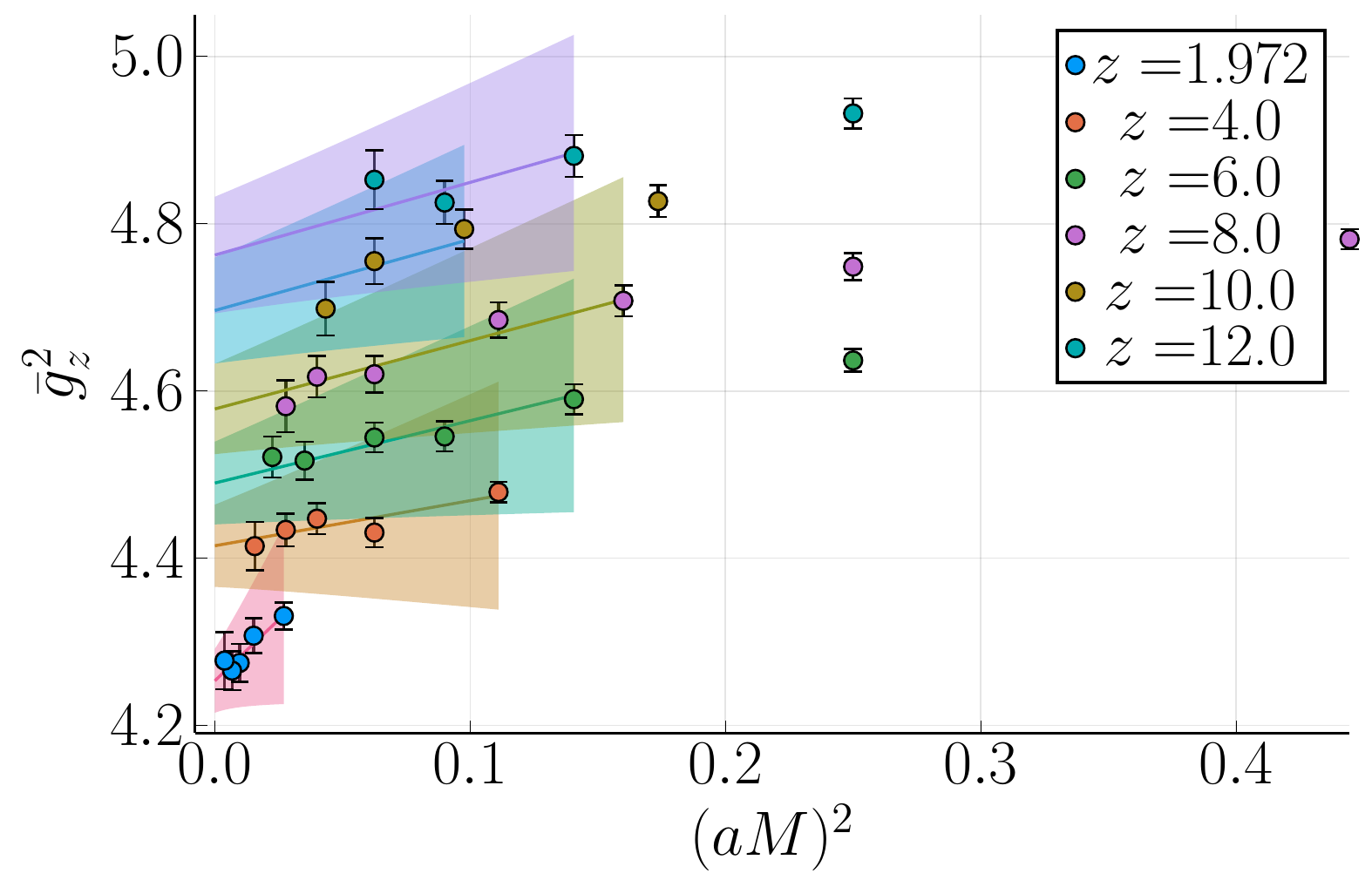}    
    \caption{$(aM)^2<0.16$.}

  \end{subfigure}
  \begin{subfigure}{0.475\textwidth}
  \includegraphics[width=\textwidth]{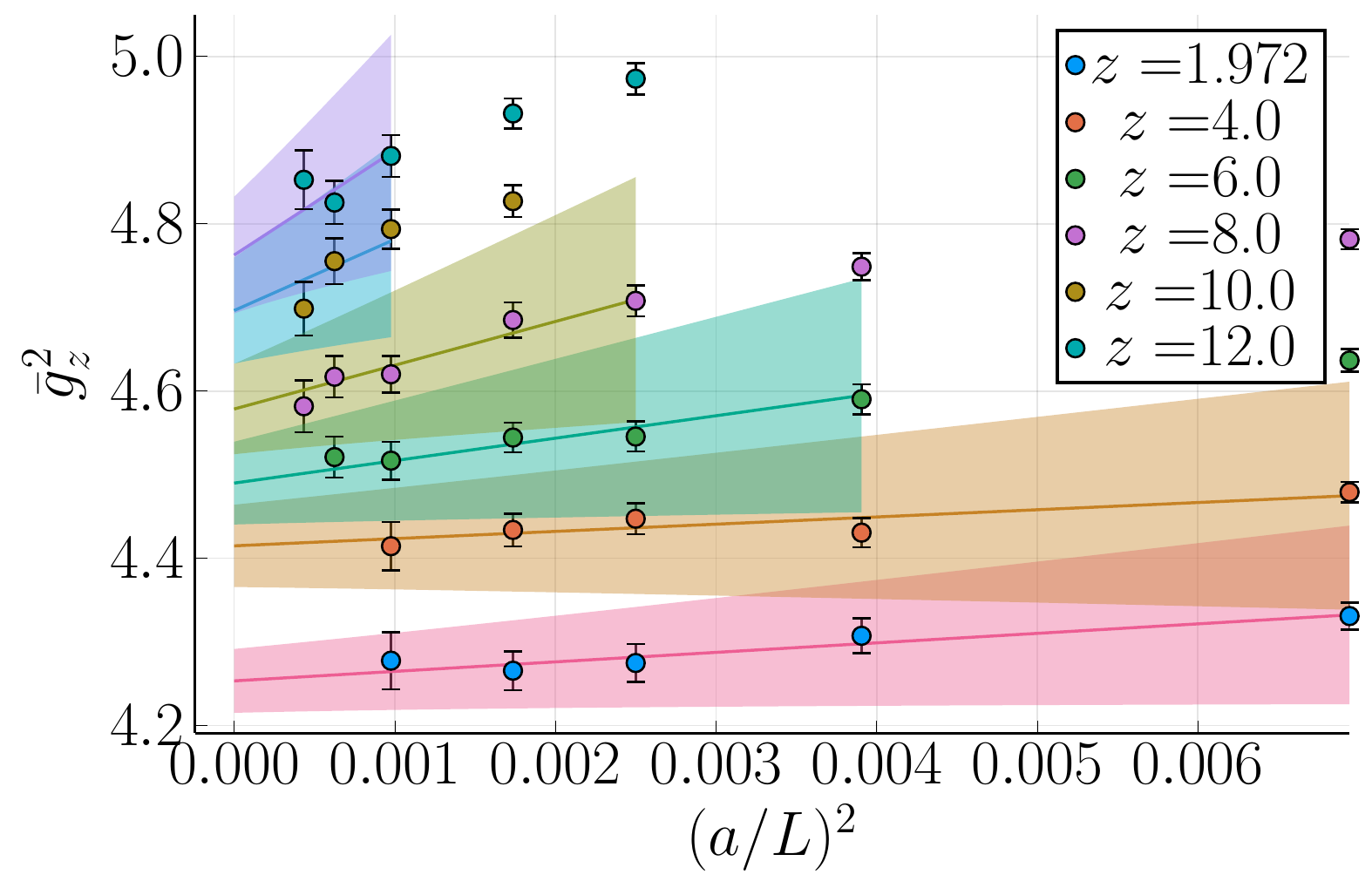}    
    \caption{$(aM)^2<0.16$.}
  \end{subfigure}
  \caption{Global continuum fit of our data for $c=0.3$ ($\bar{g}^2_z\equiv \bar{g}^2_{\rm 0.3}(z)$ in the plots)
    and two values of the mass cuts. Note that the assumed 100\% uncertainty 
    of $\bg$ is {\em not} included in the error bars of the points. However, 
    it is propagated into the uncertainties of the global fit shown by the 
    shaded areas.} 
  \label{fig:aextr030}
\end{figure}

Given these considerations, we opt for two approaches to
obtain the continuum coupling $\bar g^2_{\rm c}(z_i,0) \equiv \bar g^2_{\rm GFT,c}(\mu_{\rm
  dec},M,0)$ from the values $\bar g^2_{\rm c}(z_i,a) \equiv \bar g^2_{\rm GFT,c}(\mu_{\rm dec},M, a\mu_{\rm dec})$ at non-zero lattice spacing.%
\footnote{Note that in the following we shall often use the more compact notation $\bar g^2_{\rm c}(z)$ for 
the massive coupling. Whether we are referring to the coupling at finite $a$ or in the continuum should be clear from the context.}
\begin{description}
\item[Extrapolations at fixed $z$:] The measured values of $\bar g^2_{\rm c}(z_i,a)$ for each value of
  $z_i=M_i/\mu_{\rm dec}$ are extrapolated with the ansatz
  \begin{equation}
    \label{eq:ubyu}
    \bar g^2_{\rm c}(z_i,a) = 
    C_i(c) + p_i(c) \,
    [\alpha_\msbar(a^{-1})]^{\hat \Gamma} {(a\mu_{\rm dec})^2}\,,
  \end{equation}
  where $C_i(c),\, p_i(c)$ are independent fit parameters for each value 
  $z_i$ (with the continuum limits being $\bar g^2_{\rm c}(z_i,0)=C_i(c)$), and we use
  $\hat \Gamma\in[-1,1]$.  
  
\item[Global extrapolations: ] The measurements of the coupling for all $z_i,a\mu_{\rm
    dec}$ at a fixed $c$ are
  combined in a single fit using the  ansatz
  \begin{equation}
    \label{eq:global}
    \bar g^2_{\rm c}(z_i,a) =
    C_i(c) +
    p_1(c) [\alpha_\msbar(a^{-1})]^{\hat \Gamma} ({a\mu_{\rm dec}})^2 +
    p_2(c) [\alpha_\msbar(a^{-1})]^{\hat \Gamma'} (aM_i)^2 \,.
  \end{equation}
  In this case the fit parameters are the continuum values $C_i$ and
  the two parameters $p_{1,2}$, while we consider $\hat\Gamma\in[-1,1]$, and $\hat\Gamma'\in[-1/9,1]$.
  This simple form is the result of expanding the $a^2$-terms of the Symanzik effective 
  theory in $1/M$ and dropping $\rmO(1/M^2)$ corrections  
  (see Section~\ref{sec:corr-oa2m2-oa2}). We therefore
  need to check which values of $z$ are large enough to be included in
  the global fit.   
\end{description}

Due to the shifts to the proper LCP, the data are slightly
correlated across different values of $z$. We performed uncorrelated
fits, but judged the quality of the fits  
from the value of $\chi^2$ computed from the known covariance matrix~\cite{chiexp:inprep}, which however 
was in all cases very close to the naive number of d.o.f. 
Using data with $(aM)^2 > 0.35$ leads to
biased results and fits with bad $\chi^2$. Therefore we use only two
mass cuts $(aM)^2 \le 0.25, 0.16$ in the following analysis.

\begin{figure}[ht]
  \centering
  \begin{subfigure}{0.475\textwidth}
    \includegraphics[width=\textwidth]{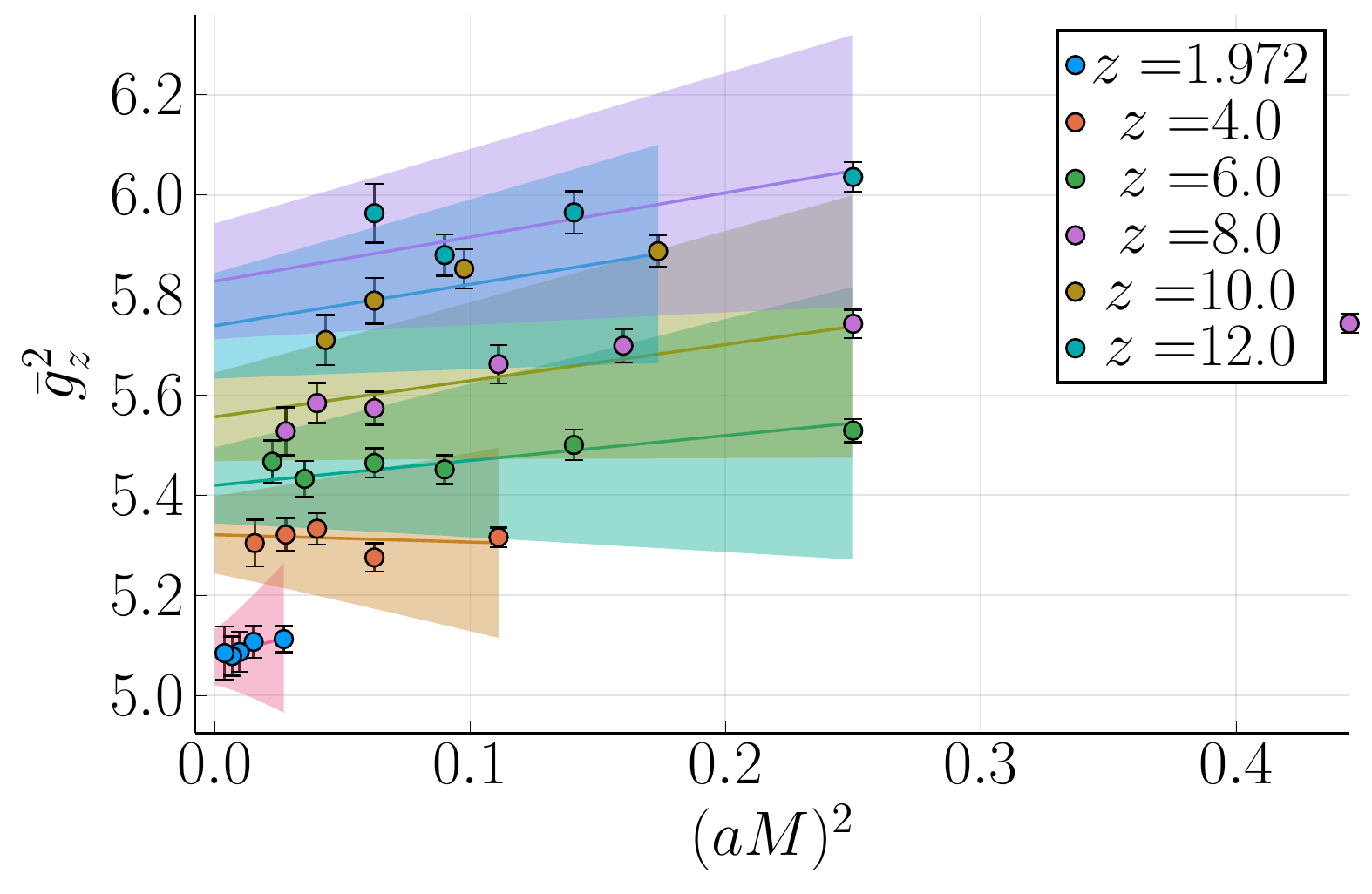}
    \caption{$(aM)^2<0.25$.}
  \end{subfigure}
  \begin{subfigure}{0.475\textwidth}
    \includegraphics[width=\textwidth]{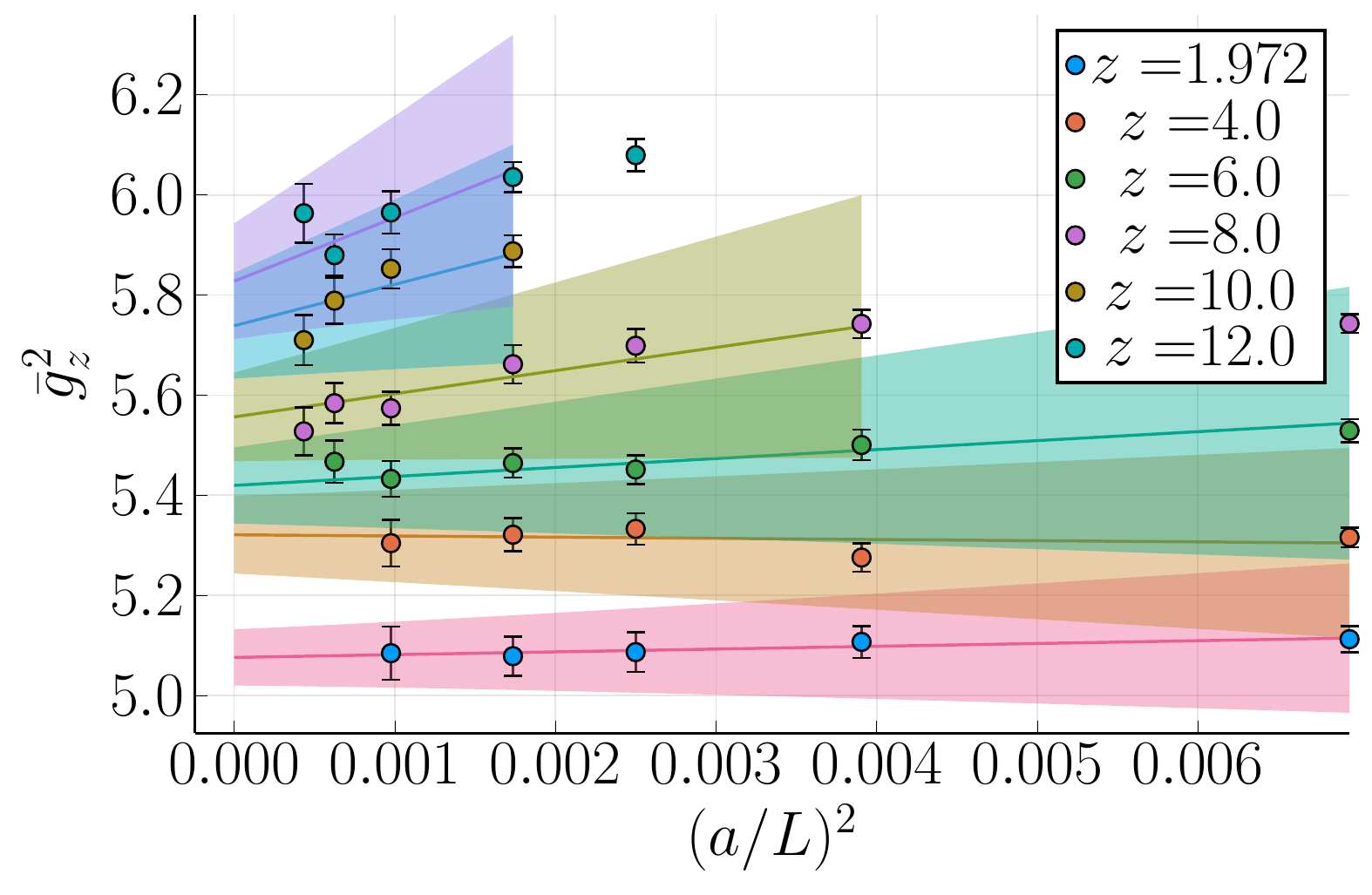}    
      \caption{$(aM)^2<0.25$.}
  \end{subfigure}

  \begin{subfigure}{0.475\textwidth}
  \includegraphics[width=\textwidth]{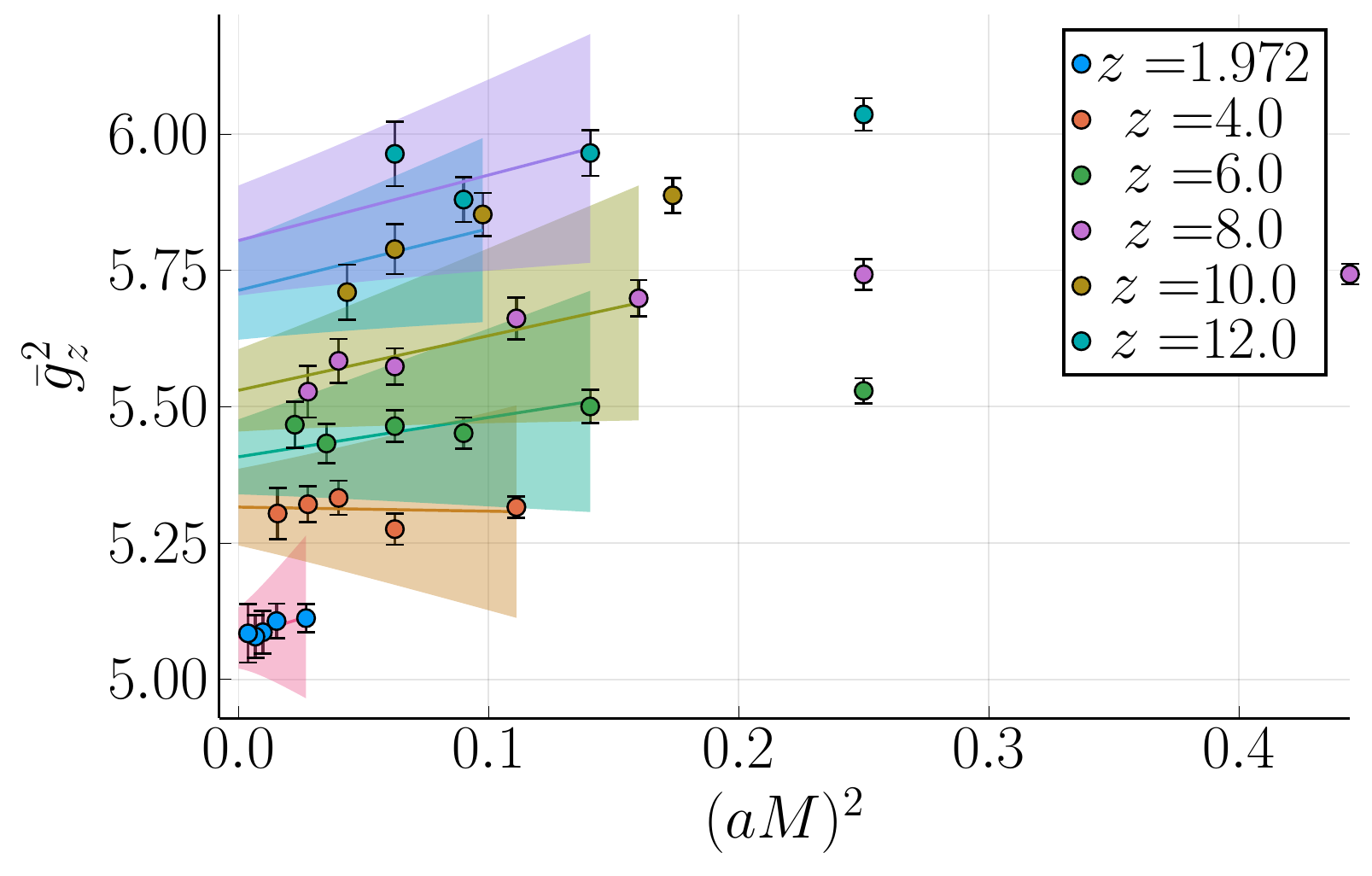}    
    \caption{$(aM)^2<0.16$.}

  \end{subfigure}
  \begin{subfigure}{0.475\textwidth}
  \includegraphics[width=\textwidth]{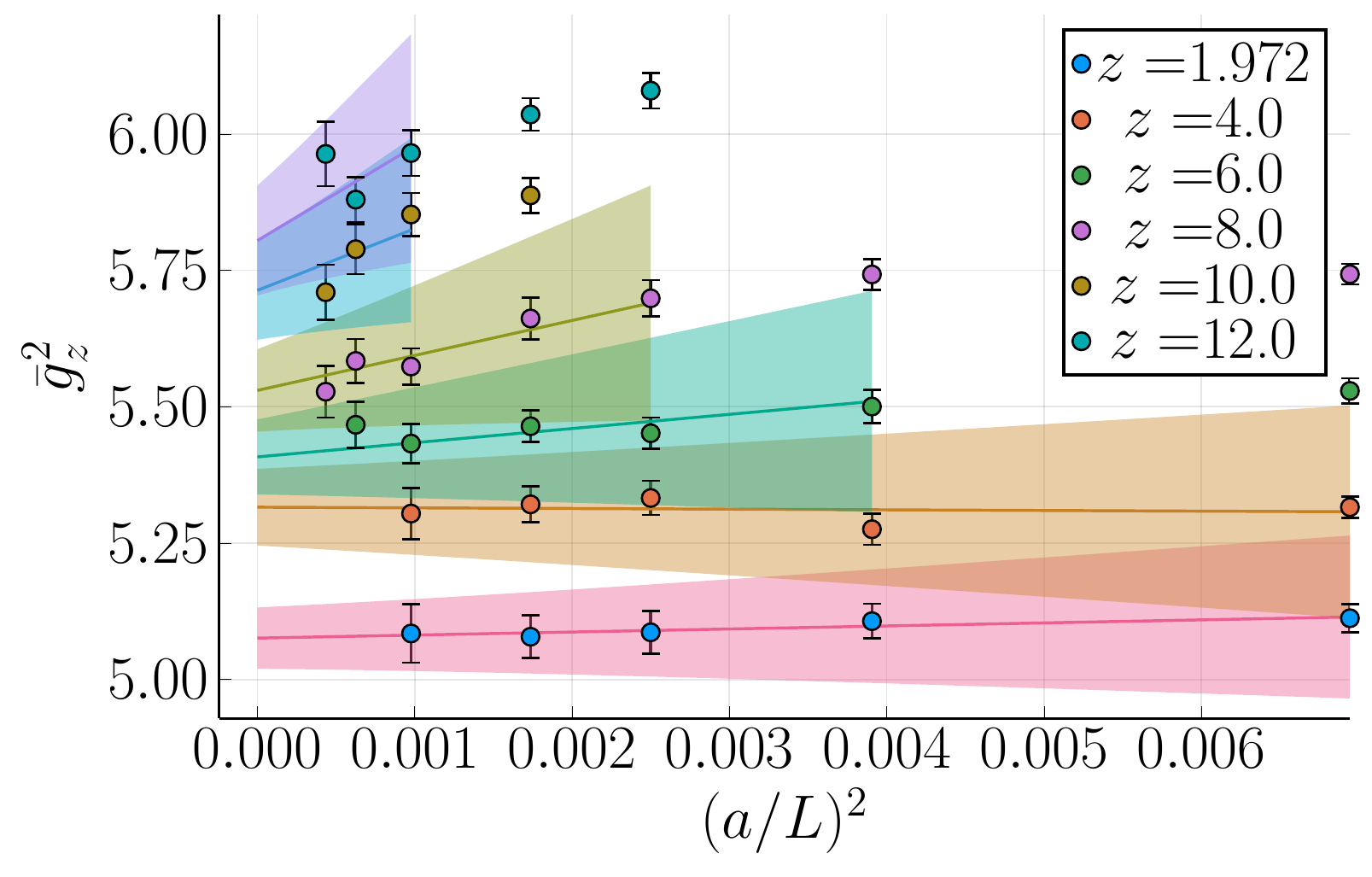}    
    \caption{$(aM)^2<0.16$.}
  \end{subfigure}
\caption{Comparison of global and fixed-$z$ extrapolations for
  $c=0.36$. Details as in figure \ref{fig:aextr030}. }
  \label{fig:aextr036}
\end{figure}

Figures~\ref{fig:aextr030}, \ref{fig:aextr036} show the different
extrapolations for $c=0.30$ and $c=0.36$, respectively. We make the
following observations concerning the fits.  
\begin{itemize}
\item Discretization effects proportional to $(a\mu_{\rm dec})^2$ are very small. 
  For the case $c=0.30$ the fit coefficient $p_1$ in the global
  analysis is very   small, well compatible with zero.  
  For $c\ne 0.3$, these scaling 
  violations are slightly larger, but still all our lattices are large
  enough to be included in the fit. This justifies that we will only
  discuss cuts in $aM$. 

\item The data at $c=0.3$ shows a very different behavior in
$(aM)^2$ for our smallest value of the mass $z=1.972$ (see figure~\ref{fig:aextr030}). 
This suggests that $z>2$ is needed for the large mass expansion to be reliable. 
For $c=0.36$ (see figure~\ref{fig:aextr036}) the behavior in $(aM)^2$
even for $z=1.972$ looks well compatible with the behavior at $z\ge 4$. 
Since the effective decoupling scale is smaller in this case, the data
at $c=0.36$ is closer to the large mass limit. 
In any case, to be on the safe side, we only include in the global analysis
$z\ge 4$ for all values of $c$, while the data with $z=1.972$ is
always fitted with an independent slope.
\end{itemize}
 
The extrapolations at fixed values of $z$ and the global analysis always lead
to compatible results. 
Also the uncertainties of the continuum limits are very 
similar except for the case $z=12$, where the error in the extrapolation
at fixed $z$ (that only uses two points) is twice as large as the
result from the global analysis. Given that our global formula is
theoretically sound, particularly so at 
large values of $z$, we have no reason to suspect that the results of
the global analysis are not accurate for $z=12$. 
Figures~\ref{fig:aextr030},~\ref{fig:aextr036} shows the results of
the global analysis with two cuts $(aM)^2
\le 0.16,0.25$.  
Results are compatible, with the extrapolations using only data with
$(aM)^2 \le 0.16$ resulting in slightly larger uncertainties. 

We shall now discuss the logarithmic corrections to scaling. 
We have tried several values of $\hat \Gamma\in[-1,1], \hat \Gamma'\in[-1/9,1]$ in the
continuum extrapolations.  
We see deviations much smaller than our uncertainties.
Our analysis shows that the logarithmic corrections have little
influence in our case.\footnote{The statement rests on the simplified model that we use for the fits. We have chosen a single term with exponent $\hat \Gamma'$ in eq.~\eqref{e:symanzikfinal1} and a single combined term with exponent $\hat\Gamma$ in the combination of eqs.~\eqref{e:symanzikfinal2} and \eqref{eq:asquare_sublead}.} 
This can be understood from the fact that in our finite volume setup
we reach very small lattice spacings, i.e.~in the range $a\in[0.006,0.015]\, {\rm fm}$. 
At the high scales $1/a$ defined by these lattice spacings, the coupling runs very slowly, rendering the effect of
the logarithmic corrections very small. This feature should be considered a virtue of our  strategy.

These considerations lead us to quote as final values for the
continuum extrapolation the results of the global fit with $aM\le0.4$
and $\hat \Gamma = \hat \Gamma' = 0$. 
This particular analysis has larger or similar uncertainties than
other choices, and provides an excellent description of our data for
$z\ge 4$. 
Table~\ref{tab:cont_res} shows the data entering the analysis together with the
final results of the extrapolations. Thanks to the
use of large lattices, the continuum extrapolations are under reasonable control. The deviation of our finest lattice spacing data from the
continuum values is at most two standard deviations. 
Obviously this ``gap'' grows with increasing $z$. Given
the importance of large $z$ for the extrapolation of $\Lambda_{\overline{\rm   MS}}$ to $z\to\infty$, 
it would be worthwhile to close the gap further by simulating even larger 
values of $L/a$ when an improved overall precision is desired.

\begin{table}
  \centering
\small
\begin{tabular}{rrrrrrrrr}
  \toprule
  &         &           &        & \multicolumn{5}{c}{$\bar g^2_{\rm c}(z)$} \\
  \cmidrule(rl){5-9}
\(z\) & \(L/a\) & \(\beta\) & \(aM\) & \(c=0.3\) & \(c=0.33\) & \(c=0.36\) & \(c=0.39\) & \(c=0.42\)\\
\midrule
1.97 & 12 & 4.3177 & 0.1643 &        4.331(16) &        4.679(20) &        5.112(26) &        5.655(34) &               6.337(44)  \\
1.97 & 16 & 4.4770 & 0.1232 &        4.307(21) &        4.666(26) &        5.107(32) &        5.656(40) &               6.343(50)  \\
1.97 & 20 & 4.6102 & 0.0986 &        4.275(23) &        4.639(30) &        5.087(39) &        5.643(51) &               6.337(67)  \\
1.97 & 24 & 4.7235 & 0.0822 &        4.266(23) &        4.631(30) &        5.078(39) &        5.634(51) &               6.327(68)  \\
1.97 & 32 & 4.9051 & 0.0616 &        4.278(34) &        4.641(43) &        5.084(53) &        5.632(67) &               6.314(87)  \\
1.97 & \(\infty\) & -- & 0.0 &        4.253(38) &        4.624(46) &        5.076(56) &        5.634(69) &        6.327(86)  \\
\midrule
4.00 & 12 & 4.3337 & 0.3333 &        4.479(12) &        4.854(16) &        5.316(20) &        5.889(25) &               6.602(33)  \\
4.00 & 16 & 4.4886 & 0.2500 &        4.431(17) &        4.811(22) &        5.276(29) &        5.849(36) &               6.559(47)  \\
4.00 & 20 & 4.6192 & 0.2000 &        4.447(19) &        4.846(24) &        5.333(31) &        5.934(41) &               6.681(54)  \\
4.00 & 24 & 4.7308 & 0.1667 &        4.434(20) &        4.834(25) &        5.321(33) &        5.921(43) &               6.663(57)  \\
4.00 & 32 & 4.9104 & 0.1250 &        4.414(29) &        4.816(37) &        5.304(47) &        5.906(59) &               6.651(76)  \\
4.00 & \(\infty\) & -- & 0.0 &        4.415(49) &        4.822(59) &       5.316(70)  &        5.923(85) &               6.68(10) \\
&\multicolumn{3}{c}{$\bg$ uncertainty:} & \hfill  (45) &\hfill  (53) &\hfill  (62) &\hfill  (74) &\hfill  (9) \\
\midrule
6.00 & 12 & 4.3508 & 0.5000 &        4.636(14) &        5.038(17) &        5.529(23) &        6.134(30) &                6.881(39) \\
6.00 & 16 & 4.5008 & 0.3750 &        4.590(18) &        5.001(23) &        5.500(31) &        6.113(40) &                6.870(53) \\
6.00 & 20 & 4.6286 & 0.3000 &        4.546(18) &        4.955(22) &        5.451(29) &        6.061(37) &                6.814(48) \\
6.00 & 24 & 4.7383 & 0.2500 &        4.544(18) &        4.961(23) &        5.464(29) &        6.081(37) &                6.841(48) \\
6.00 & 32 & 4.9159 & 0.1875 &        4.517(23) &        4.932(28) &        5.433(36) &        6.046(45) &                6.803(62) \\
6.00 & 40 & 5.0624 & 0.1500 &        4.521(25) &        4.948(33) &        5.467(42) &        6.104(55) &                6.896(72) \\
6.00 & \(\infty\) & -- & 0.0 &        4.490(50) &        4.906(58) &        5.408(69) &        6.021(82) &        6.779(100)  \\
&\multicolumn{3}{c}{$\bg$ uncertainty:} & \hfill  (44) &\hfill  (52) &\hfill  (60) &\hfill  (71) &\hfill  (85) \\
\midrule
8.00 & 12 & 4.3694 & 0.6667 &        4.781(12) &        5.215(15) &        5.743(19) &        6.393(23) &               7.197(29)  \\
8.00 & 16 & 4.5139 & 0.5000 &        4.749(16) &        5.197(22) &        5.743(28) &        6.414(37) &               7.246(48)  \\
8.00 & 20 & 4.6384 & 0.4000 &        4.708(19) &        5.156(25) &        5.699(33) &        6.366(44) &               7.191(59)  \\
8.00 & 24 & 4.7462 & 0.3333 &        4.685(21) &        5.128(28) &        5.662(38) &        6.315(59) &               7.119(83)  \\
8.00 & 32 & 4.9215 & 0.2500 &        4.620(22) &        5.053(27) &        5.574(33) &        6.210(41) &               6.992(53)  \\
8.00 & 40 & 5.0669 & 0.2000 &        4.617(25) &        5.055(31) &        5.584(40) &        6.227(52) &               7.023(68)  \\
8.00 & 48 & 5.1880 & 0.2000 &        4.582(31) &        5.012(38) &        5.528(48) &        6.156(59) &               6.926(74)  \\
8.00 & \(\infty\) & -- & 0.0 &        4.578(54) &        5.011(64) &        5.530(76) &        6.161(90) &          6.94(11)  \\
&\multicolumn{3}{c}{$\bg$ uncertainty:} & \hfill  (45) &\hfill  (53) &\hfill  (61) &\hfill  (73) &\hfill  (9) \\
\midrule
10.00 & 24 & 4.7545 & 0.5000 &        4.827(19) &        5.307(24) &        5.887(32) &        6.601(42) &               7.480(55)  \\
10.00 & 32 & 4.9274 & 0.3750 &        4.794(23) &        5.273(30) &        5.852(39) &        6.564(52) &               7.443(68)  \\
10.00 & 40 & 5.0710 & 0.3000 &        4.755(27) &        5.224(35) &        5.789(46) &        6.472(59) &               7.312(79)  \\
10.00 & 48 & 5.1916 & 0.3000 &        4.698(32) &        5.157(40) &        5.710(50) &        6.389(63) &               7.228(83)  \\
10.00 & \(\infty\) & -- & 0.0 &        4.696(63) &        5.157(76) &        5.713(90) &         6.39(11) &          7.23(14) \\
&\multicolumn{3}{c}{$\bg$ uncertainty:} & \hfill  (52) &\hfill  (61) & \hfill  (72) &\hfill  (9) &\hfill  (10)  \\
\midrule
12.00 & 20 & 4.6596 & 0.6000 &        4.973(19) &        5.475(25) &        6.079(32) &        6.823(42) &               7.739(55)  \\
12.00 & 24 & 4.7630 & 0.5000 &        4.932(18) &        5.432(23) &        6.036(30) &        6.778(38) &               7.694(49)  \\
12.00 & 32 & 4.9335 & 0.3750 &        4.881(25) &        5.372(32) &        5.965(42) &        6.690(55) &               7.584(71)  \\
12.00 & 40 & 5.0760 & 0.3000 &        4.825(26) &        5.304(32) &        5.880(41) &        6.576(52) &               7.437(66)  \\
12.00 & 48 & 5.1953 & 0.3000 &        4.852(35) &        5.355(45) &        5.963(59) &        6.711(76) &                7.64(10)  \\
12.00 & \(\infty\) & -- & 0.0 &        4.762(70) &        5.235(84) &         5.80(10) &         6.50(12) &         7.35(15) \\
&\multicolumn{3}{c}{$\bg$ uncertainty:} & \hfill  (54) & \hfill  (64) &\hfill  (74) &\hfill  (9) &\hfill  (11) \\
\bottomrule
\end{tabular}

   \caption{Values of the massive coupling $\bar g^2_{\rm c}(z)$ and its
    continuum extrapolated values. 
  The results quoted for the continuum extrapolations correspond to a
  global fit of the data with $z\ge 4$ and 
  $(aM)^2\le 0.16$ and fixing $\hat \Gamma = \hat \Gamma' = 0$. 
At finite lattice spacings, the uncertainty in $\bg$ is omitted, but the continuum values include it. 
We also given just the $\bg$-uncertainty with 100\% correlation across all data. }
  \label{tab:cont_res}
\end{table}

\subsection{Large mass extrapolations and the determination of
  $\Lambda^{(3)}_{\overline{\rm MS}}$}

\label{sec:large-mass-extr}

\subsubsection{Estimates of the three flavor $\Lambda$-parameter}
\label{sec:estim-three-flav}

\begin{table}
  \centering
\footnotesize
\begin{tabular}{lllllllll}
  \toprule
  &\multicolumn{4}{c}{$c=0.3$} & \multicolumn{4}{c}{$c=0.36$}\\
  \cmidrule(lr){2-5}  \cmidrule(lr){6-9}
  $z$ & $\bar g^2_{\rm c}(z)$ & $[\bar{g}^{(0)}_\text{GF}(\mudec)]^2$ & $\rho$ & $\Lambda^{(3)}_{\overline{\rm MS},\rm eff}$ &  $\bar g^2_{\rm c}(z)$ & $[\bar{g}^{(0)}_\text{GF}(\mudec)]^2$ & $\rho$ & $\Lambda^{(3)}_{\overline{\rm MS},\rm eff }$\\
  \cmidrule(lr){1-1} \cmidrule(lr){2-5}  \cmidrule(lr){6-9}
1.972&   4.253(38) &        3.962(33)  &        0.547(14) &  432(14) &        5.076(56)  &        3.935(36) &              0.540(14)   &          426(14) \\
  4  &   4.415(49) &        4.101(42)  &        0.496(13) &  391(13) &        5.316(70)  &        4.084(44) &              0.492(14)   &          388(13) \\
  6  &   4.490(50) &        4.165(43)  &        0.465(12) &  367(12) &        5.408(69)  &        4.140(42) &              0.460(12)   &          363(12) \\
  8  &   4.578(54) &        4.241(46)  &        0.450(12) &  355(12) &        5.530(76)  &        4.215(46) &              0.445(12)   &          351(12) \\
  10 &   4.696(63) &        4.341(54)  &        0.446(13) &  352(12) &        5.713(90)  &        4.325(54) &              0.443(13)  &          349(12) \\
  12 &   4.762(70) &        4.397(59)  &        0.438(13) &  345(12) &         5.80(10)  &        4.379(60) &              0.434(13)  &          343(12) \\
  \bottomrule
  \end{tabular}

   \caption{The massive couplings, $\bar g^2_{\rm c}(z)$, together with the associated
    pure gauge coupling, $[\bar{g}^{(0)}_\text{GF}(\mudec)]^2$, after a non-perturbative matching to the
    scheme with $T=L, c=0.3$. The 
    coupling $[\bar{g}^{(0)}_\text{GF}(\mudec)]^2$ is  used to obtain $\rho = \Lambda^{(3)}_{\overline{\rm MS},\rm eff }/\mu_{\rm dec}$ and     
    $\Lambda^{(3)}_{\overline{\rm MS},\rm eff }$, which is $\Lambda^{(3)}_{\overline{\rm MS}}$ up to power
    corrections in $1/M$. We show results for two representative values of $c=0.3, 0.36$.}
  \label{tab:largeM}
\end{table}

Once the values of the massive coupling $\bar g^2_{\rm GFT,c}(\mu_{\rm
  dec}, M)$ are known in the continuum,
decoupling tells us that the values of these couplings are the same as
in the pure gauge theory, up to heavy mass corrections. In order to make use of 
this together with the results of ref.~\cite{DallaBrida:2019wur}, 
we first need to match our coupling to the GF coupling definition of ref.~\cite{DallaBrida:2019wur}. 
The difference is our choice of $T=2L$ as well as  of
$c$-values in the massive coupling $\bar g^2_{\rm GFT,c}(\mu_{\rm  dec}, M)$  (cf.~Section~\ref{sec:note-diff-valu}), 
compared to the choice $T=L$ and $c=0.3$ made in the pure gauge theory~\cite{DallaBrida:2019wur}. 
The two different schemes can be matched non-perturbatively in the pure gauge theory.
There, the couplings $\bar{g}^{(0)}_{\rm GF}(\mu)$ (with
$T=L$ and $c=0.3$) and $\bar{g}^{(0)}_{\rm GFT, c}(\mu)$ (with $T=2L$ and
and arbitrary $c$) are related by
\begin{equation}
\bar{g}^{(0)}_{\rm GF }(\mu) = \chi_{\rm c}\left(\bar{g}^{(0)}_{\rm GFT,c }(\mu) \right) \,.
\end{equation}
The functions $\chi_{\rm c}$ for the relevant values of $c=0.30,0.33,0.36,0.39,0.42$, 
are precisely determined as described in Appendix~\ref{subsec:MatchingYM}.

We define $\bar{g}^{(0)}_\text{GF}(\mudec)$ as the values of the pure gauge coupling ($T=L$,
$c=0.3$) that correspond to the values of the massive coupling
extrapolated to the continuum, i.e. (cf.~table~\ref{tab:cont_res})
\begin{equation}
 	\bar{g}^{(0)}_\text{GF}(\mudec) = \chi_{\rm c}\left(\bar g^{(3)}_{\rm GFT,c}(\mu_{\rm dec}, M)\right)\,.
\end{equation}
Pure gauge theory results for the function $\varphi_{\rm GF}^{(0)}$ (see
Appendix~\ref{subsec:RunningYM}) then yield values for
\begin{equation}
  \frac{\Lambda^{(0)}_{\overline{\rm MS}}}{\mu_{\rm dec}} =
  \frac{\Lambda^{(0)}_{\overline{\rm MS}}}{\Lambda^{(0)}_\text{GF}} \varphi_{\rm GF}^{(0)}(\bar{g}^{(0)}_\text{GF}(\mudec))\,. 
\end{equation}
Since $z=M/\mu_{\rm dec}$ is a known input, the non-linear equation (cf.~eq.~(\ref{eq:basic}))
\begin{equation}
  \rho
  P \left( z/\rho\right) =
  \frac{\Lambda^{(0)}_{\overline{\rm MS} }}{\mu_{\rm dec}}
\end{equation}
allows us to determine $\rho = \Lambda^{(3)}_{\overline{\rm MS},\rm eff }/\mu_{\rm dec} = \Lambda^{(3)}_{\overline{\rm MS} }/\mu_{\rm dec}+\rmO(1/z)$, see table~\ref{tab:largeM}. 
With $\mu_{\rm dec} =          789(15)$ MeV
obtained in $N_{\rm f} = 3$ QCD~\cite{DallaBrida:2016kgh}, we convert these ratios to the effective three flavor $\Lambda$-parameter, again equal to $\Lambda^{(3)}_{\overline{\rm   MS}}$
up to $\rmO(1/M)$ corrections. Results are also listed in table~\ref{tab:largeM}.

\subsubsection{$M\to\infty$ extrapolation}
\label{sec:mtoinfty-extr}

According to the discussion in Section~\ref{sec:limits} we expect the
estimates of $\Lambda^{(3)}_{\overline{\rm MS} ,\rm eff}$ of
table~\ref{tab:largeM} to approach $\Lambda^{(3)}_{\overline{\rm MS}}$ with power corrections of the
form $z^{-k}$, accompanied by logarithmic corrections. 
The function $P$ is approximated by high order perturbation theory.
Since the used masses $m_\star$ are large, the associated
$\rmO(\alpha^4_{\overline{\rm MS} }(m_\star))$ uncertainties can 
be neglected.
Linear terms of $\rmO(z^{-1})$ are a consequence of our  boundary
conditions. The choice $T=2L$ suppresses their effects to a level below our statistical precision, as we were able to show
by an explicit computation (cf.~Appendix~\ref{app:boundarycontribution}). 
We therefore assume leading $1/z^2$ corrections, with logarithmic corrections as discussed in 
Section~\ref{subsec:decouplinglogcorrections}. In practice we fit the parameters $A,\,B$ in
\begin{equation}
  \label{eq:mtoinfty}
  \Lambda^{(3)}_{\overline{\rm MS}  ,\rm eff} = A + \frac{B}{z^2}[\alpha(m_\star)]^{\hat\Gamma_{m}}\,,
\end{equation}
to the data, in order to obtain $\Lambda^{(3)}_{\overline{\rm MS}}=A$. Since the leading exponent $\hat \Gamma_{m}$ is
presently not known, we  vary it in a reasonable range $\hat \Gamma_{m} \in [0,1]$ (cf.~Section~\ref{subsec:decouplinglogcorrections}).

The first issue that we have to deal with is what values of $z$ are
included in this extrapolation. 
Part of the difficulty here is that the estimates of
$\Lambda^{(3)}_{\overline{\rm MS} }$ coming from different values of
$c,z$ are very correlated. Correlations are due to many sources:  $\bg$, the running in the pure gauge theory, the scale $\mudec$, all enter in the same way for all $c,z$. There are also less obvious correlations. E.g.
the global fit performed to obtain the continuum limit has common parameters $p_1,p_2$ describing the cutoff effects. All of these correlations are precisely known -- they do not involve difficult-to-estimate correlation matrices from Monte Carlo chains. 

\begin{figure}
  \centering
  \includegraphics[width=0.9\textwidth]{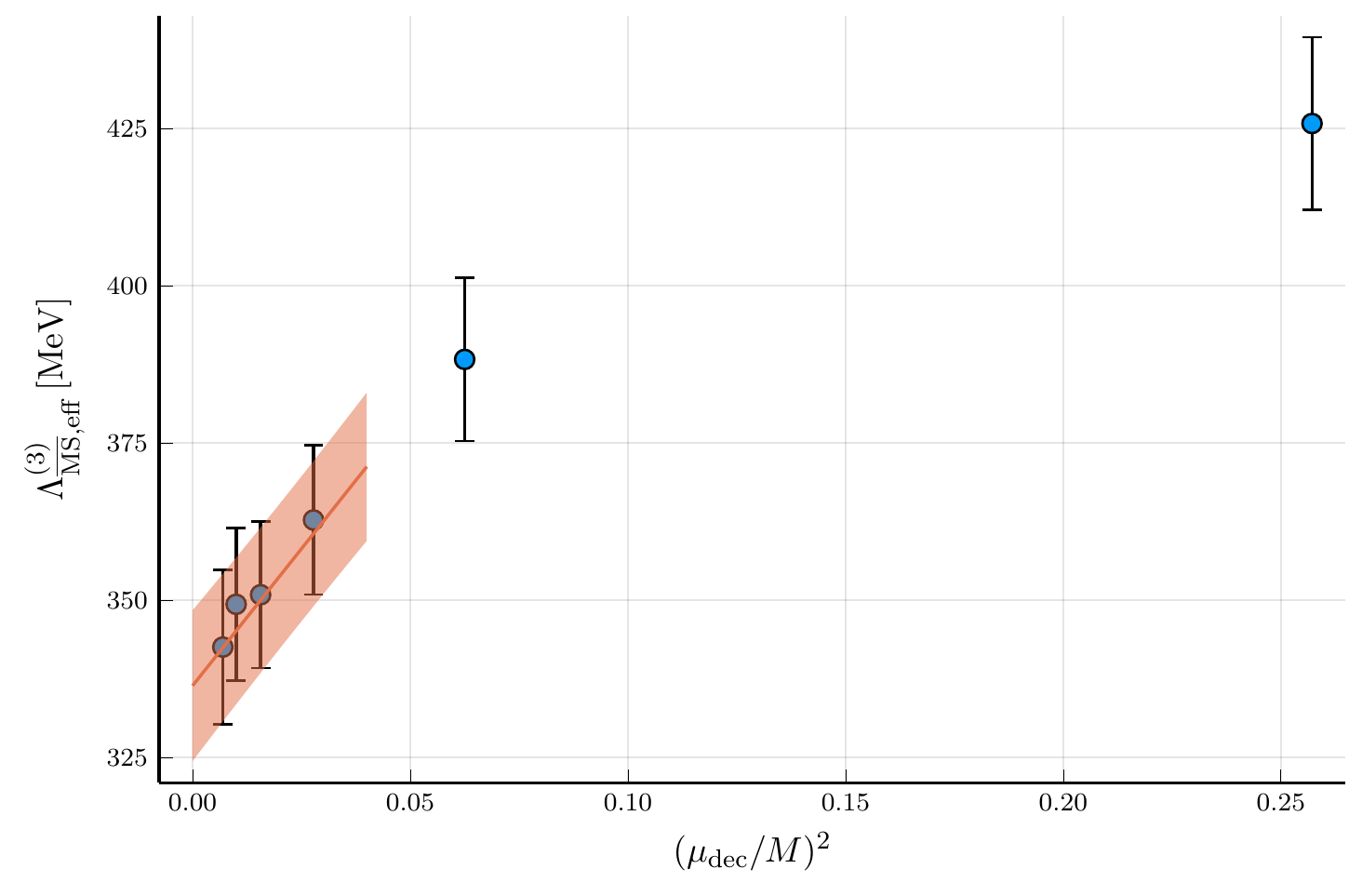} 
  \caption{Values for $\Lambda^{(3)}_{\overline{\rm MS},\rm eff }$ from \tab{tab:largeM} ($c=0.36$) and their extrapolation $M\to\infty$ using~\eq{eq:mtoinfty} with $\hat\Gamma_{ m}=0$.  \label{fig:mtoinfty}}
\end{figure}

We therefore performed correlated fits to \eq{eq:mtoinfty}.  Visually they all look very good; an example is displayed in \fig{fig:mtoinfty}. 
The $\chi^2$-values  are found above the numbers of {\rm d.o.f.}, but the quality of fit, Q, reported in table \ref{tab:final_results},  is generally good enough. Only fits including $z=4$ and the smallest values of $c$ are statistically discouraged.
As a precaution against higher order
corrections (i.e. 
$\rmO(z^{-3})$, etc.) we exclude the $z=4$ data also for the larger values of $c$ and use $c=0.36,\,z\geq 6$ as our central result.
Note that the Q-value is relatively small for the $z\geq8$ fits since they only contain one degree of freedom. The fact that Q becomes better including more data is supporting our 
choice of the $z\geq6$ fits.

As a check of this analysis we also performed uncorrelated fits, computed their 
Q-value from the known covariance matrix~\cite{chiexp:inprep} and found entirely
consistent results. 
\begin{table}[t!]
  \centering
  \begin{tabular}{llrllrllrr}
    \toprule
    \multicolumn{3}{c}{$z\ge 4$} & \multicolumn{3}{c}{$z\ge 6$} 
    & \multicolumn{3}{c}{$z\ge 8$}\\
    \cmidrule(lr){1-3} \cmidrule(lr){4-6} \cmidrule(lr){7-9}%
    $c$ & $\Lambda^{(3)}_{\overline{\rm MS}}$ & Q [\%] 
    & $c$ & $\Lambda^{(3)}_{\overline{\rm MS}}$ & Q [\%]
    & $c$ & $\Lambda^{(3)}_{\overline{\rm MS}}$ & Q [\%]
    \\
    \cmidrule(lr){1-3} \cmidrule(lr){4-6} \cmidrule(lr){7-9}
      0.30 &               349(11) &      2 &
      0.30 &               340(12) &     11 &
      0.30 &               338(13) &      4 
  \\ %
      0.33 &               345(11) &      8  &
      0.33 &               338(12) &     13  &
      0.33 &               338(13) &      4  
  \\ %
      0.36 &               342(11) &     16 &
      0.36 &               336(12) &     16 &
      0.36 &               338(13) &      6 
  \\ %
      0.39 &               339(11) &     21 &
      0.39 &               335(12) &     16 &
      0.39 &               338(13) &      7 
  \\ %
      0.42 &               336(11) &     23 &
      0.42 &               333(12) &     15 &
      0.42 &               337(13) &      7 
\\
\bottomrule
  \end{tabular}
  \caption{Estimates of $\Lambda^{(3)}_{\overline{\rm MS},\rm eff}$ (see
    table~\ref{tab:largeM}) are extrapolated to $M\to\infty$ according
  to eq.~(\ref{eq:mtoinfty}) with $\hat\Gamma_{m} = 0$. 
}
  \label{tab:final_results}
\end{table}

We  now proceed to investigate the effect of the logarithmic corrections. 
Fits with $\hat\Gamma_{m} =1$ yield only about 3~MeV higher values
for $\Lambda$ when the $z=4$ data is excluded. Further excluding 
also $z=6$ reduces these shifts to only 1-2~MeV.
We take the result with $z\ge 6$ and $c=0.36$ as our
final result, and add 3 MeV as our
estimate of the systematic effect due to the logarithmic 
corrections or higher orders in $1/M$ in the $M\to\infty$ extrapolation, see figure \ref{fig:mtoinfty}.

Taking all these points into account, we quote as our final
result
\begin{equation}
  \label{eq:finla_lam}
  \Lambda^{(3)}_{\overline{\rm MS} } = 336(10)(6)_{b_\mathrm{g}}(3)_{\Gamma_{m}}\, {\rm MeV} = 336(12)\, {\rm MeV}\,.
\end{equation}
Here the first error is statistical, the second is due to $\bg$ and the third results from the estimated uncertainty in the $z$-extrapolation. The combined error 
covers all central results that we obtained by varying the cuts in
$z$, $(aM)^2$, and the different $\hat\Gamma_{m}$ except for two  cases. These extreme cases have small $c\leq0.33$ and include $z=4$ data, where corrections to decoupling are expected to be the largest. They yield Q-values below 2\%. 

We further note that there is a significant  correlation 
of the above statistical error with the one of the previous work~\cite{Bruno:2017gxd},
\begin{equation}
	\Lambda^{(3)}_{\overline{\rm MS} } = 341(12)\,\mathrm{MeV}\, \label{eq:alpha_lam}\,,
\end{equation}
using step scaling in the three-flavor theory up to high energy. The common piece is exactly the scale $\mudec=789(15)\,\mathrm{MeV}$. The off-diagonal element of the covariance matrix of the
two determinations is
\begin{equation}
	\mathrm{Cov}(\eqref{eq:finla_lam},\eqref{eq:alpha_lam}) = 41\,\mathrm{MeV}^2 \,,
\end{equation}
compared to the diagonal ones of $144\,\mathrm{MeV}^2$, which at present happen to be about the same for each of the individual determinations. As a quantitative measurement of the compatibility of the two different determinations we note that
their difference is not significant at all: $\Lambda\eqref{eq:alpha_lam}-\Lambda\eqref{eq:finla_lam} = 5(14)\,\MeV$.

\section{Result for $\alpha_s(m_Z)$}
\label{sec:alphas}

Our result for $\Lambda^{(3)}_{\overline{\rm MS} }$
(eq.~(\ref{eq:finla_lam})) can be translated, after running across the
charm and bottom quark thresholds, into a value of the four and five flavor
$\Lambda$-parameter. Using the FLAG
values~\cite{Aoki:2021kgd} (based
on~\cite{McNeile:2010ji,Yang:2014sea,Nakayama:2016atf,Petreczky:2019ozv}) 
$m_{\rm c, \star} = 1275(5)$ MeV, 
$m_{\rm b, \star} = 4171(20)$ MeV for the charm and bottom quark
mass thresholds\footnote{The uncertainties in the quark and $Z$-boson
  masses are negligible in all quoted results.}, we obtain
the following values for the four  
and five flavor $\Lambda$-parameters
\begin{eqnarray}
  \Lambda^{(4)}_{\overline{\rm MS} } &=& 294(10)(6)_{b_{\rm g}}(3)_{\Gamma_{ m}}(0.7)_{3\to 4,{\rm PT}}(1)_{3\to 4,{\rm NP}}\, {\rm MeV} = 294(12)\, {\rm MeV}\,, \\
  \Lambda^{(5)}_{\overline{\rm MS} } &=& 211.3(8.1)(5.0)_{b_{\rm g}}(2.4)_{\Gamma_{ m}}(0.7)_{3\to 5,{\rm PT}}(0.8)_{3\to 5,{\rm NP}}\, {\rm MeV} = 211.3(9.8)\, {\rm MeV}\,.
\end{eqnarray}
where the first error is statistical, and the second represents the
uncertainty associated with the logarithmic corrections in the limit
$M\to\infty$ (see Section~\ref{sec:large-mass-extr}). 
The last two errors come instead from crossing the charm and bottom
thresholds: first a perturbative error (determined by taking
the difference in the decoupling relations and RG functions between
the last two known orders), and 
second an estimate of 0.3\% in $\Lambda^{(3)}_{\overline{\rm MS}}$ for
the non-perturbative corrections in the 
decoupling of the charm~\cite{Athenodorou:2018wpk}.

Using the experimental value $m_Z = 91187.6(2.1)$ MeV for the $Z$ boson
pole mass~\cite{ParticleDataGroup:2020ssz} we get
\begin{equation}
  \label{eq:alphas}
    \alpha_s(m_Z) = 0.11823(69)(42)_{b_{\rm g}}(20)_{\Gamma_{\rm m}}(6)_{3\to 5,{\rm PT}}(7)_{3\to 5,{\rm NP}} = 0.11823(84)\,. 
\end{equation}

\begin{figure}[t]
  \centering
  \includegraphics[width=\textwidth]{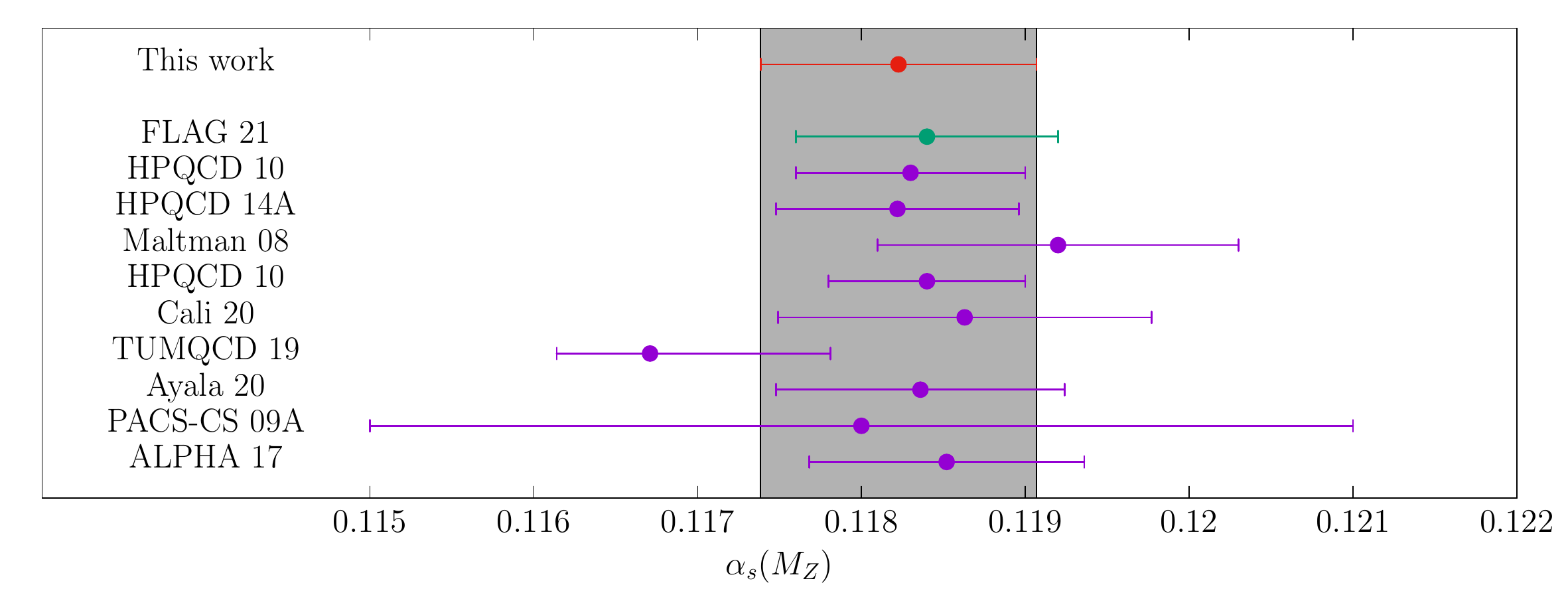}
  \caption{Our result compared with other lattice computations~\cite{Aoki:2009tf,McNeile:2010ji,Chakraborty:2014aca,Bruno:2017gxd,Bazavov:2019qoo,Ayala:2020odx,Cali:2020hrj}  that
    enter in the FLAG average~\cite{Aoki:2021kgd} (acronyms taken from the FLAG report~\cite{Aoki:2021kgd}).}
  \label{fig:comparison}
\end{figure}

Figure~\ref{fig:comparison} shows a comparison of our results with
other lattice computations~\cite{Maltman:2008bx,Aoki:2009tf,McNeile:2010ji,Chakraborty:2014aca,Bruno:2017gxd,Bazavov:2019qoo,Ayala:2020odx,Cali:2020hrj} 
that enter the FLAG average~\cite{Aoki:2021kgd}.
Our result shows a good agreement with the FLAG average, our
previous determination of the strong coupling~\cite{Bruno:2017gxd}, and
the other lattice works that enter in the FLAG average. 
It is important to point out that the result of this work is largely
independent from our previous determination~\cite{Bruno:2017gxd}. 
Only the value of $\mu_{\rm dec} = 789(15)$ MeV is shared between both
determinations of the strong coupling (see Section~\ref{sec:mtoinfty-extr}). 
This amounts to 28\% of the squared error.

\section{Conclusions and outlook}
\label{sec:conclusions}

The determination of the strong coupling on the lattice faces
particular challenges compared with low energy hadronic quantities. 
One has to connect a low energy scale with the perturbative
high energy regime of QCD. 
Due to the slow running of the coupling, perturbative scales are very large and 
these two regimes cannot be comfortably simulated on a single
lattice.  
This ``window problem'' which is due to the fact that
only a limited range of scales can be simulated on a single lattice is the reason why most lattice determinations of the strong
coupling have uncertainties dominated by the truncation
errors of the perturbative series: they apply perturbation theory at in-between energy scales
(see~\cite{DelDebbio:2021ryq} for a review). 
 One exception is the step scaling method~\cite{Luscher:1991wu}, which
was designed to cover a large  scale difference non-perturbatively. 
In practice, however, the method is quite demanding, and a reduction
of the current uncertainty in the strong coupling using this technique
is possible but requires large computational resources. 

An alternative strategy based on the decoupling of heavy quarks 
built on \cite{Athenodorou:2018wpk,Knechtli:2017xgy}
was formulated in ref.~\cite{DallaBrida:2019mqg}. 
In short, one connects the theory with physical quark masses to
the one where up, down and strange quark have masses far above the low energy QCD scales. Decoupling of the heavy quarks 
relates the theory to the pure gauge theory and we can use the
knowledge of a pure gauge intermediate energy scale, $\mudec\approx 800~\MeV$, in units of the $\Lambda_\msbar^{(0)}$ parameter. Thus the non-perturbative running between 
$\mudec$ and perturbative scales is taken from the pure gauge theory
where it is much more tractable from the numerical point
of view. The connection of $\mudec\approx 800~\MeV$ to the physical 
scales $\fpi$, $\fK$ requires only one or two step-scaling steps with light quarks; 
we could here take it from previous work \cite{Bruno:2017gxd}.

In this paper we have worked out  practical and theoretical aspects in detail and in particular demonstrated how systematic effects of various kinds can be 
controlled by numerical extrapolations and/or explicit computations. 
This is far from trivial, since a very good precision is required 
in all steps to reach the desired  accuracy of the 
strong coupling. For practical reasons intermediate scales of the theory are always defined by values of associated renormalized couplings and those are defined in the Schr\"odinger functional. 
We then need to control corrections to the continuum limit and
the decoupling limit of order $a$ and $1/M$ besides the ones of order $a^2$ and $1/M^2$
present also when space-time has no boundaries. 
We showed how the decoupling effective theory can be used to 
remove the $1/M$ corrections, again by non-perturbative information in the pure gauge theory, and how
Symanzik and decoupling effective theories, applied in that order, 
restrict the form of a combined continuum and $1/M^2$ extrapolation 
of the couplings in the massive theory. Together with the high accuracy~\cite{Athenodorou:2018wpk} of
perturbation theory~\cite{Gerlach:2018hen,vanRitbergen:1997va,Czakon:2004bu,Baikov:2016tgj,Luthe:2016ima,Herzog:2017ohr} in the relation $P=\Lambda_\msbar^{(\nf)}/\Lambda_\msbar^{(\nl)} $, which we use for $\nf=3,\nl=0$, this is a key to the 
precision reached in the result.

Building on these important theoretical steps,
we have shown that precise results can be obtained
using the decoupling strategy: our result, $\alpha_s(M_Z) = 0.11823(84)$, is among
the most precise determinations existing so far. The
error is still statistically 
dominated, with negligible perturbative uncertainties. This also
opens the way to further reduce the current
uncertainty in the strong coupling with moderate additional effort. The main sources of uncertainty
are first the single step-scaling step in QCD, second the missing knowledge of
the improvement parameter $b_{\rm g}$, a parameter that affects the
continuum extrapolations of our massive couplings, and third the pure gauge theory
non-perturbative running at high energies. 
The first and third source of uncertainty are statistical in nature
and can be substantially improved at a modest cost with existing  
techniques. Lastly, a non-perturbative determination of $b_{\rm g}$
would completely eliminate the second largest source of uncertainty on
our result. 

Finally it is worth mentioning that the good agreement between the
result of this work, and the previous determination by the ALPHA
collaboration~\cite{Bruno:2017gxd} (using the step-scaling method in
three flavor QCD), represents a highly non-trivial cross-check of the
methods.

We expect that the use of heavy quarks as a tool for non-perturbative
renormalization will have more applications in the future. 
For example, the determination of the strong coupling directly in large volume is possible, in principle~\cite{DallaBrida:2019mqg}. 
The idea can straightforwardly be applied to the determination of quark masses. Other renormalization problems, such as the determination of RGI 4-fermion operators may be tractable, 
but there remains work to be done in continuum perturbation theory: the high accuracy available for the perturbative
decoupling of the QCD parameters~~\cite{Gerlach:2018hen,vanRitbergen:1997va,Czakon:2004bu,Baikov:2016tgj,Luthe:2016ima,Herzog:2017ohr} needs to be extended also to such operators.

\section*{Acknowledgements}

We are grateful to our colleagues in the ALPHA-collaboration for discussions and the sharing of code as well as intermediate un-published results. 
In particular we thank P.~Fritzsch, J.~Heitger, S.~Kuberski for preliminary results of the HQET project \cite{Fritzsch:2018yag}. 
RH was supported by the Deutsche Forschungsgemeinschaft in the SFB/TRR55.
SS and RS acknowledge funding by the H2020 program in the  {\em
  Europlex} training network, grant agreement No. 813942. AR
acknowledges financial support from the Generalitat Valenciana 
(genT program CIDEGENT/2019/040) and the Spanish Ministerio de Ciencia e
Innovacion (PID2020-113644GB-I00).  
Generous computing resources were supplied by the North-German
Supercomputing Alliance (HLRN, project bep00072), the High Performance
Computing Center in Stuttgart (HLRS) under PRACE project 5422 and by
the John von 
Neumann Institute for Computing (NIC) at DESY, Zeuthen.

The authors are grateful for the hospitality extended to them at the IFIC Valencia
during the final stage of this project.
\begin{appendices}

\section{Boundary O($1/m$) contributions}
\label{app:boundary}

\subsection{Perturbative determination of $\omega_{{\rm b}}$ at leading order}
\label{app:omegab}

In Sect.~\ref{subsec:boundary}, we anticipated how relations analogous
to eq.~(\ref{eq:LeadingO1MCorr}) may be used as matching conditions
to determine the coefficient $\omega_{{\rm b}}$ appearing in the 
effective action, eq.~(\ref{eq:SdecSF}). As also mentioned there, a 
convenient quantity to consider for this application is the Schr\"odinger 
functional coupling, $\bar{g}^2_{\rm SF}(\mu)$~\cite{Luscher:1992an,Luscher:1993gh,Sint:1995ch}. 
We refer the reader to these references for a detailed definition of this coupling. 
For the present discussion, we recall that $\bar{g}^2_{\rm SF}(\mu)$ is related, 
up to a normalization constant $k$, to the expectation value of the $\eta$-derivative 
of the action, where the parameter $\eta$ enters the definition of the spatially 
constant Abelian boundary fields defining the SF. In formulas (cf.~ref.~\cite{Sint:1995ch}),
\begin{equation}
	{k\over \bar{g}^2_{\rm SF}(\mu,\bar{z})} = \langle S'\rangle\,,
	\qquad
	S'={\rmd S\over \rmd \eta}\,,
	\qquad
	\bar{z}=\overline{m}(\mu)/\mu\,,
	\qquad
	\mu=L^{-1}\,,
\end{equation}
where $L$ is the spatial extent of the finite volume, and $\langle\cdot\rangle$
indicates the expectation value in QCD with $\Nf$ flavours of quarks with 
(renormalized) mass $\overline{m}(\mu)$ and given SF boundary conditions.
A specific choice of scheme for the quark masses is not necessary for the
following discussion.

Given these definitions, in  the large quark mass limit, an analogous relation to
eq.~(\ref{eq:LeadingO1MCorr}) holds for $S'$, as there are no O($1/m$)
corrections besides those coming from the effective action. More precisely,
\begin{equation}
	\label{eq:MatchingSF}
	\langle S'\rangle 
	= \langle S'\rangle_{\rm dec} 
	- \frac1{m_\star L} \omega_{{\rm b}}(\alpha_\star) L\langle S'S_{\rm 1,dec}\rangle_{\rm dec} 
	+{\rm O}((m_\star L)^{-2})\,,
\end{equation}
where we indicated with $\langle\cdot\rangle_{\rm dec}$ the (connected) SF correlation functions
in the effective, pure Yang-Mills theory. In order to determine the lowest order coefficient in 
the expansion
\begin{equation}
	\omega_{{\rm b}}(\alpha_\star)= 
	\omega_{\rm b}^{(1)}\,\alpha_\star + 
	\omega_{\rm b}^{(2)}\,\alpha_\star^2 + 
	\rmO(\alpha_\star^3)\,,
	\qquad
	\alpha_\star\equiv g^2_\star/(4\pi)=\alpha^{(\Nf)}_{\MSbar}(m_\star)\,,
\end{equation}
we first need the lowest order perturbative results 
\begin{align}
	\label{eq:S'}
	\langle S'\rangle_{\rm dec}
	&={k\over [\gbar^{(0)}_\msbar(\mu)]^2}
	\bigg[1+y_1\,[\gbar^{(0)}_\msbar(\mu)]^2+\rmO([\gbar^{(0)}_\msbar]^4)\bigg]\,,
	\\
	\label{eq:S'S1dec}
	-L\langle S'S_{\rm 1,dec} \rangle_{\rm dec}
	&={2k\over [\gbar^{(0)}_\msbar(\mu)]^2}
	\bigg[1+c_1[\gbar^{(0)}_\msbar(\mu)]^2 +\rmO([\gbar^{(0)}_\msbar]^4)\bigg]\,,   
\end{align}
where $\gbar^{(0)}_\msbar(\mu=1/L)$ is the coupling of the pure gauge theory
in the $\overline{\rm MS}$-scheme. These results can be easily inferred from the
pure Yang-Mills computations of refs.~\cite{Bode:1998hd,Bode:1999sm}.%
\footnote{In fact, the result in eq.~(\ref{eq:S'}) is trivial, while that
		  of eq.~(\ref{eq:S'S1dec}) can be deduced from, e.g.~eq.~(6.23) of
 		  ref.~\cite{Husung:2019ytz}.} 
The constants $y_1$ and $c_1$ are determined through a next-to-leading 
order calculation, but  their actual value is not relevant at the order we 
are interested in. Secondly, we need the expansion 
\begin{equation}
	\label{eq:gbarsfexp}
	\langle S'\rangle= {k\over [\gbar^{(\Nf)}_\msbar(\mu)]^2}
	\bigg[1+ f_1(\bar{z}) [\gbar^{(\Nf)}_\msbar(\mu)]^2+  \rmO([\gbar^{(\Nf)}_\msbar]^4)\bigg]\,,
\end{equation}
in QCD with $\Nf$ quark flavours. In particular, we are interested in the limit 
of large quark masses, $\bar{z}\to\infty$, for which (cf.~ref.~\cite{Sint:1995ch}),
\begin{equation}
	\label{eq:f1z}
	f_1(\bar{z}) = y_1 -\frac1{12\pi^2}\nf \log(\bar{z}) + f_{11} \frac1{\bar{z}} + \rmO(\bar{z}^{-2})\,,
\end{equation}
where $y_1$ is the pure Yang-Mills result introduced earlier. As we shall see
shortly, the matching coefficient $\omega_{\rm b}^{(1)}$ only depends on $f_{11}$,
while the first two terms in the above equation are reabsorbed into the relation between 
the $\msbar$-couplings of the fundamental and effective theory. The two couplings need 
in fact to be matched. At the order in the perturbative expansion we are interested
in, we may consider the relation (see e.g.~ref.~\cite{Kniehl:2006bg}),
\begin{equation}
	\label{eq:MatchingMSbar}
	{1\over [\bar{g}^{(0)}_{\overline{\rm MS}}({\mu})]^{2}}
	={1\over [\bar{g}^{(\Nf)}_{\overline{\rm MS}}({\mu})]^{2}}
	\bigg[1+h_1({\bar{z}})[\bar{g}^{(\Nf)}_{\overline{\rm MS}}({\mu})]^{2}+
	{\rm O}([\bar{g}^{(\Nf)}_{\overline{\rm MS}}]^{4})\bigg]\,,
	\quad
	h_1({\bar{z}})=-{1\over 12\pi^2}\Nf\log(\bar{z})\,.
\end{equation}
We can now combine the results of eqs.~(\ref{eq:MatchingSF})-(\ref{eq:MatchingMSbar}).
Noticing that: $[\bar{g}^{(0)}_{\overline{\rm MS}}(\mu))]^2=[\bar{g}^{(\Nf)}_{\overline{\rm MS}}(\mu)]^2+
{\rm O}([\bar{g}^{(\Nf)}_{\overline{\rm MS}}]^4)
=g_\star^2+\rmO(g_\star^4)$ and $\overline{m}(\mu)=m_\star(1+\rmO(g_\star^2))$,
neglecting higher-order terms in $1/\bar{z}$ and $g_\star^2$, 
we arrive at the sought after relation
\begin{equation}
	\omega_{\rm b}^{(1)} = 2\pi f_{11}\,.
\end{equation}
The value of $f_{11}$ can be inferred from the 1-loop calculation of the 
massive SF coupling of ref.~\cite{Sint:1995ch}. More precisely, eqs.~(3.17)-(3.18) 
of this reference give $f_{11}=q_1/(4\pi)$. For definiteness, we take the
result $q_1=-0.10822$, obtained for $\theta=\pi/5$, as this coupling definition 
shows significantly smaller O($1/\bar{z}^2$) corrections than the definition with
$\theta=0$. We can in fact use the difference between the results for 
$\theta=0,\pi/5$ as an estimate of the systematic uncertainties involved in the
extraction of $q_1$. In conclusion, we find:
\begin{equation}
	\omega_{{\rm b}}^{(1)}=-0.0541(5)N_{\rm f} \,.	
\end{equation}

\subsection{Estimates of the O($1/m$) boundary effects }
\label{app:boundarycontribution}

\subsubsection{Definitions}

Given $\omega_{{\rm b}}$ at leading order in perturbation theory, we can now
employ eq.~(\ref{eq:LeadingO1MCorr}) to get an estimate of the O($1/M$)
corrections to the massive GFT-coupling $\bar{g}^{(3)}_{\rm GFT, c}(\mu_{\rm dec},M)$, 
defined in the SF with temporal extent $T=2L$, and evaluated at the renormalization
scale $\mu_{\rm dec}=1/L_{\rm dec}$. (As usual, $M$ is the RGI mass of the degenerate 
heavy quarks in the $\Nf=3$ theory and $c=\sqrt{8t}/L$, with $t$ the flow time at 
which the coupling is measured.) To this end, we define,
\begin{equation}
	\Delta_{\rm c}(z)\equiv 
	[\bar{g}^{(3)}_{\rm GFT,c}(\mu_{\rm dec},M)]^2-[\bar{g}^{(0)}_{\rm GFT,c}(\mu_{\rm dec})]^2\,,
	\qquad
	z=M/\mu_{\rm dec}\,,
\end{equation}
where we implicitly assume that the $\Lambda$-parameters in the two
theories have been properly matched. Applying eq.~(\ref{eq:LeadingO1MCorr})
to the case at hand we have,
\begin{equation}
	\label{eq:O1MCorrection}
	\Delta_{\rm c}(z)\overset{z\to\infty}{\approx}
	-{\omega_{{\rm b}}(\alpha_\star)\over z}\bigg({M\over m_\star}\bigg)\lim_{a\to 0}\,
	L_{\rm dec}\,\mathcal{N}^{-1} {\langle t^2 E_{\rm mag}(t,T/2) \hat\delta(Q)S_{\rm 1,dec}^R
	\rangle_{\rm dec}\over \langle  \hat\delta(Q)\rangle_{\rm dec}}
	+ {\rm O}\big({z^{-2}}\big)\,,
\end{equation}
where, we recall, $\alpha_\star\equiv g^2_\star/(4\pi)$ with $g_\star\equiv\bar{g}_{\overline{\rm MS}}(m_\star)$
and $m_\star=\overline{m}_{\overline{\rm MS}}(m_\star)$, and we take the connected part of
the (ratio of) correlation functions. In the above equation we use the short hand notation 
(cf.~eq.~(\ref{eq:GFT}))
\begin{equation}
	[\bar{g}^{(0)}_{\rm GFT,c}(\mu)]^2=
	\mathcal{N}^{-1}  
	{\langle t^2E_{\rm mag}(t,T/2)\hat\delta(Q)\rangle_{\rm dec}
	\over \langle  \hat\delta(Q)\rangle_{\rm dec}}\,,
	\quad
	\mu=1/L\,,
	\quad
	T=2L\,,
	\quad
	c=\sqrt{8t}/L\,,
\end{equation}
for the definition of the ${\rm GFT}$-coupling in the pure-gauge theory. 
The field $E_{\rm mag}(t,x_0)$ represents our chosen discretization for the magnetic 
component of the energy density at positive flow time. For the latter, we consider the 
Zeuthen flow for the discretization of the flow equations~\cite{Ramos:2015baa}, while for the observable  
we take the specific combination of plaquette and clover discretizations of the flow energy 
density proposed in ref.~\cite{Ramos:2015baa}, which is O($a^2$)-improved. 
The quantity $\hat \delta(Q)$ is instead a discretization of the continuum $\delta$-function 
that projects to the topological $Q=0$ sector. It is zero whenever 
$|Q|>0.5$ and one otherwise, with $Q$ the topological charge computed with 
the clover discretization of the field strength tensor built from gauge fields 
flowed at time $\sqrt{8t}=cL$ using the Zeuthen flow (cf.~e.g.~ref.~\cite{DallaBrida:2016kgh}). 
In order to evaluate the correlation function in eq.~(\ref{eq:O1MCorrection}) 
on the lattice, we also need a viable discretization for $S_{\rm 1,dec}$. We choose,
\begin{equation}
	S^R_{\rm 1,dec}=
	a^3\sum_{\mathbf{x}}\,
	\big[\mathcal{O}^R_{{\rm b}}(0,\boldsymbol{x})+  \mathcal{O}^R_{{\rm b}}(T-a,\boldsymbol{x})\big]\,,
	\quad
	\mathcal{O}^R_{\rm b}(x_0,\boldsymbol{x}) = 	
	Z_{\rm b}(g_0)
	\mathcal{O}_{\rm b}(x_0,\boldsymbol{x})\,,
\end{equation}
where
\begin{equation}
	a^4\mathcal{O}_{\rm b}(x_0,\boldsymbol{x})=
	{2\over g_0^2} 
	\sum_{k=1}^3 
	{\rm Re}\,{\rm tr}\big[1-U_{0k}(x_0,\boldsymbol{x})\big]\,,
	\quad
	x_0=0,T-a\,,
\end{equation}
with $U_{\mu\nu}(x)$ being the plaquette at $x$ in the $\mu,\nu$-direction.
As discussed in Sect.~\ref{subsec:boundary}, on the lattice, the boundary field $\mathcal{O}_{\rm b}$
requires a finite renormalization. In the following, we consider the 1-loop
approximation~\cite{Karsch:1982ve}
\begin{equation}
	\label{eq:Zb}
	Z_{\rm b}(g_0)=1-0.13194\, g_0^2+{\rm O}(g_0^4)\,,
\end{equation}
which can be readily inferred from the renormalization of the energy density in 
the pure-gauge theory obtained in this reference. 

In practice, we can evaluate the correlator in eq.~(\ref{eq:O1MCorrection}) by 
exploiting the identity (note the appearance of the \emph{bare} boundary action)
\begin{equation}
	\label{eq:ctderv}
	-L\,\mathcal{N}^{-1}{ \langle t^2 E_{\rm mag}(t,T/2)\hat\delta(Q)
	S_{\rm 1,dec}\rangle_{\rm dec} \over 
	\langle \hat\delta(Q)\rangle_{\rm dec}} \,	
	= 
	{L\over a} {\rmd [\bar{g}_{\rm GFT,c}^{(0)}(\mu)]^2\over
	\rmd c_{\rm t}}\,.
\end{equation}
Thus, the connected correlator of interest can be computed by varying 
the boundary O($a$) improvement coefficient $c_{\rm t}$ in simulations 
(cf.~Appendix \ref{sec:simulations}). Given the result in eq.~(\ref{eq:O1MCorrection}), 
and defining 
\begin{equation}
	\label{eq:p1}
	p_{\rm c}(\bar{g})\equiv \bigg[\lim_{a\to0} Z_{\rm b}(g_0) {L\over a} 
	{\rmd [\bar{g}_{\rm GFT,c}^{(0)}(\mu)]^2\over\rmd c_{\rm t}}\bigg]\bigg|_{\bar{g}}\,,
\end{equation}
we have that the leading-order (LO) estimate for the relative O($1/M$) 
corrections to the GFT-coupling can be written as,
\begin{equation}
	\label{eq:MasterEq}
	{\Delta_{\rm c}(z)\over \bar{g}^2_{\rm c}(z) }\Bigg|_{\rm LO}
	={1\over (4\pi)}{\omega_{\rm b}^{(1)} \over z}\bigg({M\over m_\star}\bigg)
	\Bigg({g_\star^2\over \bar{g}^2_{\rm c}(z) }\Bigg)
	p_{\rm c}(\bar{g}^2_{\rm c}(z))
	+ {\rm O}\bigg(g_\star^2,{1\over z^2}\bigg)\,,
\end{equation}
where we introduced the shorthand notation: 
$\bar{g}_{\rm c}(z)\equiv \bar{g}^{(3)}_{\rm GFT,c}(\mu_{\rm dec},M)$.

\subsubsection{Simulation results}

\begin{table}[hpbt]
	\centering
	\begin{tabular}{lllll}
		\toprule
		$c$ & $L/a$ & $\beta$ & $[\bar{g}^{(0)}_{\rm GFT,c}]^2$ &  $\frac{L}{a}\frac{\rmd[\bar{g}^{(0)}_{\rm GFT,c}]^2}{\rmd c_{\rm t}}$ \\
		\midrule
		\multirow{3}{*}{$0.3$} 
		& $10$ & $6.2556$ & $4.8513(22)$ & $-0.46(14)$ \\
		& $10$ & $6.3400$ & $4.4462(19)$ & $-0.45(12)$ \\
		& $10$ & $6.4200$ & $4.1359(17)$ & $-0.43(11)$ \\
		\midrule
		\multirow{3}{*}{$0.3$} 
		& $12$ & $6.4200$ & $4.6805(20)$ & $-0.45(15)$ \\
		& $12$ & $6.4630$ & $4.4895(19)$ & $-0.54(15)$ \\
		& $12$ & $6.5619$ & $4.1094(17)$ & $-0.28(13)$ \\
		\midrule
		\multirow{4}{*}{$0.3$} 
		& $16$ & $6.6669$ & $4.7627(14)$ & $-0.42(09)$ \\
		& $16$ & $6.6920$ & $4.6515(19)$ & $-0.55(12)$ \\
		& $16$ & $6.7140$ & $4.5641(19)$ & $-0.52(12)$ \\
		& $16$ & $6.7859$ & $4.2942(12)$ & $-0.25(08)$ \\
		\midrule
		\midrule
		\multirow{3}{*}{$0.42$} 
		& $10$ & $6.2556$ & $7.5116(61)$ & $-2.91(38)$ \\
		& $10$ & $6.3400$ & $6.6790(52)$ & $-2.35(32)$ \\		
		& $10$ & $6.4200$ & $6.0807(44)$ & $-1.98(28)$ \\
		\midrule
		\multirow{3}{*}{$0.42$} 
		& $12$ & $6.4200$ & $7.1857(57)$ & $-2.36(43)$ \\
		& $12$ & $6.4630$ & $6.7975(52)$ & $-2.42(40)$ \\   
		& $12$ & $6.5619$ & $6.0584(44)$ & $-1.68(34)$ \\   
		\midrule
		\multirow{4}{*}{$0.42$} 
		& $16$ & $6.6669$ & $7.1159(39)$ & $-2.21(25)$	\\
		& $16$ & $6.6920$ & $6.8934(51)$ & $-2.29(33)$	\\
		& $16$ & $6.7140$ & $6.7275(50)$ & $-2.19(32)$	\\
		& $16$ & $6.7859$ & $6.2173(32)$ & $-1.52(21)$	\\
		\bottomrule
	\end{tabular}
	\caption{Results for the $\Delta c_{\rm t}\to0$ extrapolations,
		eq.~(\ref{eq:Dct2zLimit}), for different values of $L/a$,
		$c$, and couplings $[\bar{g}^{(0)}_{\rm GFT,c}]^2$. The results refer
		to the magnetic coupling discretized using the Zeuthen flow and O($a^2$)-improved 
		plaquette + clover definition of the flow energy density~\cite{Ramos:2015baa}. 
		We considered linear extrapolations in $(\Delta c_{\rm t})^2$ 
		using all available $\Delta c_{\rm t}$ values.}
	\label{tab:dgsqdct}
\end{table}

In order to estimate the relevant $c_{\rm t}$-derivatives in eq.~(\ref{eq:ctderv}),
we simulated lattices with $L/a=10,12,16$ using the Wilson-plaquette gauge action,
and varied $c_{\rm t}$ around its two-loop value $c_{\rm t}^\star$~\cite{Bode:1999sm}, i.e.~$c_{\rm t}^\pm=
c_{\rm t}^\star\pm\Delta c_{\rm t}$. For $L/a=10,12$, we considered variations 
$\Delta c_{\rm t}=\{0.075,0.1,0.15,0.2\}$, while for $L/a=16$ we took
$\Delta c_{\rm t}=\{0.1,0.15,0.2,0.3\}$. For $L/a=10,12$ and $L/a=16$
we simulated 3 and 4 values of $\beta$, respectively, in order to cover 
the relevant range of values for $[\bar{g}^{(0)}_{\rm GFT,c}]^2$; this for two schemes 
$c=0.3,0.42$. In Table \ref{tab:dgsqdct} we collect the results for%
\begin{equation}
	\label{eq:Dct2zLimit}
	{L\over a} {\rmd [\bar{g}_{\rm GFT,c}^{(0)}(\mu)]^2\over
		\rmd c_{\rm t}} = 
	\lim_{\Delta c_{\rm t}\to0} {L\over a}{\Delta[\bar{g}^{(0)}_{\rm GFT,c}]^2\over2\Delta c_{\rm t}},
	\quad
	\Delta[\bar{g}^{(0)}_{\rm GFT,c}]^2 \equiv 
	[\bar{g}^{(0)}_{\rm GFT,c}(\mu)]^2|_{c_{\rm t}^+}-[\bar{g}^{(0)}_{\rm GFT,c}(\mu)]^2|_{c_{\rm t}^-}\,,
\end{equation}
obtained as linear extrapolations in $(\Delta c_{\rm t})^2$ using all available 
$\Delta c_{\rm t}$. The sum, $\chi^2_{\rm tot}$, of the $\chi^2$'s of all fits is good, 
i.e.~$\chi_{\rm tot}^2/{\rm d.o.f.}_{\rm tot}\approx 1$, with ${\rm d.o.f.}_{\rm tot}$
the total number of degrees of freedom considering all fits. At the level of individual 
fits, these have a $\chi^2$ up to $\approx 7$ for two degrees of freedom.

In Table \ref{tab:dgsqdct_interpol} we report the results for the $c_{\rm t}$-derivatives 
of the $\Nf=0$ coupling interpolated at the values of the coupling of interest. The latter 
are specified by the results in the $\Nf=3$-theory (cf.~Table  \ref{tab:cont_res}):
\begin{equation}
	\label{eq:TargetCouplings}
	\bar{g}^{2}_{\rm 0.30}(z=6)=4.490(50)\,,
	\qquad
	\bar{g}^{2}_{\rm 0.42}(z=4)=6.68(10)\,.
\end{equation}
For the interpolations we considered the following functional form,
\begin{equation}
	\label{eq:Interpol}
	{L\over a}{\rmd[\bar{g}^{(0)}_{\rm GFT,c}]^2\over \rmd c_{\rm t}}=
	\sum_{k=1}^3 c_k\,[\bar{g}^{(0)}_{\rm GFT,c}]^{2k}\,,
\end{equation}	
where the value of $c_1$ is fixed to its tree-level result (cf.~Table~\ref{tab:TreeLevel}).

\begin{table}[hpbt]
	\centering
	\begin{tabular}{llll}
		\toprule
		$c$ & $L/a$ & ${1\over g_0^2}\frac{L}{a}\frac{\rmd[\bar{g}^{(0)}_{\rm GFT,c}]^2}{\rmd c_{\rm t}}\times 10^{4}$ \\
		\midrule
		\multirow{3}{*}{$0.3$} &
		$10$ & $-2.6042383805$ 		\\
		& $12$ & $-2.4106092056$ 	\\
		& $16$ & $-2.2186159754$ 	\\
		\midrule	
		\midrule
		\multirow{3}{*}{$0.42$} &
		$10$ & $-12.918755710$ 		\\
		& $12$ & $-12.000133367$	\\ 
		& $16$ & $-11.123348145$ 	\\  
		\bottomrule
	\end{tabular}
	\caption{Results for the $c_{\rm t}$-derivative of the GFT-coupling
		at tree-level in lattice perturbation theory. The results refer
		to the magnetic coupling discretized using the Zeuthen flow
		and O($a^2$)-improved plaquette + clover definition of the flow energy density~\cite{Ramos:2015baa}.}
	\label{tab:TreeLevel}
\end{table}

\begin{figure}[hbpt]
	\centering
	\includegraphics[width=0.75\textwidth]{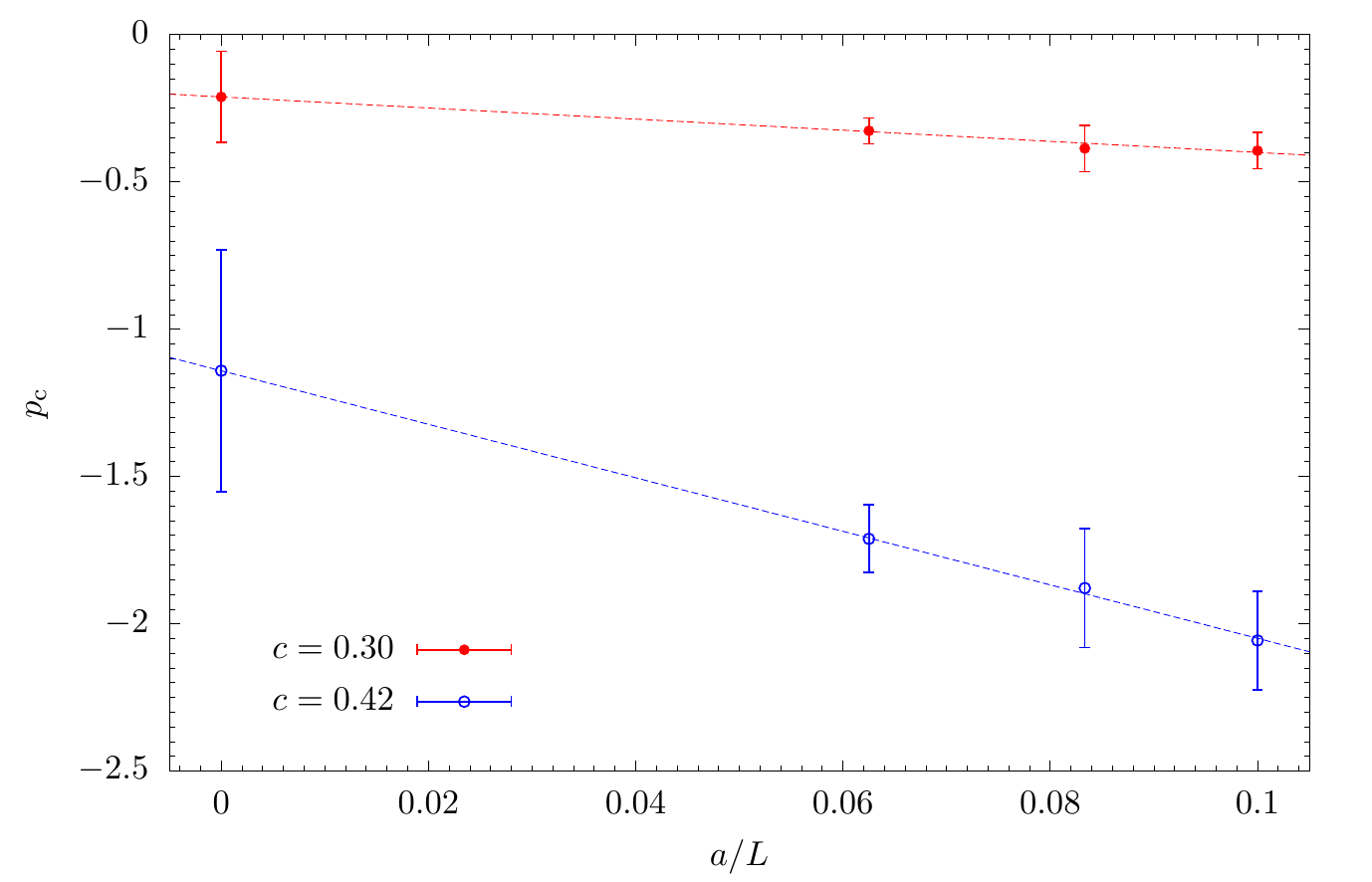}\\
	\caption{Illustrative continuum limit extrapolations for the $p_{\rm c}$ 
		results of Table \ref{tab:dgsqdct_cont}.} 
	\label{fig:contlim}
\end{figure}

Once the results for the different $L/a$ were interpolated to a fixed value 
of the coupling and properly renormalized with the 1-loop value of $Z_{\rm b}$,
eq.~(\ref{eq:Zb}), we performed an extrapolation to $a/L\to0$, assuming 
leading O($a$) effects. The results are reported in Table
\ref{tab:dgsqdct_cont}, while Figure \ref{fig:contlim} illustrates the
corresponding extrapolations. The results for $p_{\rm c}(\bar{g})$ 
for $c=0.42$ tend to be larger (in module) than those for $c=0.30$. This is expected, as the
footprint of the flow energy density defining the GFT-coupling extends
closer to the SF boundaries for $c=0.42$, therefore increasing the sensitivity 
to the $\rmO(1/M)$ counterterms.

\begin{table}[hbt]
	\centering
	\begin{tabular}{lllll}
		\toprule
		$c$ & $L/a$ &  $\beta$ & $[\bar{g}^{(0)}_{\rm GFT,c}]^2$ & {$\frac{L}{a}\frac{\rmd[\bar{g}^{(0)}_{\rm GFT,c}]^2}{\rmd c_{\rm t}}$} \\
		\midrule
		\multirow{3}{*}{$0.3$} 
      & $10$ & $6.3299(4)$ & $4.490$ & $-0.45(7)$ \\
      & $12$ & $6.4629(5)$ & $4.490$ & $-0.44(9)$ \\
      & $16$ & $6.7324(4)$ & $4.490$ & $-0.37(5)$ \\
		\midrule
		\midrule	
		\multirow{3}{*}{$0.42$} 
      & $10$ & $6.3399(6)$ & $6.68$ & $-2.35(19)$ \\	
      & $12$ & $6.4771(7)$ & $6.68$ & $-2.14(23)$ \\ 
      & $16$ & $6.7196(6)$ & $6.68$ & $-1.94(13)$ \\     
		\bottomrule
	\end{tabular}
	\caption{Results of the interpolations eq.~(\ref{eq:Interpol}) using 
		the values in Table \ref{tab:dgsqdct} and the target couplings,
		eq.~(\ref{eq:TargetCouplings}). (Note that we did not propagate the 
		error on $[\bar{g}^{(0)}_{\rm GFT,c}]^2$ to the $c_{\rm t}$-derivatives.) All 
		fits have good or acceptable $\chi^2$. The values of $\beta$ have 
		been obtained by a quadratic interpolation of the results for
		$[\bar{g}^{(0)}_{\rm GFT,c}]^{-2}$ from Table \ref{tab:dgsqdct} as a function
		of $g_0^{-2}$.}
	\label{tab:dgsqdct_interpol}
\end{table}

\begin{table}[hbt]
	\centering
	\begin{tabular}{llll}
		\toprule
		$c$ & $[\bar{g}^{(0)}_{\rm GFT,c}]^2$ & $L/a$ & $p_{\rm c}$ \\
		\midrule
		\multirow{3}{*}{$0.3$} &
		\multirow{3}{*}{$4.490$} 
		&   $10$ & $-0.39(6)$ 	\\
		& & $12$ & $-0.39(8)$ 	\\
		& & $16$ & $-0.33(4)$ 	\\
		\midrule
		& & $\infty$ & $-0.21(16)$ \\	    
		\midrule	
		\midrule
		\multirow{3}{*}{$0.42$} &
		\multirow{3}{*}{$6.68$} 
		&   $10$ & $-2.06(17)$ 	\\
		& & $12$ & $-1.88(20)$	\\ 
		& & $16$ & $-1.71(11)$ 	\\   
		\midrule  
		& & $\infty$ & $-1.14(41)$ \\	    
		\bottomrule
	\end{tabular}
	\caption{Results for $p_{\rm c}$ at finite lattice spacing (cf.~eq.~(\ref{eq:p1})) 
		obtained using the values for the $c_{\rm t}$-derivatives of the GFT-coupling 
		in Table \ref{tab:dgsqdct_interpol} and the 1-loop approximation for $Z_{\rm b}$, 
		eq.~(\ref{eq:Zb}). The results of a $L/a\to\infty$ extrapolation linear in 
		$a/L$ are also given.}
	\label{tab:dgsqdct_cont}
\end{table}

\subsubsection{O($1/M$) corrections: LO estimates}

We first estimate $\Delta_{\rm c}(z)$ corresponding to the coupling
at $z=6$ and $c=0.3$ (cf.~eq.~(\ref{eq:TargetCouplings})). 
In this case we have,
\begin{equation}
	\bar{g}^2_{0.3}(z=6)=4.490(50)\,,
	\qquad
	g^2_\star\approx 3\,,
	\qquad
	M/m_\star\approx 1.5\,,
\end{equation}
where to compute $g_\star$ we used
\begin{equation}
	{M\over\Lambda^{(\Nf=3)}_{\overline{\rm MS}}}
	=z\times {\mu_{\rm dec}\over\Lambda^{(\Nf=3)}_{\overline{\rm MS}}}\,,
\end{equation}
with $\mu_{\rm  dec}=789\,\MeV$ and $\Lambda^{(\Nf=3)}_{\overline{\rm MS}}=341\,\MeV$.
For the $\rmO(1/M)$ counterterm $p_{\rm c}$ we take the result (cf.~Table \ref{tab:dgsqdct_cont}),
\begin{equation}
	p_{0.30}(\bar{g}^{2}_{0.30}(z=6))= -0.21\,.
\end{equation}
Putting the numbers together we obtain for the LO estimate,
eq.~(\ref{eq:MasterEq}),
\begin{equation}
	{\Delta_{0.3}(z=6)\over \bar{g}^{2}_{0.30}(z=6)}\Bigg|_{\rm LO}
	\approx 5\times 10^{-4}\,.
\end{equation}
This is about a factor 20 smaller than the statistical 
error on the coupling.

As a second case, we estimate $\Delta_{\rm c}(z)$ for $z=4$ and 
$c=0.42$. In this case we have,
\begin{equation}
	\bar{g}^2_{0.42}(z=4)=6.68(10)\,,
	\qquad
	g^2_\star\approx 3.5\,,
	\qquad
	M/m_\star\approx 1.4\,.
\end{equation}
For the O($1/M$) counterterm $p_{\rm c}$ we take the result (cf.~Table \ref{tab:dgsqdct_cont})
\begin{equation}
	p_{0.42}(\bar{g}^{2}_{0.42}(z=4))=-1.14\,.
\end{equation}
Putting the numbers together we obtain for the LO estimate,
eq.~(\ref{eq:MasterEq}),
\begin{equation}
	{\Delta_{\rm 0.42}(z=4)\over \bar{g}^{2}_{0.42}(z=4)}\Bigg|_{\rm LO}
	\approx
	3\times 10^{-3}\,,
\end{equation}
which is about a factor 5-6 smaller than the statistical error on the coupling.
A slightly more conservative estimate may be obtained in this case by taking $p_{0.42}=-1.71$, 
which corresponds to the result at the smallest lattice spacing we simulated, 
i.e.~$L/a=16$. In this case, the systematic O($1/M$) effects are about
3-4 times smaller than the statistical error on the coupling. 
The smallest value of $z$ entering the analysis of Sect.~\ref{sec:mtoinfty-extr}
is $z=6$. Considering that these effects decrease linearly
with increasing $z$ and that the results for $c=0.42$ offer an upper-bound for 
values of $c<0.42$, it appears to be safe to assume that $\rmO(1/M)$ effects 
can be neglected altogether.
 
\clearpage

\section{Summary of pure-gauge results}
\label{sec:summary-pure-gauge}

\subsection{Matching GF and GFT schemes at the decoupling scale}
\label{subsec:MatchingYM}

\begin{figure}
  \centering
  \includegraphics[width=0.7\textwidth]{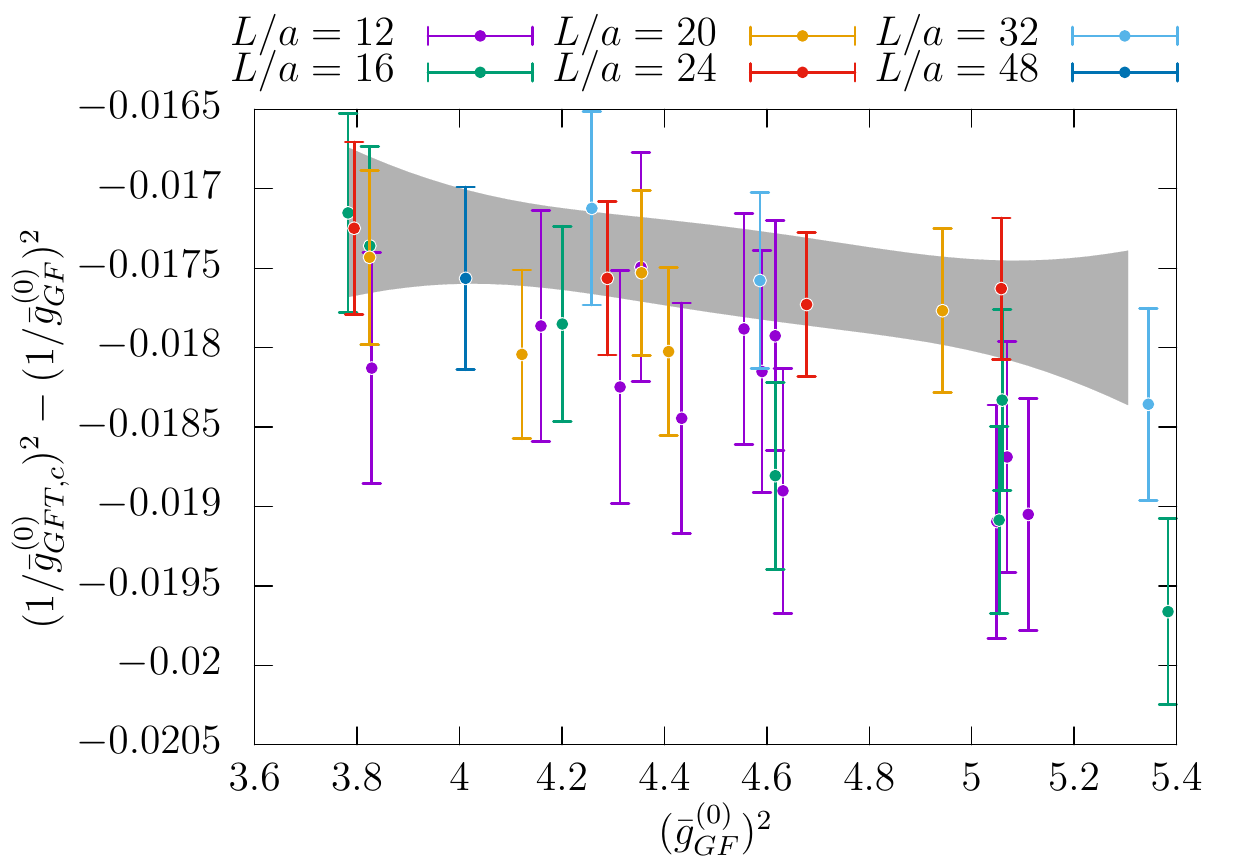}
  \caption{Fit to match the schemes  $\bar g^{(0)}_\text{GF}$ and
  $\bar g^{(0)}_\text{\rm GFT,c}$ for $c=0.3$.}
  \label{fig:T2L}
\end{figure}

As explained in Sect.~\ref{subsec:RGI}, we need the continuum relation 
\begin{equation}
	\bar{g}^{(0)}_{\rm GF }(\mu) = \chi_{\rm c}\left(\bar{g}^{(0)}_{\rm GFT,c }(\mu) \right) 
	\label{e:chi}
\end{equation}
for $\mu$ around $\mudec$, i.e. $\bar{g}^{(0)}_{\rm GFT,c }=\bar{g}^{(3)}_{\rm GFT,c }(\mu_{\rm  dec},M)$ is in the range of
Tables~\ref{tab:cont_res} and  \ref{tab:largeM}.
On a finite lattice, eq.~\eqref{e:chi} is obtained
by computing both couplings at the same $L/a=1/(\mu a)$ and $\beta$ and the relation contains discretization errors of order $(a/L)^2$. 
Results of our simulations for $L/a=12,\ldots,48$ and $\beta\in[6.2,7.6]$  are 
reported in Table~\ref{tab:T2L}. We parameterize the data by
\begin{equation}
	\label{eq:GFT2GF}
  \frac{1}{[\bar g^{(0)}_\text{GFT,c}]^2} -  \frac{1}{[\bar g^{(0)}_\text{\rm GF}]^2} =
  P^{\rm c}_{n_p}([\bar g^{(0)}_\text{\rm GF}]^2) + \Big({a\over L}\Big)^2 Q_{n_q}^{\rm c}([\bar g^{(0)}_\text{\rm GF}]^2)\,,
\end{equation}
where $P_{n_p}^{\rm c}(g^2)$, $Q_{n_q}^c(g^2)$ are polynomials of degree $n_p$ and $n_q$, respectively. The sought-after continuum relation (\ref{e:chi}) is then 
implicitly given in terms of
\begin{equation}
  P^{\rm c}_{n_p}(g^2)=\sum_{k=0}^{n_p} p^{\rm c}_k\, g^{2k}\,,
\end{equation}
through,
\begin{equation}
	[\chi_{\rm c}^{-1}(g)]^2=\frac{g^2}{1+g^2 P^{\rm c}_{n_p}(g^2)}\,,
\end{equation}
where $\chi_{\rm c}^{-1}$ is the inverse of the function $\chi_{\rm c}$.
For all values of $c=0.3-0.42$, taking $n_p=n_q=2$  already yields good 
fits. As an illustrative example, we give here the results for $P^{\rm c}_2$ for 
the case $c=0.3$, where we find ($p_k\equiv p^{0.3}_k$)
\begin{eqnarray}
  p_0 &=& -1.6168053320710727 \times 10^{-2}\,,  \\
  p_1 &=& -1.6086523109487365 \times 10^{-4}\,, \\
  p_2 &=& -3.0358986840266259 \times 10^{-5}\,,
\end{eqnarray}
with covariance
\begin{equation}
\label{e:covP}
  \mathrm{cov}(p_i, p_j) =\left(
  \begin{array}{lll}
     2.92499121\times 10^{-4} & -1.29098204\times 10^{-4} &  1.41308163\times 10^{-5}\\
    -1.29098204\times 10^{-4} &  5.73202489\times 10^{-5} & -6.27896098\times 10^{-6}\\
     1.41308163\times 10^{-5} & -6.27896098\times 10^{-6} &  6.91910100\times 10^{-7}\\
    \end{array}
  \right)\,.
\end{equation}
Figure~\ref{fig:T2L} demonstrates that the discretization errors are small and our largest lattice results are in agreement with the continuum band. 
Note that, the error of $P^{\rm c}_2$, computable in terms of \eqref{e:covP}
for the case of $c=0.3$, is negligible 
in the final $\Lambda^{(3)}_{\overline{\rm MS}}/\mu_{\rm dec}$
determination for all values of $c$ considered.  

\begin{table}
  \centering
  \begin{tabular}{rrrrrrrr}
    \toprule
     $L/a$ & $\beta$ & $[\bar g^{(0)}_\text{GF}]^2$ & $[\bar g^{(0)}_\text{GFT,0.30}]^2$  & 
    $[\bar g^{(0)}_\text{GFT,0.33}]^2$  & $[\bar g^{(0)}_\text{GFT,0.36}]^2$ & 
    $[\bar g^{(0)}_\text{GFT,0.39}]^2$ & $[\bar g^{(0)}_\text{GFT,0.42}]^2$ \\ 
    \midrule  
    12 & 6.2556 & 5.1104(49) &  5.6616(89) &   6.329(12) &    7.141(16) &   8.144(22) &   9.393(30)  \\
    12 & 6.2643 & 5.0690(50) &  5.5995(84) &   6.251(11) &    7.042(15) &   8.018(20) &   9.232(27)  \\
    12 & 6.2654 & 5.0489(49) &  5.5876(90) &   6.232(12) &    7.013(16) &   7.977(21) &   9.175(29)  \\
    12 & 6.3451 & 4.6316(71) &  5.0759(72) &   5.6147(94) &   6.264(12) &   7.060(16) &   8.043(21)  \\
    12 & 6.3509 & 4.6164(43) &  5.0329(72) &   5.5618(93) &   6.199(12) &   6.980(16) &   7.942(20)  \\
    12 & 6.3560 & 4.5905(68) &  5.0078(69) &   5.5318(90) &   6.163(12) &   6.936(16) &   7.889(21)  \\
    12 & 6.3642 & 4.5554(44) &  4.9594(69) &   5.4718(88) &   6.088(11) &   6.842(15) &   7.771(20)  \\
    12 & 6.3894 & 4.4339(44) &  4.8288(65) &   5.3162(86) &   5.901(11) &   6.613(15) &   7.490(20)  \\
    12 & 6.4133 & 4.3543(40) &  4.7133(63) &   5.1798(82) &   5.739(10) &   6.421(13) &   7.258(17)  \\
    12 & 6.4200 & 4.3136(40) &  4.6821(68) &   5.1419(87) &   5.694(11) &   6.366(14) &   7.192(18)  \\
    12 & 6.4630 & 4.1590(38) &  4.4928(63) &   4.9196(78) &   5.4311(99) &  6.053(13) &   6.816(16)  \\
    12 & 6.5619 & 3.8287(34) &  4.1143(55) &   4.4749(69) &   4.9051(88) &  5.426(11) &   6.063(14)  \\
    16 & 6.4200 & 5.3833(70) &  6.0206(99) &   6.781(13) &    7.710(19) &   8.864(26) &   10.313(35) \\
    16 & 6.4740 & 5.0535(67) &  5.5929(85) &   6.247(11) &    7.040(15) &   8.018(20) &   9.233(27)  \\
    16 & 6.4741 & 5.0594(49) &  5.5766(89) &   6.230(12) &    7.023(16) &   7.999(21) &   9.214(28)  \\
    16 & 6.5619 & 4.6165(56) &  5.0554(71) &   5.5943(94) &   6.242(12) &   7.034(16) &   8.009(21)  \\
    16 & 6.6669 & 4.2007(56) &  4.5412(61) &   4.9771(78) &   5.498(10) &   6.131(13) &   6.908(17)  \\
    16 & 6.7859 & 3.8249(45) &  4.0969(56) &   4.4595(76) &   4.8915(97) &  5.415(12) &   6.054(16)  \\
    16 & 6.8000 & 3.7825(44) &  4.0449(56) &   4.3998(70) &   4.8238(88) &  5.338(11) &   5.968(14)  \\
    20 & 6.6669 & 4.9431(61) &  5.4190(86) &   6.037(11) &    6.785(15) &   7.705(19) &   8.846(26)  \\
    20 & 6.7859 & 4.4082(54) &  4.7887(68) &   5.2779(89) &   5.865(11) &   6.581(15) &   7.462(19)  \\
    20 & 6.8000 & 4.3553(53) &  4.7153(64) &   5.1890(83) &   5.757(11) &   6.449(14) &   7.299(18)  \\
    20 & 6.8637 & 4.1219(48) &  4.4531(60) &   4.8781(77) &   5.3855(99) &  6.001(13) &   6.756(16)  \\
    20 & 6.9595 & 3.8245(44) &  4.0977(54) &   4.4637(69) &   4.9001(87) &  5.429(11) &   6.076(14)  \\
    24 & 6.7859 & 5.0578(61) &  5.5530(79) &   6.206(10) &    6.999(14) &   7.978(19) &   9.196(25)  \\
    24 & 6.8637 & 4.6776(56) &  5.1006(66) &   5.6520(86) &   6.316(11) &   7.130(15) &   8.135(19)  \\
    24 & 6.9595 & 4.2884(52) &  4.6378(61) &   5.0988(79) &   5.651(10) &   6.324(13) &   7.150(17)  \\
    24 & 7.1146 & 3.7942(53) &  4.0599(50) &   4.4194(63) &   4.8479(80) &  5.367(10) &   6.002(13)  \\
    32 & 6.9595 & 5.345(14) &  5.926(11) &    6.666(14) &    7.570(19) &   8.694(26) &   10.103(35)  \\
    32 & 7.1146 & 4.5866(93) &  4.9887(70) &   5.5195(91) &   6.158(12) &   6.937(16) &   7.898(20)  \\
    32 & 7.2000 & 4.2584(90) &  4.5933(63) &   5.0480(82) &   5.592(10) &   6.255(13) &   7.069(17)  \\
    48 & 7.6000 & 4.0120(74) &  4.3161(61) &   4.7213(79) &   5.205(10) &   5.792(13) &   6.512(17)  \\
    \bottomrule
  \end{tabular}
  \caption{Data used for the matching of the $T=L$ and $T=2L$ schemes.}
  \label{tab:T2L}
\end{table}

\subsection{High-energy running}
\label{subsec:RunningYM}

In this appendix we collect the main ingredients for the determination of 
\begin{equation}
	\label{eq:Lambda0Mudec}
	\frac{\Lambda^{(0)}_{\overline{\rm MS} }}{\mu_{\rm dec}} = 
	\frac{\Lambda^{(0)}_{\overline{\rm MS} }}{\Lambda^{(0)}_{\rm GF}} \varphi_{\rm GF}^{(0)}(\bar{g}^{(0)}_\text{GF}(\mudec))\,,
\end{equation}
where we recall that it will be used with $\bar{g}^{(0)}_\text{GF}(\mudec)=\chi_{\rm c}(\bar{g}^{(3)}_{\rm GFT,c }(\mu_{\rm  dec},M))$, cf.~Table~\ref{tab:cont_res}. Following \cite{DallaBrida:2019wur,Nada:2020jay}, we rewrite eq.~(\ref{eq:Lambda0Mudec}) as 
\begin{equation}
	\frac{\Lambda^{(0)}_{\overline{\rm MS} }}{\mu_{\rm dec}} = 
	\frac{\Lambda^{(0)}_{\overline{\rm MS} }}{\mu_{\rm swi}} \times{\mu_{\rm swi}\over{\mu_{\rm dec}}}
	\quad
	\text{where}
	\quad
	[\bar{g}^{(0)}_{\rm GF}(\mu_{\rm swi})]^2 = {4\pi\over 5}\,,
\end{equation}
and insert~\cite{DallaBrida:2019wur}%
\footnote{When comparing with the results of ref.~\cite{DallaBrida:2019wur} note that 
$\mu_{\rm swi}=0.3\,\mu_{\rm ref}$, of that reference.  The difference comes from 
the different relation between $\mu$ and $L$ used in ref.~\cite{DallaBrida:2019wur}, 
where $\mu=1/\sqrt{8t}=1/(0.3L)$ as opposed to $\mu=1/L$ adopted in the current paper.}  
\begin{equation}
	\label{eq:LambdaMuSwi}
	\frac{\Lambda^{(0)}_{\overline{\rm MS} }}{\mu_{\rm swi}}=0.2658(36)\,.
\end{equation}
With given $\bar{g}^{(0)}_{\rm GF}(\mu_{\rm dec})$, the  
ratio of scales ${\mu_{\rm swi}/{\mu_{\rm dec}}}$ can be obtained 
through the relation
\begin{equation}
	\label{eq:MuSwiMuDec}
	 \ln\bigg({\mu_{\rm swi}\over{\mu_{\rm dec}}}\bigg)=
	 \int^{\bar{g}^{(0)}_{\rm GF}(\mu_{\rm swi})}_{\bar{g}^{(0)}_{\rm GF}(\mu_{\rm dec})}
	 {\rmd g\over \beta_{\rm GF}^{(0)}(g)}\,,
\end{equation}
where $\beta_{\rm GF}^{(0)}$ is the non-perturbative $\beta$-function for $\bar{g}^{(0)}_{\rm GF}(\mu)$. 
In the region of couplings, $\bar{g}^{(0)}_{\rm GF}\in [2,11]$, a convenient parametrization 
for the needed $\beta$-function is given by 
\begin{equation}
\label{e:betaGF0param}
	\beta_{\rm GF}^{(0)}(\bar{g})=-{\bar{g}^3\over \sum_{k=0}^3 p_k\bar{g}^{2k}}\,,
\end{equation}
with coefficients 
 \begin{align}
 	p_0 = 14.93613381 \,,
 	\quad 
 	p_1 =  -1.03947429 \,,
 	\quad
 	p_2 = 0.18007512 \,,
 	\quad
 	p_3 = -0.01437036 \,,
 \end{align}
and covariance
\begin{multline}
 	{\rm cov}(p_i,p_j) = \\[1ex]
 	\left(
 	\begin{array}{cccc}
 		5.24669327\times 10^{\text{-1}} &   -3.26120586\times 10^{\text{-1}} &  6.03484522\times 10^{\text{-2}} &   -3.33454413\times 10^{\text{-3}}\\
 		-3.26120586\times 10^{\text{-1}} &   2.07627940\times 10^{\text{-1}} &  -3.91082685\times 10^{\text{-2}} &   2.19046893\times 10^{\text{-3}}\\
 		6.03484522\times 10^{\text{-2}} &  -3.91082685\times 10^{\text{-2}} &   7.47684098\times 10^{\text{-3}} &  -4.23948184\times 10^{\text{-4}}\\
 		-3.33454413\times 10^{\text{-3}} &   2.19046893\times 10^{\text{-3}} &  -4.23948184\times 10^{\text{-4}} &   2.42972883\times 10^{\text{-5}}\\
 	\end{array}
	\right)\,.
\end{multline}
In ref.~\cite{DallaBrida:2019wur} a similar representation of a 
pure gauge theory $\beta$-function is given which applies to 
the coupling defined by the electric components of the flow energy density instead of our choice eq.~\eqref{eq:GF}.
For completeness, we collect in Table~\ref{tab:Lambda0MuDec} the values for 
${\Lambda^{(0)}_{\overline{\rm MS} }}/{\mu_{\rm dec}}$ obtained by combining 
the results for the massive couplings in Table~\ref{tab:cont_res}, the matching 
relation, eq.~(\ref{eq:GFT2GF}), and the high-energy running, 
eqs.~(\ref{eq:LambdaMuSwi})-(\ref{eq:MuSwiMuDec}).

\begin{table}
	\centering
	\begin{tabular}{rrrrrr}
		\toprule
		& \multicolumn{5}{c}{$\Lambda_{\overline{\rm MS}}^{(0)}/\mu_{\rm  dec}$} \\
		\cmidrule(lr){2-6}	
		$z$ & $c=0.30$ & $c=0.33$  & $c=0.36$ & $c=0.39$ & $c=0.42$ \\ 
		\midrule  

          1.972  &  0.685(13) & 0.681(13) & 0.678(14)  & 0.677(14) & 0.678(14) \\ 
          4      &  0.723(15) & 0.720(15) & 0.719(16)  & 0.718(16) & 0.719(16) \\ 
          6      &  0.740(16) & 0.736(16) & 0.734(16)  & 0.732(16) & 0.731(16) \\ 
          8      &  0.760(18) & 0.756(18) & 0.753(17)  & 0.750(17) & 0.748(17) \\ 
          10     &  0.787(20) & 0.785(20) & 0.783(19)  & 0.781(19) & 0.780(19) \\ 
          12     &  0.802(21) & 0.799(21) & 0.797(21)  & 0.794(21) & 0.793(21) \\ 
		\bottomrule
	\end{tabular}
	\caption{Results for $\Lambda_{\overline{\rm MS}}^{(0)}/\mu_{\rm  dec}$ corresponding to
		different values of $\bar{g}^{(3)}_{\rm GFT,c }(\mu_{\rm  dec},M)$ of Table~\ref{tab:cont_res}.}
	\label{tab:Lambda0MuDec}
\end{table}

\clearpage

\section{Simulations}
\label{sec:simulations}
We use a variant of the open-source (GPL v2) package openQCD version 1.6~\cite{openQCD,Luscher:2012av} in plain C with MPI parallelization. This software has been successfully used in various large-scale projects. The simulations are performed on lattices of size $\frac{T}{a}\times \left(\frac{L}{a}\right)^3$. 
We impose Schr\"odinger Functional (SF) boundary conditions on the
gauge and fermion fields~\cite{Luscher:1992an,Sint:1993un}, {i.e.},
Dirichlet boundary conditions in Euclidean time at $x_0=0,T$, and
periodic (up to a phase $e^{i\theta}$ for fermionic fields, with $\theta=0.5$) boundary conditions with period $L$ in the three spatial directions. We use one-loop boundary improvement coefficients $c_{\rm t}$ and $\tilde c_{\rm t}$~\cite{Takeda:2003he,DallaBrida:2016kgh}.   
Our lattice setup uses non-perturbatively $\rmO(a)$-improved
Wilson fermions and the tree-level Symanzik $\rmO(a^2)$-improved gauge action~\cite{Bulava:2013cta};
we refer the reader to ref.~\cite{DallaBrida:2016kgh} for more details.
The bare (linearly divergent) quark mass is denoted by $m_0$ and
the pure gauge action has a prefactor  $\beta=6/g_0^2$.
The ensemble generation proceeds according to a variant of the Hybrid Monte-Carlo (HMC) algorithm~\cite{Duane:1987de}. The classical equations of motion are solved numerically for trajectories of length $\tau = 2$ in all simulations, leading to Metropolis proposals. In order to reduce the computational cost and obtain a high acceptance rate, we split the action and corresponding forces as follows. 
For the u/d quark doublet we use an even--odd preconditioned~\cite{DeGrand:1988vx} Dirac operator and Hasenbusch’s
mass factorization~\cite{Hasenbusch:2001ne} with  twisted masses~\cite{Hasenbusch:2002ai} of increasing values $\mu_0<\mu_1\ldots$ roughly set at equal
distances on a logarithmic scale as suggested in Ref.~\cite{Schaefer:2012tq}. 
The strange quark is simulated with the rational hybrid Monte-Carlo (RHMC) algorithm~\cite{Kennedy:1998cu,Clark:2006fx}, decomposing the fermion determinant into a reweighting factor
and a Zolotarev optimal rational approximation of $(\hat D^\dagger\hat D)^{-1/2}$~\cite{Achiezer:1992} 
in the spectral range $[r_a,r_b]$ of $\hat D^\dagger\hat D$ with a number of poles $N_p$. The algorithmic simulation parameters are summarized in \tab{tab:param}. The measurements (gradient flow observables, correlation functions and reweighting factors) are done during ensemble generation.

\begin{table}[h!]
\centering
\begin{tabular}{l l l}
\toprule
ensemble      & massless           & massive\\
\midrule
Force 0, level  & gauge, 0       & gauge, 0    \\
Force 1, level  & TM1-EO-SDET, 1$\qquad$ & TM1-EO-SDET, 1 \\
Force 2, level  & TM2-EO, 1            & TM2-EO, 1    \\
Force 3, level  & TM2-EO, 1            & TM2-EO, 2      \\
Force 4, level  & RAT-SDET, 1        & TM2-EO, 1      \\
Force 5, level  & -  & RAT, 2      \\
\midrule
Level 0, nstep & OMF4, 1        & OMF4, 1         \\
Level 1, nstep & OMF4, 8-10        & OMF4, 1       \\
Level 2, nstep & -       & OMF2, 6       \\
\midrule 
Solver 0, res & CGNE  $10^{-12}$      & CGNE $10^{-11}$        \\
Solver 1, res & DFL\_SAP\_GCR  $10^{-12}$      & DFL\_SAP\_GCR $10^{-11}$      \\
Solver 2, res & MSCG $10^{-12}$      & MSCG $10^{-11}$      \\
Solver 3, res & -        & CGNE $10^{-10}$        \\
Solver 4, res & -        & DFL\_SAP\_GCR $10^{-10}$      \\
Solver 5, res & -       & MSCG $10^{-10}$      \\
\midrule 
$a\mu_0$    & 0.0 & 0.0 \\
$a\mu_1$    & 0.1 & 0.1 \\
$a\mu_2$    & 1.5 & 1.0 \\
$N_p, [r_a,r_b]$ & 10-12, [0.01-0.1,7.0] & 8-9, [0.1,6.0]     \\
\bottomrule
\end{tabular}
\caption{Parameters of the algorithm: We give the forces with their integration levels, the integrators for the different levels and number of steps (nstep) per trajectory resp. outer level, the solvers for the various levels and their residue, the twisted-mass parameters $a\mu_i$, the number of poles $N_\mathrm{p}$ and the ranges $[r_a,r_b]$ used in the RHMC. For the force and solver name abbreviations we refer the interested reader to the openQCD documentation~\cite{openQCD}.}\label{tab:param}
\end{table}

\subsection{$\Nf=3$ renormalization runs}  
\label{app:massless}

For the determination of massive simulation parameters detailed in~\sect{app:renormass} we perform massless renormalization runs along the line of constant physics defined in~\eq{eq:LCP1}, { i.e.}, $\gGF^2(\Ldec) = 3.949$ and $\overline m = 0$, corresponding to a scale $\mudec = 789(15)\,\MeV$ or  
fixed physical volume with $L_\text{dec}=0.250(5)$ fm~\cite{Bruno:2017gxd,DallaBrida:2016kgh}. We generate a large set of new ensembles around $Lm_{\rm PCAC}=0$ for $L/a=12, 16, 20, 24, 32, 40$ and $48$, where we measure reweighting factors and the following observables, as well as the covariance between them:
\begin{itemize}
\item the gradient flow coupling $\gGF^2$ defined in \eq{eq:GF}, using the magnetic components of the energy density
\beq
   \gGF^2(L) = \mathcal{N}^{-1}t^2 \left.\frac{\langle \hat\delta(Q) E_{\rm mag}(t,T/2)\rangle}{\langle \hat \delta(Q)\rangle}\right|_{\sqrt{8t}=cL}\,.%
\eeq
The quantity $\hat \delta(Q)$ is zero whenever $|Q|>0.5$ and one otherwise, and $Q$ is the topological charge computed with the clover discretization of the field strength tensor built from gauge fields at flow time $\sqrt{8t}=cL$. At $L=L_{\rm dec}$, the volume is small enough for non-zero topological sectors to be highly suppressed. In practice, we do not see any configuration with $Q$ different from zero and the projection onto the $Q=0$ sector has no effect in our simulations. 
\item the current quark mass $m_{\rm PCAC}$, defined through
the partially conserved axial current (PCAC) relation~\cite{Gell-Mann:1960mvl}, in its $\Oa$-improved definition
\begin{gather}
	\label{eq:mpcac} m_{\rm PCAC}(g_0^2,m_0) =
\left.\frac{\half(\partial_0^*+\partial_0)\fA(x_0)+c_A
a\partial_0^*\partial_0\fP(x_0)}{2\fP(x_0)}\right|_{x_0=T/2}\,,
\end{gather}
with an axial current improvement coefficient $c_A$ from~\cite{Bulava:2015bxa} and SF correlation functions
$\fA$ and $\fP$ that are given by the correlation between a pseudo-scalar boundary operator and a local iso-vector axial current or pseudoscalar density, projected to zero spatial momentum at time $x_0$ and $Q=0$ (see above).
\item the pseudoscalar renormalisation constant $\ZP$ in the SF-scheme~\cite{Capitani:1998mq}, 
defined through the renormalization condition
\bea
\ZP(\mu=1/L) \frac{\fP(L/2)}{\sqrt{3f_1}}=\left.\frac{\fP(L/2)}{\sqrt{3f_1}}\right|_{\rm tree\text{-}level}, 
\eea
where $f_1$ is the ``boundary-to-boundary'' correlator, again projected to $Q=0$.\\
A precise definition of all SF correlation functions can be found for instance in \cite{Capitani:1998mq}.
\end{itemize}
The renormalization runs are summarized in Table~\ref{tab:massless} together with all measured observables and their integrated autocorrelation times $\tau_{\rm int}$  in units of measurements (every five trajectories, {\it i.e.}, 10 MDUs). 

\begin{table}[h!]
\footnotesize
\thisfloatpagestyle{empty}
\begin{tabular}{ccccccccccc}
\toprule  
$L/a$  &$ \beta   $&$ \kappa $&$ N_{\rm rep}$&$ N_{\rm ms}$ & $Lm_{\rm PCAC}$ & $\tau_{\rm int}$ & $\gGF^2$& $\tau_{\rm int}$ & $\ZP$& $\tau_{\rm int}$ \\
\midrule
12 	&  4.3020 	& 0.1324810	& 3	& 5400	& 1.3164(9)	& 0.89	& 4.167(8)		& 1.1	& 0.6400(8	& 1.6 \\
	&		& 0.13364602 	& 3	& 3014	& 0.8825(10)	& 0.84	& 4.080(11)	& 1.3	& 0.6190(9)	& 1.5 \\
       	&	     	& 0.13481160  	& 3	& 3000 	& 0.4486(11)	& 0.87	& 4.031(11)	& 1.4	& 0.5990(10)	& 1.6 \\
       	&	    	& 0.13540205	& 5	& 3748 	& 0.2260(8) 	& 0.64 	& 3.982(9) 	& 1.3 	& 0.5890(10) 	& 1.5 \\
       	&           	& 0.13575881	& 5	& 4159 	& 0.0890(8) 	& 0.66 	& 3.970(8) 	& 1.1 	& 0.5816(10) 	& 1.6 \\
       	&           	& 0.1359977	& 6	& 2637 	& -0.0016(11) 	& 0.65 	& 3.950(14) 	& 1.4 	& 0.5770(16) 	& 1.8 \\
       	&           	& 0.13623743	& 5	& 2215 	& -0.0915(11) 	& 0.50 	& 3.919(11) 	& 1.0 	& 0.5724(21) 	& 2.7 \\
       	&            	& 0.13659861	& 5	& 1865 	& -0.2289(15) 	& 0.71 	& 3.911(13) 	& 1.3 	& 0.5645(21) 	& 2.1 \\
       	&           	& 0.13720485	& 5	& 1613 	& -0.4710(16) 	& 0.67	& 3.878(15)	& 1.0	& 0.5504(30)	& 2.0 \\
\midrule
16  	& 4.4662	& 0.13222210	& 6	& 2176	& 1.6815(9)	& 0.74	& 4.179(17)	& 2.3	& 0.6231(15)	& 3.4 \\
	&		& 0.13333741 	& 6	& 6235	& 1.1316(6)	& 0.79	& 4.082(9)		& 1.8	& 0.6055(8)	& 2.5 \\
	&		& 0.13445845 	& 6	& 5401	& 0.5768(6)	& 0.67	& 4.017(10)	& 2.1	& 0.5887(9)	& 2.8 \\
       	&           	& 0.13503421 	& 9	& 5006 	& 0.2898(6) 	& 0.68 	& 3.981(10) 	& 2.0 	& 0.5784(11) 	& 2.8 \\
	&		& 0.13536894 	& 6	& 5115	& 0.1213(6)	& 0.73	& 3.933(10)	& 2.3	& 0.5734(11)	& 2.6 \\
	&		& 0.13582883	& 6	& 4871	& -0.1115(7)	& 0.79	& 3.936(10)	& 2.0	& 0.5622(13)	& 2.9 \\
	&		& 0.13617580	& 6	& 4752	& -0.2898(7)	& 0.72	& 3.908(9)		& 1.6	& 0.5545(13)	& 2.2 \\
\midrule
20   	& 4.5997 	& 0.13309781 	& 12	& 1715	& 1.3609(9)	& 0.75	& 4.142(27)	& 5	& 0.5947(20)	& 5 \\
	&	  	& 0.13418441 	& 12	& 1296	& 0.6924(11)	& 0.88	& 4.035(33)	& 4	& 0.5821(20)	& 3 \\
	&		& 0.13473439	& 12	& 1798	& 0.3483(9)	& 0.68	& 3.976(29)	& 6	& 0.5731(25)	& 6 \\
       	&           	& 0.13506654	& 10	& 1385	& 0.1422(9)	& 0.60	& 3.983(30)	& 5	& 0.5646(46)	& 11 \\
       	&           	& 0.13506654	& 10	& 1370	& 0.1402(10)	& 0.69	& 3.958(21)	& 3	& 0.5680(21)	& 3 \\
       	&           	& 0.13551198	& 10	& 1193	& -0.1387(12)	& 0.88	& 3.963(29)	& 4	& 0.5542(34)	& 4 \\
       	&           	& 0.13551198	& 10	& 1271	& -0.1375(12)	& 0.84	& 3.969(23)	& 3	& 0.5597(29)	& 3 \\
       	&           	& 0.13584798	& 10	& 1004	& -0.3545(14)	& 0.96	& 3.918(21)	& 2.4	& 0.5499(31) 	& 3 \\
\midrule
24   	& 4.7141 	& 0.13289106	& 11	& 2662	& 1.5854(6)	& 0.66	& 4.174(25)	& 5	& 0.5863(13)	& 4 \\
	&	  	& 0.13394806 	& 11	& 2805	& 0.8048(6)	& 0.75	& 4.058(21)	& 5	& 0.5755(17)	& 7 \\
	&	    	& 0.13448288	& 11	& 2907 	& 0.4066(6) 	& 0.64 	& 3.989(22) 	& 5 	& 0.5638(23) 	& 8 \\
      	&           	& 0.13480584	& 10	& 2564 	& 0.1642(8) 	& 0.90 	& 4.007(23) 	& 4 	& 0.5619(25) 	& 8 \\
       	&           	& 0.13523886	& 10	& 1996	& -0.1628(8) 	& 0.75 	& 3.959(29) 	& 6 	& 0.5507(27) 	& 5 \\
       	&           	& 0.13556545	& 10	& 1906	& -0.4101(10)	& 0.96	& 3.998(28)	& 6	& 0.5455(42)	& 9 \\
\midrule
32  	& 4.9000 	& 0.13256312 	& 20	& 2709	& 2.0170(5)	& 0.69	& 4.150(24)	& 6	& 0.5714(20)	& 9 \\
	&	  	& 0.13357336 	& 10	& 2045	& 1.0261(6)	& 0.70	& 4.075(44)	& 11	& 0.5663(28)	& 12 \\
	&	  	& 0.13357336 	& 10	& 660	& 1.0261(11)	& 0.82	& 4.128(69)	& 10	& 0.5687(40)	& 8 \\
	&	    	& 0.13408427 	& 10	& 2084	& 0.5197(6)	& 0.77	& 3.970(39) 	& 10	& 0.5604(25) 	& 11 \\
	&	    	& 0.13408427 	& 10	& 603	& 0.5208(26)	& 0.51	& 4.063(80) 	& 12	& 0.5665(29) 	& 5 \\
	&           	& 0.13439270 	& 16	& 2666	& 0.2120(5)	& 0.66	& 3.948(29)	& 10	& 0.5685(27)	& 12 \\
       	&           	& 0.13439270 	& 10	& 909	& 0.2141(10)	& 0.87	& 3.930(31)	& 4	& 0.5517(28)	& 6 \\
       	&           	& 0.13480615 	& 16	& 1044	& -0.2016(8)	& 0.70	& 3.920(42)	& 9	& 0.5382(57)	& 9 \\
       	&           	& 0.13511791 	& 10	& 610	& -0.5188(12)	& 0.76	& 3.917(64)	& 10	& 0.5337(88)	& 10 \\
\midrule
40 & 5.0671 & 0.132280 & 4 & 1255 & 2.4246(6) & 0.65 & 4.134(42) & 11 & 0.5600(59) & 33 \\
& & 0.133250 & 4 & 860 & 1.2345(7) & 0.68 & 4.069(62) & 16 & 0.5585(28) & 7 \\
& & 0.133740 & 4 & 746 & 0.6297(7) & 0.53 & 3.938(60) & 11 & 0.5664(57) & 20\\
& & 0.134036 & 4 & 800 & 0.2618(8) & 0.67 & 3.955(44) & 6 & 0.5511(37) & 7 \\
& & 0.134234 & 4 & 845 & 0.0146(9) & 0.81 & 3.852(87) & 23 & 0.5544(80) & 15 \\
& & 0.134433 & 4 & 738 & -0.2360(9) & 0.78 & 3.979(43) & 6 & 0.5330(10) & 17 \\
& & 0.134732 & 4 & 981 & -0.6150(9) & 0.91 & 3.899(42) & 8 & 0.5449(97) & 19 \\
& & 0.135233 & 4 & 564 & -1.2538(12) & 0.60 & 3.804(47) & 6 & 0.5423(169) & 12 \\
\midrule
48 & 5.1739 & 0.13211537 & 3 & 569 & 2.8273(10) & 0.65 & 4.221(60) & 8 & 0.5525(64) & 16 \\
   &        & 0.13306215 & 3 & 562 & 1.4350(11) & 0.65 & 4.066(61) & 6 & 0.5472(61) & 14 \\
   &        & 0.13354065 & 3 & 596 & 0.7253(12) & 0.82 & 3.983(70) & 11 & 0.5274(62) & 15 \\
   &        & 0.13382941 & 3 & 610 & 0.2946(12) & 0.82 & 4.391(136) & 17 & 0.5365(10) & 16 \\
   &        & 0.13402261 & 3 & 616 & 0.0017(11) & 0.74 & 4.147(140) & 16 & 0.5543(71) & 13 \\
   &        & 0.13421637 & 3 & 575 & -0.2863(10) & 0.43 & 3.819(74) & 7 & 0.5320(13) & 15 \\
   &        & 0.13450805 & 3 & 579 & -0.7291(18) & 1.07 & 3.965(63) & 6 & 0.5564(95) & 7 \\
   &        & 0.13499703 & 2 & 200 & -1.4751(25) & 1.43 & 3.911(65) & 6 & 0.5830(272) & 8 \\
\bottomrule
\end{tabular}
 \caption{Results for $Lm_{\rm PCAC}$, $\gGF^2$ and $\ZP$ from massless renormalization runs and their integrated autocorrelation times $\tau_{\rm int}$  in units of measurements (every five trajectories, {i.e.}, 10 MDUs).}
 \label{tab:massless}
\end{table}

\subsection{$\Nf=3$ massive}
\label{app:massive}

We simulate massive quarks with $M\approx1.6, 3.2, 4.7, 6.3, 7.9,9.5$ GeV, corresponding to $z=LM=1.972, 4, 6, 8, 10, 12$ on $T\times L^3$  and lattices with $L/a=12, 16, 20, 24, 32, 40, 48$ and $T=2L$.
The determination of simulation parameters for a given $z$ value is discussed in the next section. 
The results for the GFT couplings using the Zeuthen flow were given in \tab{tab:cont_res}. 
When approaching small lattice spacings, the HMC algorithm is known to suffer from critical slowing down.
Since we have chosen the trajectory length constant in all our runs, our flow measurements exhibit the expected Langevin scaling $\tau_{\rm int}\propto a^{-2}$, which however is still manageable on our finest lattice. The lattice size $L_{\rm dec}$ is small enough for non-zero topological sectors to be highly suppressed.
For our error analysis we use the $\Gamma$-method \cite{Wolff:2003sm}. 

\begin{figure}[h]
\centering
\includegraphics[width=.75\textwidth]{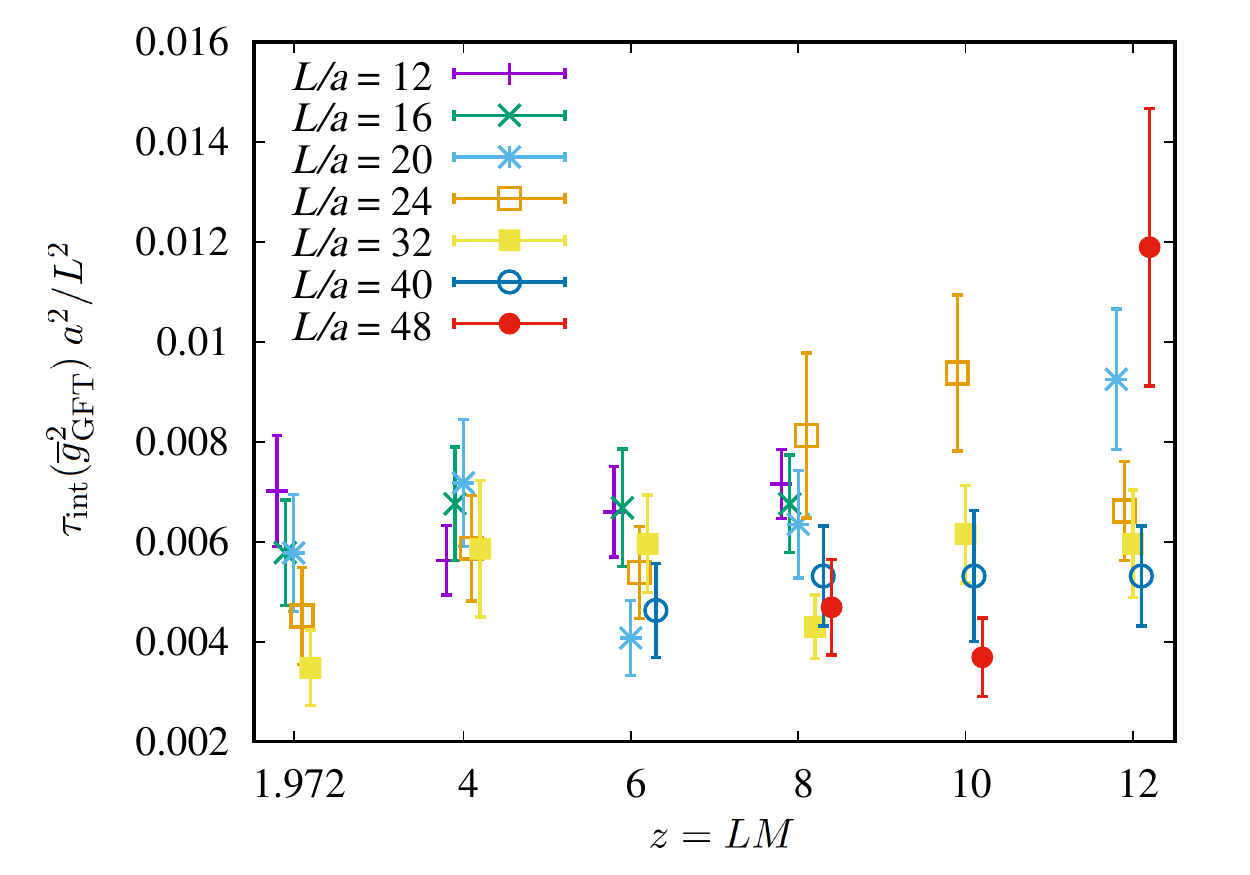}
\caption{Integrated autocorrelation times $\tau_{\rm int}(\gGFT^2)$ of the gradient flow coupling  in units of measurements (every five trajectories, {i.e.}, 10 MDUs) scaled with $a^2/L^2$ vs. the dimensionless quantity $z=LM$ of our massive simulations. The points are slightly shifted from actual $z$ values for better visibility.}\label{fig:tauint}
\end{figure}

\clearpage

\section{Mistuning Corrections}
\label{app:betashifts}
To correct the mistuning of the LCP, the slope
\begin{equation}
   S = \left.\frac{\partial \gGF^2}{\partial \tilde g_0^2}\right|_{L/a} = \frac{\gGF \beta^{(3)}_{\rm GF}(\gGF)}{g_0\beta^{(3)}_0(g_0)}
      + \rmO(a^2)\, ,
\end{equation}
is required. To compute it, the non-perturbative $\beta_{\rm GF}$-function of the gradient flow coupling at $\gGF^2=3.949$ and the non-perturbative 
bare $\beta_0$-function 
at the bare couplings of our LCP are needed. The first has been determined to high precision in~\cite{DallaBrida:2016kgh} and is given by
\begin{equation}
	\beta_{\rm GF}^{(3)}(\sqrt{3.949}) = -0.471 \pm 0.004\, .
\end{equation}
The bare $\beta_0$-function can be estimated by parameterizing the $L/a\ge 20$ data in the first two columns of \tab{tab:lcp} according to
\begin{equation}
   \ln\left(\frac{L}{a}\right) \approx {\rm const.} + \frac{1}{2k_3 g_0^2}\, ,
\end{equation}
the $\beta_0$ function in this short range of couplings is then given by
\begin{equation}
   \beta_0^{(3)} \approx -k_3 g_0^3, \qquad \text{with}\qquad k_3 = 0.054\, .
\end{equation}
With this slope the $\beta_{\rm LCP}$ values in the last column of \tab{tab:lcp} are determined. From these then the 
desired simulation parameters for the massive runs follow, as described in \sect{sec:massive-line-const-phys}. 

In practice, the massive runs were
performed at slightly different parameters, and need to be corrected as well.
The relevant slope here is given by the dependence of the massive coupling on the bare one
\begin{equation}
   \left.\frac{\partial \gGFT^2}{\partial \tilde g_0^2}\right|_{z,L/a}
	= \left.\frac{\partial \gGFT^2}{\partial \ln(L)}\right|_{z,L/a} \left( \frac{{\rm d}\tilde g_0^2}{{\rm d}\,\ln(a)} \right)^{-1}\, .
\end{equation}
The last factor is again related to the bare $\beta_0$-function, and the same parameterization as above can be used. The first
factor, however, 
has never been determined but can be approximated based on decoupling. Matching of the three-flavor and the pure gauge theories means, that
$\Lambda^{(0)}(M) = \Lambda^{(3)} \, P(M/\Lambda^{(3)})$. For physical couplings in the continuum limit we then have
\begin{equation}
   [\gGFT^{(3)}(L,z)]^2 = [\gGFT^{(0)}(L)]^2 + \rmO(z^{-2}).
\end{equation}
The $\Nf=0$ coupling and $\Lambda^{(0)}$ are related 
\begin{eqnarray}
   [\gGFT^{(0)}(L)]^2 &=& F(L \Lambda^{(0)}) \\
	              &=& F\left(L \Lambda^{(3)} P\left(\frac{z}{L\Lambda^{(3)}}\right)\right) \\
		      &=& [\gGFT^{(3)}(L,z)]^2 + \rmO(z^{-2})\, .
\end{eqnarray}
The exact form of the function $F$ will not be needed, but it is essentially the inverted relation \eq{eq:LRGI1}
together with a scheme change of the $\Lambda$ parameter. Taking the $L$-derivative of the second line and then
using decoupling to express everything in terms of $\Nf=0$ quantities results in
\begin{equation}
	L\frac{\partial [\gGFT^{(3)}]^2}{\partial L} = -2 \gGFT^{(0)} \beta^{(0)}_{\rm GFT} (\gGFT^{(0)}) \left(1-\eta^M \right)  + \rmO(z^{-2})\, ,
\end{equation}
with $\eta^M = \frac{M}{P} \, \left.\frac{\partial P}{\partial M}\right|_{\Lambda^{(3)}}= 2/11 + \rmO(\alpha(m_\star))$, where the $\rmO(\alpha(m_\star))$
corrections are very small \cite{Athenodorou:2018wpk} and can safely be dropped.

The shifts are applied to couplings at finite $z$ and finite $a$. The leading corrections to the $z\to\infty$, $a\to 0$
limit are built into the extrapolation formulae \eq{eq:global} and \eq{eq:mtoinfty}, which is why $p_1$, $p_2$ and $B$ are known in
\begin{eqnarray}
	[\gGFT^{(3)}]^2 &=& F\left(L \Lambda^{(3)}_{\rm eff} P\left(\frac{z}{L\Lambda^{(3)}_{\rm eff}}\right)\right) + (p_1+p_2z^2) \frac{a^2}{L^2}\, , \\
        \Lambda^{(3)}_{\rm eff} &=& \Lambda^{(3)} + \frac{B}{z^2}\, .
\end{eqnarray}
The $L$-derivative of the coupling at finite $z$ and $a$ has then corrections $R_{\rm a}$ and $R_{\rm z}$ of the form
\begin{equation}
	L\frac{\partial [\gGFT^{(3)}]^2}{\partial L} = -2 \gGFT^{(0)} \beta^{(0)}_{\rm GFT} (\gGFT^{(0)}) \left(1-\eta^M \right) \left[ 1+ R_{\rm z} + R_{\rm a}\right]\, ,
\end{equation}
which to leading order are given by
\begin{eqnarray}
   R_{\rm z} &\approx& \frac{B/\Lambda}{z^2}4 [\gGFT^{(0)}]^2\, k_0\, , \\
   R_{\rm a} &\approx& \frac{-2(p_1 +p_2 z^2)\frac{a^2}{L^2}}{-2 \gGFT^{(0)} \beta^{(0)}_{\rm GFT} (\gGFT^{(0)} \left(1-\eta^M \right)}\, .
\end{eqnarray}
Finally, $k_0$ parameterizes the $\beta^{(0)}_{\rm GFT}$-function in the relevant range of couplings
\begin{equation}
   \beta^{(0)}_{\rm GFT}(\gGFT) \approx -k_0\, \gGFT^3\, . 
\end{equation}
Its value is $0.079$ and $0.076$ for $c=0.3$ and $c=0.36$, respectively.
 
\section{Mass renormalization}
\label{app:renormass}
To compute the strong coupling from decoupling relations, simulations at constant
renormalized mass are necessary. This appendix describes, how a given value of the 
RGI quark mass in GeV translates into a hopping parameter at a given $\beta$ value. 

In a first step the RGI mass $M$ is translated into a renormalized mass $\overline{m}(\mu_{\rm dec})$ in the SF scheme
\begin{equation}
	\overline{m}(\mu_{\rm dec}) = M\times \frac{\overline m (\mu_0/2)}{M} \times \frac{\overline{m}(\mu_{\rm dec})}{\overline{m}(\mu_0/2)}\, .
\end{equation}
Here ${M}/{\overline{m}(\mu_0/2)} = 1.7505(89)$ has been
determined in~\cite{Campos:2018ahf} for a scale $\mu_0$ 
defined by $\overline{g}^2_{\rm SF}(\mu_0)=2.0120$, which corresponds to $\gGF^2(\mu_0/2) = 2.6723(64)$. For the
(short) running of the mass
\begin{equation}
	\frac{\overline{m}(\mu_{\rm dec})}{\overline{m}(\mu_0/2)} = 
	 \exp\left\{-\int\limits_{\overline{g}(\mu_{\rm dec})}^{\overline{g}(\mu_0/2)} \frac{\tau(x)}{\beta(x)}\, dx \right\}\, ,
\end{equation}
a parametrization of non-perturbatively determined~\cite{Campos:2018ahf} $\tau/\beta$ data is used
\begin{equation}
        \frac{\tau(\gGF)}{\beta(\gGF)} = \frac{1}{\gGF}\left(f_0 + f_1 \gGF^2 +f_2 \gGF^4 + f_3\gGF^6 \right)\, ,
\end{equation}
with $f_0=1.28493, f_1=-0.292465, f_2=0.0606401, f_3=-0.00291921$. The parameters are correlated, namely
\begin{equation}
  \begin{split}
    {\rm cov}(f_i,f_j) =
    &\left(
      {\footnotesize
      \begin{array}{cccc}
  \phantom{+}2.33798\times 10^{-2} & -1.47011\times 10^{-2} &  \phantom{+}2.81966\times
                                                   10^{-3} &
                                                             -1.66404\times
                                                             10^{-4} \\
 -1.47011\times 10^{-2} &  \phantom{+}9.54563\times 10^{-3} & -1.87752\times
                                                   10^{-3} &
                                                             \phantom{+}1.12962\times
                                                           10^{-4} \\
  \phantom{+}2.81966\times 10^{-3} & -1.87752\times 10^{-3} &  \phantom{+}3.78680\times
                                                   10^{-4} &
                                                             -2.32927\times
                                                             10^{-5} \\
 -1.66404\times 10^{-4} &  \phantom{+}1.12962\times 10^{-4} & -2.32927\times
                                                   10^{-5} &
                                                             \phantom{+}1.46553\times
                                                           10^{-6} \\
    \end{array}
  }
  \right)\,.
  \end{split}
\end{equation}
This allows to obtain
\begin{equation}
  \frac{M}{\overline{m}(\mudec)} = 1.474(11).
\end{equation}
This renormalized mass can then be related to either the bare subtracted quark mass $\mq(g_0^2,m_0) = m_0 - m_{\rm crit}(g_0^2)$ or 
to the bare PCAC mass $m_{\rm PCAC}(g_0^2,m_0)$.
\begin{eqnarray}\label{eq:mbar}
	\overline m(\mu_{\rm dec}) &=& m_{\rm PCAC}(g_0^2,m_0) \frac{\ZA(\tilde g_0^2)}{\ZP(\tilde g_0^2,a\mu_{\rm dec})} \left[1+ (\bA(g_0^2)-\bP(g_0^2)) a \mq(g_0^2,m_0) \right] \label{eq:mbarmpcac}\\
			 &=& \mq(g_0^2,m_0)  \Zm(\tilde g_0^2,a\mu_{\rm dec}) \left[1+ \brm(g_0^2) a \mq(g_0^2,m_0) \right]\, ,
\end{eqnarray}
valid only up to cutoff effects of $O(a^2)$ with the improved bare coupling~\cite{Luscher:1996sc}
\begin{equation}
\tilde g_0^2 = g_0^2 \left(1 + \bg(g_0^2)a \mq\right)\, .
\end{equation}
The second relation is more useful, since it connects $\overline m$ directly to the simulation parameter $\kappa$:
$a\mq = am_0-am_{\rm crit}(g_0^2) = \frac{1}{2\kappa}-\frac{1}{2\kappa_c}$. On the other hand, the renormalization factors
in the first relation are easier to compute, or already known. The missing piece is then the relation
between $m_{\rm PCAC}$ and $\mq$,
\bea
	&&m_{\rm PCAC}(g_0^2,m_0) = \label{e:zhatbhat}\\ && \; \underbrace{\frac{\ZP(\tilde g_0^2,a\mu_{\rm dec}) \Zm(\tilde g_0^2,a\mu_{\rm dec})}{\ZA(\tilde g_0^2)}}_{\equiv \hat Z(\tilde g_0^2)} \mq(g_0^2,m_0) \left[1+ \underbrace{[\brm(g_0^2)+\bP(g_0^2)-\bA(g_0^2)]}_{\equiv \hat b(g_0^2)} a \mq(g_0^2,m_0)\right]\, ,\non
\eea
with renormalization factor $\hat Z(\tilde g_0^2)$ and improvement coefficient $\hat b(g_0^2)$, which
can both be obtained non-perturbatively from our close-to-massless simulations. 
More precisely, since these runs vary $m_0$ around $m_{\rm crit}$ while keeping $g_0$ constant (instead of $\tilde g_0$), what we have access to
are $\hat Z$ and $\hat b^{\rm eff}$ in
\begin{equation}
	m_{\rm PCAC}(g_0^2,m_0) = \hat Z(g_0^2) \mq(g_0^2,m_0) \left[1 + 
	\underbrace{\left(\hat b(g_0^2) + g_0^2 \bg(g_0^2)\frac{\hat Z'(g_0^2)}{\hat Z(g_0^2)}\right)}_{\equiv \hat b^{\rm eff}(g_0^2)}a\mq(g_0^2,m_0) \right]\, .
\end{equation}
First, we perform a simultaneous linear fit to the data in~\tab{tab:massless} at fixed $L/a$
	 \begin{eqnarray}
	    Lm_{\rm PCAC}   &=& c_0 + c_1 Lm_0 + c_2 (Lm_0)^2 \\
            \ZP             &=& c_3 + c_4 Lm_0 + c_5 (Lm_0)^2 \\
	    \gGF^2          &=& c_6 + c_7 Lm_0 + c_8 (Lm_0)^2 \, .
         \end{eqnarray}
Then $am_{\rm crit}$,  $\hat Z$, $\hat b^{\rm eff}$,  $\ZP$, $\gGF^2$ and $\rmd\,\gGF^2/\rmd\,am_0$
at zero quark mass, as well as their covariances are obtained from the fit parameters $c_i$.
The results of these fits can be found in Table~\ref{tab:fitpar}.

%
\begin{table}[h]
\centering
\begin{tabular}{ccccccc}
\toprule
$L/a$& $\beta$  & $am_{\rm crit}$        & $\hat Z$      & $\hat b^{\rm eff}$    & $Z_P$                 & $Z_A$ \\
\midrule
  12 & 4.3020 & -0.323417(38) & 1.1707(31) & -0.408(26) & 0.57708(82) & 0.8322(26) \\
  16 & 4.4662 & -0.312928(23) & 1.1613(22) & -0.486(29) & 0.56759(82) & 0.8432(34) \\
  20 & 4.5997 & -0.304289(24) & 1.1492(28) & -0.461(39) & 0.5628(15) & 0.8515(41) \\
  24 & 4.7141 & -0.296941(14) & 1.1455(12) & -0.497(18) & 0.5554(17) & 0.8582(47) \\
  32 & 4.9000 & -0.285427(12) & 1.1374(13) & -0.608(45) & 0.5446(35) & 0.8681(55) \\
  40 & 5.0671 & -0.275473(11) & 1.1259(12) & -0.490(19) & 0.5459(42) & 0.8762(62) \\
  48 & 5.1739 & -0.2693605(82) & 1.1233(10) & -0.517(17) & 0.5427(49) & 0.8810(66) \\
\bottomrule
\end{tabular}
 \caption{Results for $am_{\rm crit}$, $\hat Z, \hat b^{\rm eff}$ and $Z_P$ from fits and extrapolations of $Z_A$.}
 \label{tab:fitpar}
\end{table}
 
\begin{figure}
   \includegraphics[width=\linewidth]{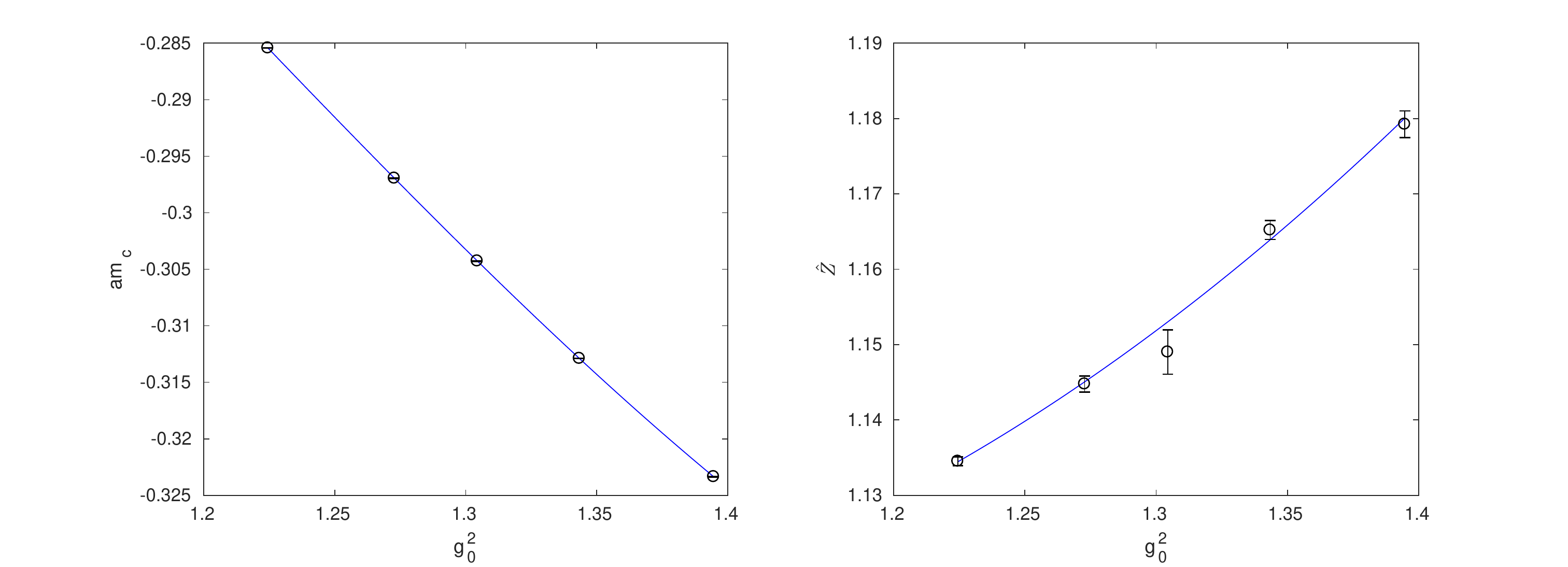}
   \caption{Critical mass and $\hat Z$ as a function of $g_0^2$.}\label{fig:amcZhat}
\end{figure}

To read off $\hat Z'/\hat Z$ to compute $\hat b$ from $\hat b^{\rm eff}$, and to obtain
the value $am_{\rm crit}(g_0^2)$ from data at slightly shifted $\tilde g_0^2$, we 
carry out another linear fit of the results in~\tab{tab:fitpar}
         \begin{eqnarray}
            am_{\rm crit} &=& d_0 + d_1 g_0^2 + d_2 g_0^4 + d_3 g_0^6 \\
	    \hat Z        &=& d_4 + d_5 g_0^2 + d_6 g_0^4\, .
         \end{eqnarray}
The result is shown in Fig.~\ref{fig:amcZhat}, asymptotic PT behavior is not built into
these fits.

\begin{figure}[h!]
   \centering
   \includegraphics[width=0.55\linewidth]{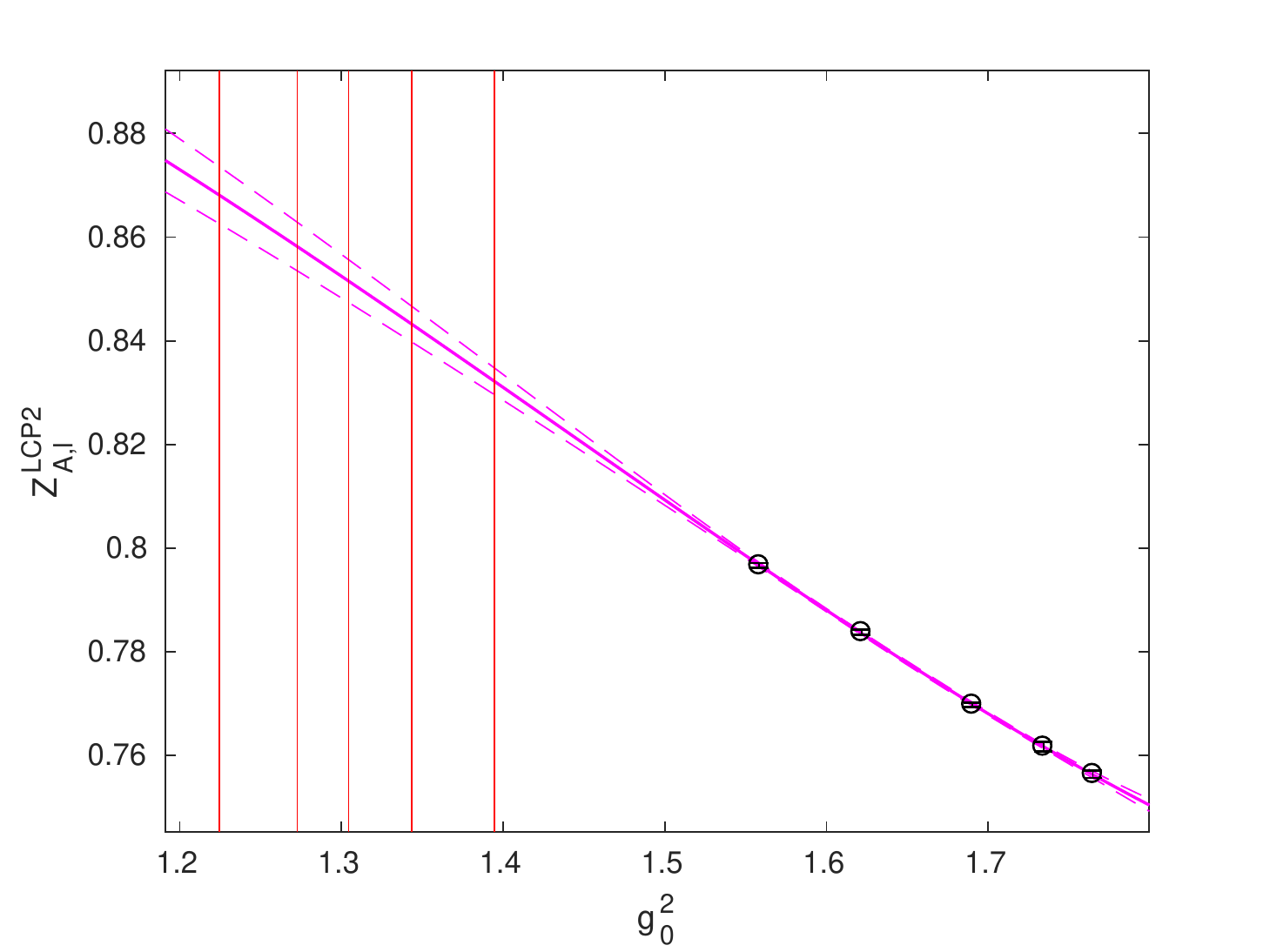}
   \caption{Extrapolations of $\ZA$.}\label{fig:ZA}
\end{figure}

The axial current renormalization $\ZA$ was computed very precisely based on universal relations between correlators in a chirally rotated Schr\"odinger Functional~\cite{DallaBrida:2018tpn}. The calculation was carried out
at the $\beta$ values at which also the large volume CLS simulations are carried out. 
These are coarser lattices, than the ones needed here, which means that an
extrapolation of the data is necessary. A fit, with restriction to 1-loop perturbation theory for
$g_0\to 0$, as proposed by the authors is
\begin{equation}
   \ZA = 1-0.090488 g_0^2 + c_1 g_0^4 + c_2 g_0^6 + c_3 g_0^8\, ,
   \label{e:zafit1}
\end{equation}
with $c_1 = 0.127163, c_2=-0.178785, c_3=0.051814$ and
\begin{equation}
   {\rm cov}(c_i,c_j) = 10^{-2} \times \begin{pmatrix} 
	   0.29841165 & -0.36050066 &  0.10868891\\
	  -0.36050066 &  0.43567202 & -0.13140137\\
	   0.10868891 & -0.13140137 &  0.03964605
   \end{pmatrix}\, .
\end{equation}
The right column of~\tab{tab:fitpar} lists the extrapolated renormalization factors at the necessary $\beta$ values, see also~\fig{fig:ZA}.

With known $\ZA$, $\ZP$ and $\hat Z$, the $\\ZmZm$ in the relation between the bare and the renormalized mass is determined non-perturbatively. 
The only missing pieces are the improvement 
coefficients $\bA$, $\bP$ and $\bg$, for which we take their 1 loop approximations~\cite{Sint:1995ch, Sint:1997jx,Aoki:1998qd}
\begin{eqnarray}
	\bA &=& \phantom{-}\,1 + 0.0881(13) \times \CF \ g_0^2\, ,\\
	\bP &=& \phantom{-}\,1 + 0.0889(14) \times \CF \ g_0^2\, ,\\
	\brm &=& -\frac{1}{2} - 0.0576(11) \times \CF \ g_0^2\, , \\
	\bg &=& \phantom{-}\,0 + 0.012000(2)\times  \Nf \ g_0^2\, .
\end{eqnarray}
The perturbative result for $\brm$ will not be used, except for testing PT, $\brm$ is instead obtained as
$\brm = \hat b-\bP+\bA$.

If $\tilde g_0$ denotes the bare coupling of the LCP defined by $\overline m = 0, \gGF^2=3.949$,
then the massive simulations need to be carried out at a slightly different bare coupling, in order to keep $\tilde g_0$ and hence the
volume constant. If we denote this slightly shifted coupling by $g_0^2$, we find that the bare parameters of the massive simulation,
$g_0^2$ and $m_0$, have to be chosen such that
\begin{eqnarray}
	g_0^2 &=& \tilde g_0^2 \left(1 - \bg(\tilde g_0^2) a \mq(g_0^2,m_0)\right)\\
	L\overline m(\mu_{\rm dec}) &=&
		      L\mq(g_0^2,m_0)  \Zm(\tilde g_0^2) \left[1+ \brm(\tilde g_0^2) a \mq(g_0^2,m_0) \right]\, 
\end{eqnarray}
hold.
Note that $b_x(g_0^2)a\mq = b_x(\tilde g_0^2)a\mq + O(a^2)$ was used to derive
these expressions. These equations can be explicitly solved for the simulation parameters in three steps
\begin{eqnarray}
	a\mq  &=& \frac{1}{2\brm}\left( \pm \sqrt{1+\frac{4 \brm L\overline m}{L/a\ \Zm}}-1 \right)\, , \label{eq:mpar1}\\
        g_0^2 &=& \tilde g_0^2 / \left(1 + \bg a \mq\right) \, \\
	am_0       &=& a\mq + am_{\rm crit}(g_0^2) \, .\label{eq:mpar3}
\end{eqnarray}

\begin{figure}[h]
\centering
\includegraphics[width=.695\textwidth]{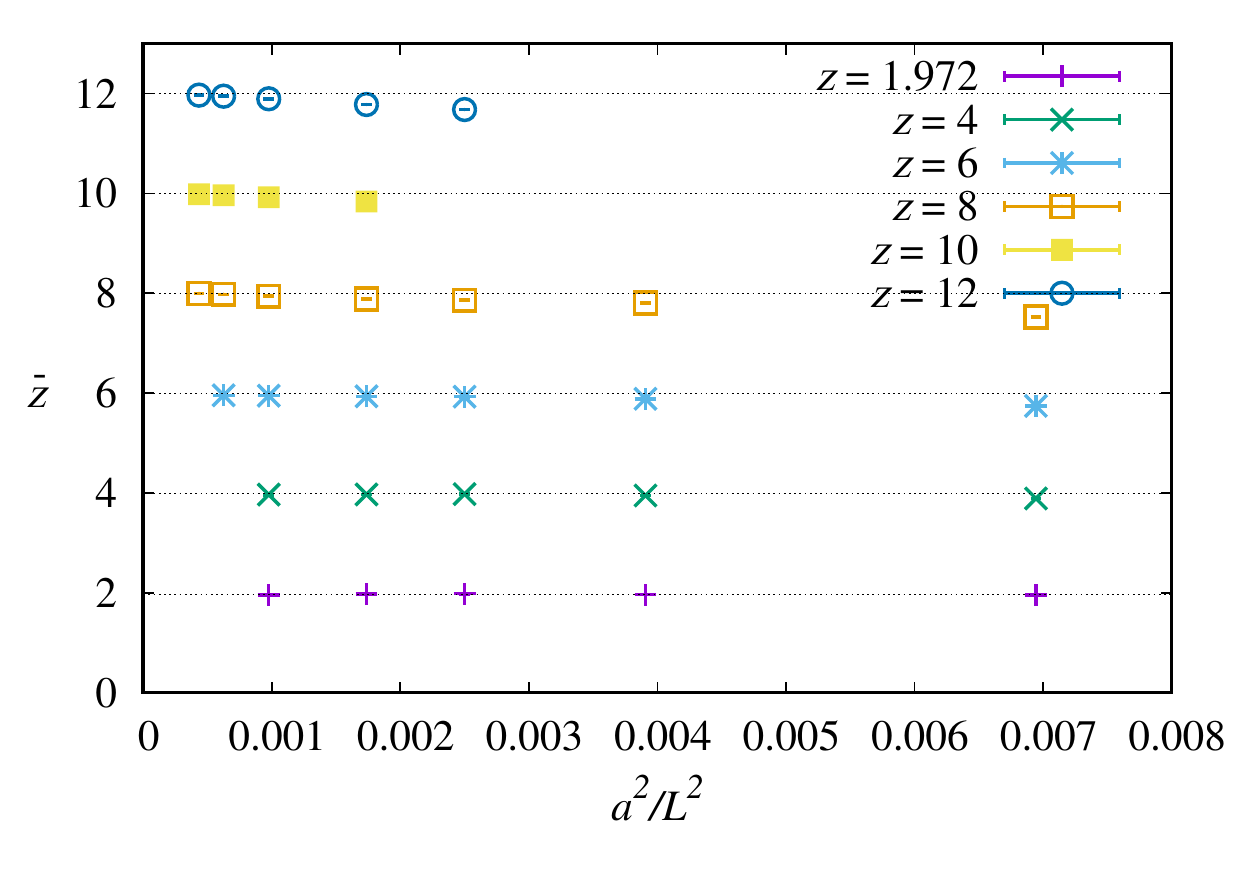}	
\caption{Overview of our massive simulations for $M=1.6\ldots9.5$ GeV, $T=2L$, showing the dimensionless RGI mass $\tilde z$, eq.~\eqref{e:ztilde} which agrees well with the input $z$-values prescribed in terms of the bare mass $\mq$.}\label{fig:lmz}
\end{figure}

The uncertainties in all quantities entering these equations are known\footnote{We treat
the difference between tree-level and 1-loop values as uncertainties of the improvement coefficients, that are known only perturbatively.}, 
which means that we know the precision of the massive simulation 
parameters, and conversely we control the precision that we can expect to reach for $z$. How an error on $z$ propagates
into the determination of $\Lambda$ is discussed below, but first, let us consider a cross-check of demonstrating that $z$ is fixed correctly.
Once the massive simulations have been carried out, we measure the PCAC 
masses\footnote{For the evaluation of the PCAC mass on the massive ensembles we use the improved derivative defined in~\cite{Guagnelli:2000jw}.}
and compute 
\begin{equation}
	\tilde z=Lm_{\rm PCAC}\frac{M}{\bar m(\mu)}\frac{\ZA(\tilde g_0^2)}{\ZP(\tilde g_0^2,a\mu)}[1+(
\bA-\bP)a\mq] = z+\rmO(a^2)
\label{e:ztilde}
\end{equation}
The results are shown in \fig{fig:lmz}. The precision is satisfactory, and the $\rmO(a^2)$ effects are reasonably small.

\clearpage

\section{Propagating the error of $z$}
\label{app:errz}
One would like to estimate which precision is necessary in fixing $M$ 
in order to achieve a certain precision in $\gGF^2(\mu,M) $ and then
in $\Lambda^{(3)}$ obtained from $\Lambda^{(0)}$. 

The most relevant place where
an uncertainty in $M$ matters is the $P$ function (to be precise $P_{0,3}$ in the
notation of \cite{Athenodorou:2018wpk}), which relates the $\Lambda$ parameters as
\begin{eqnarray}
 \Lambda^{(0)}_{\overline{\rm MS}} = P(M/\Lambda^{(3)}_{\overline{\rm MS}}) \, \Lambda^{(3)}_{\overline{\rm MS}}\,.
\end{eqnarray}
One can simply use this noting that 
\begin{eqnarray}
 \left.M\partial_M \Lambda^{(3)}_{\overline{\rm MS}}\right|_{\Lambda^{(0)}_{\overline{\rm MS}}} = 
	M\partial_M \frac{ \Lambda^{(0)}_{\overline{\rm MS}}}{P(M/\Lambda^{(3)}_{\overline{\rm MS}})} = - \Lambda^{(3)}_{\overline{\rm MS}}\eta^M \,,
\end{eqnarray}
and insert the known perturbation theory for $\eta^M = \frac{M}{P}\partial_M P$.
The function $\eta^M$ is known to 4 loops and the first two perturbative terms are (cf. \cite{Athenodorou:2018wpk})
\begin{equation}
        \eta^M \sim \frac{6}{33} + \frac{1926}{(4\pi)^2 33^2} g_\star^2
        + \ldots \,,
\end{equation}
where $g_\star=\bar{g}^{}_{\msbar}(m_\star)$ and $\overline{m}^{}_{\msbar}(m_\star)=m_\star$.
This is enough to estimate the required precision in the tuning of $M$.
 
Another place where $M$ mistunings need to be accounted for, is the continuum extrapolation
of the massive coupling. To propagate the $M$ uncertainty onto $\gbar^2$, its derivative with 
respect to the mass must be known.
 
For this we are going to use the fundamental formula Eq.~(\ref{eq:basic})
\begin{equation}
      {\Lambda^{(3)}_{\overline{\rm MS}}} =
    \mu_{\rm dec}\times \varphi_{\rm GF}^{(0)}(\bar g_{\rm GF}) \times
    \frac{\Lambda^{(0)}_{\overline{\rm MS}}}{\Lambda^{(0)}_{\rm GF}}
    \times
    P(M/\Lambda^{(3)}_{\overline{\rm MS}}) \,.
\end{equation}
where ${\bar g_{\rm GF}^2}$ is the massive coupling in the GF scheme in Eq.~(\ref{eq:GF}). 
First note that
\begin{equation}
  \varphi_{\rm GF}^{(0)}({\bar g_{\rm GF}}) = \exp\left\{
    -\int^{\bar g_{\rm GF}}\frac{{\rm d} x}{\beta_{\rm GF}^{(0)}(x)}
  \right\} \times \text{constant}\,,
\end{equation}
with 
$z=M/\mu_{\rm dec}$ only appearing in the first term. Now we can take the logarithm
\begin{equation}
  \log\Lambda_{\overline{\rm MS}}^{(3)} = -\int^{\bar
    g_{\rm GF}}\frac{{\rm d} x}{\beta_{\rm GF}^{(0)}(x)}
  + \log(P(\mu_{\rm dec}z/\Lambda_{\overline{\rm MS}}^{(3)})) + \text{$z$ independent terms}\,.
\end{equation}
Taking a derivative with respect to $z$ gives
\begin{equation}
  \frac{\partial \bar g _{\rm GF}}{\partial z} =
  -\beta_{\rm GF}^{(0)}(\bar g_{\rm GF})\partial_z \log P
\end{equation}
and finally 
\begin{equation}
  \label{eq:dzgs}
  \frac{\partial \bar g _{\rm GF}^2}{\partial z} = \frac{\eta^{M}}{z} \times (-2 \bar g _{\rm GF} \beta_{\rm GF}^{(0)}(\bar g_{\rm GF}))\,,
\end{equation}
where we have used $\partial_z \log P=\eta^{M}/z$.

\begin{figure}
    \centering
    \includegraphics[width=0.49\textwidth]{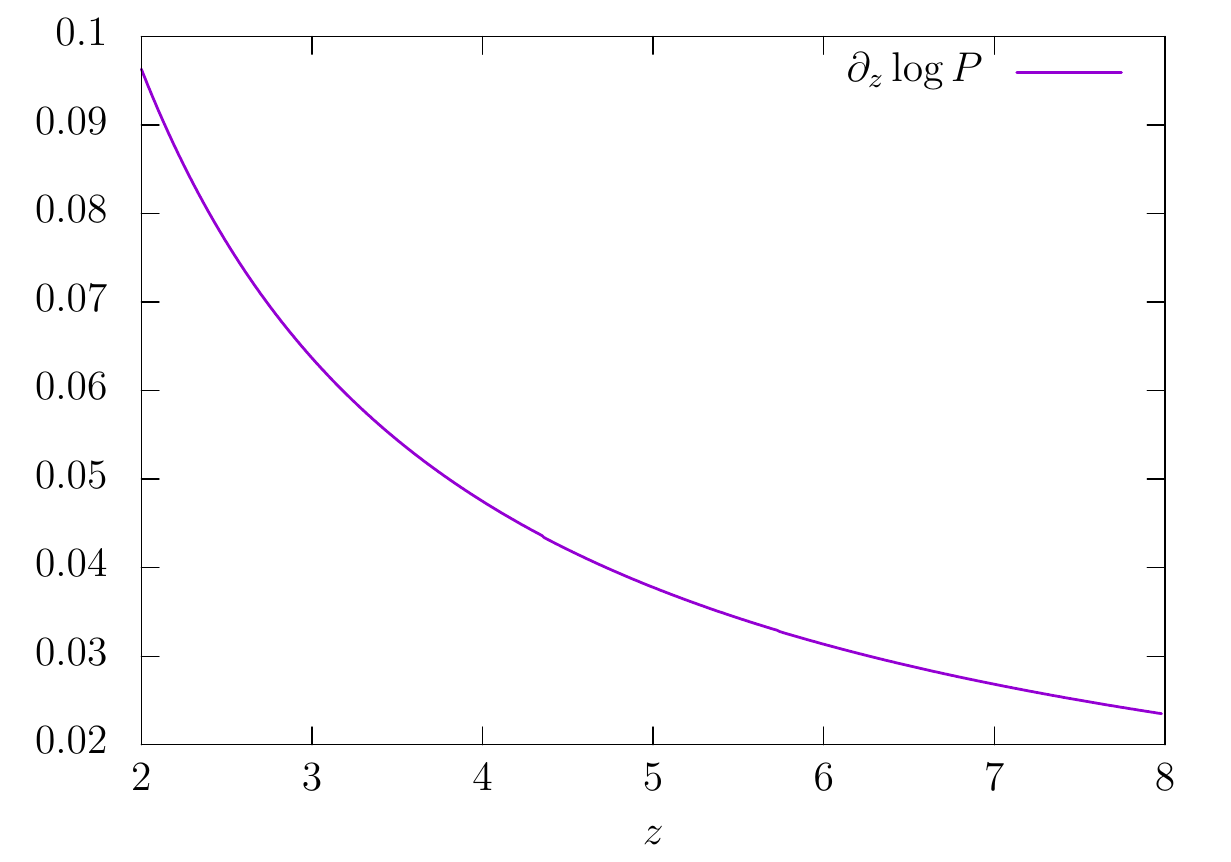}
    \includegraphics[width=0.49\textwidth]{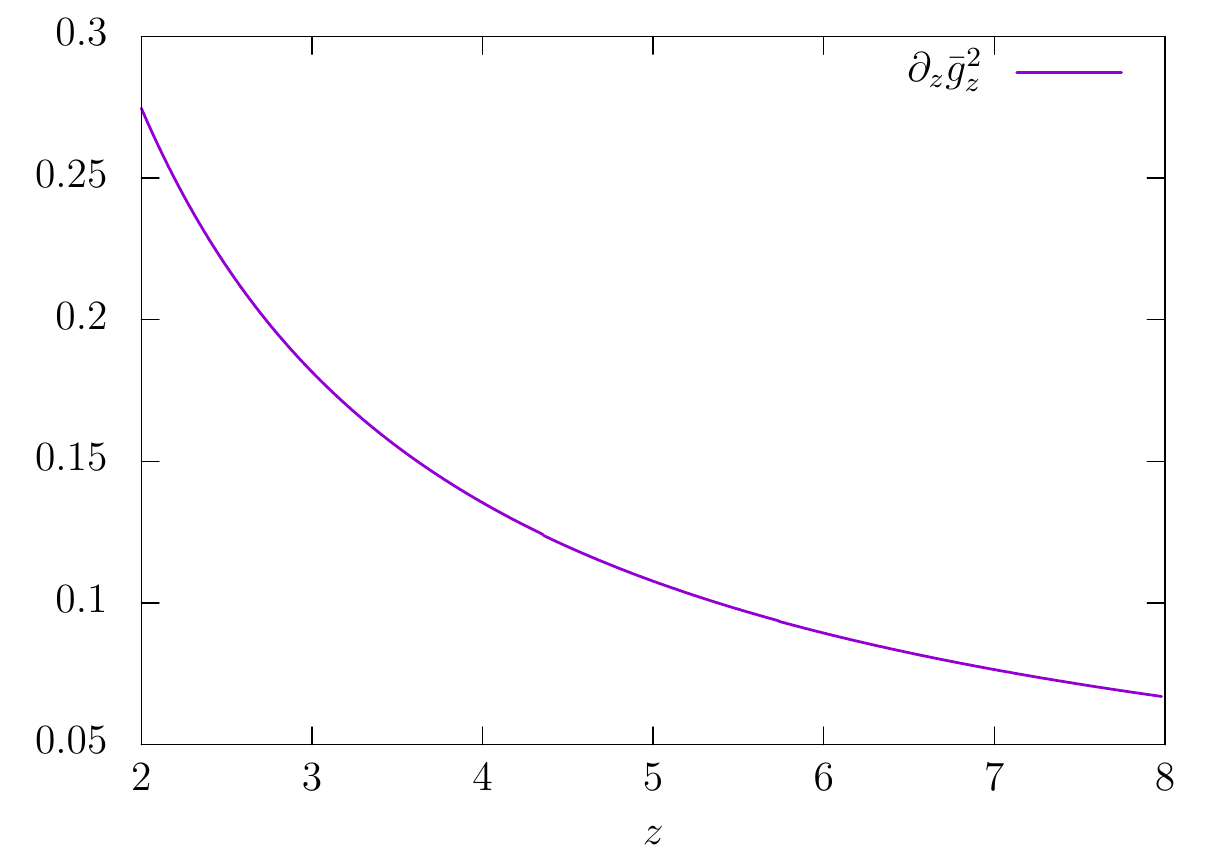}
  \caption{(a) The derivative $\partial_z\log P=\eta^M/z$. (b) The full
    derivative of the coupling Eq.~(\ref{eq:dzgs}).}
  \label{fig:dz}
\end{figure}

Figs.~\ref{fig:dz} show the results. The derivative is not that small,
especially at $z=4$ ($z=2$ is irrelevant).

Numerically we find with excellent approximation that 
\begin{equation}
  z\partial_z \log P \approx 0.19
\end{equation}
so that we will use for the error propagation the formula
\begin{equation}
  \frac{1}{\bar g _{\rm GF}^2}\frac{\partial \bar g ^2_{\rm GF}}{\partial z} =
  -\frac{0.38}{z \bar g _{\rm GF}}\beta_{\rm GF}^{(0)}(\bar g _{\rm GF})\,, 
\end{equation}
so that
\begin{equation}
  {\delta \bar g _{\rm GF}^2} =
  {0.38}{\bar g _{\rm GF}}\beta_{\rm GF}^{(0)}(\bar g _{\rm GF}) \frac{\delta z}{z} \,. 
\end{equation}

Finally, note that the massive coupling $\bar g_\text{GFT}^2$ is
different from $\bar g _{\rm GF}^2$ (because of $T=2L$, the mass effects and possibly the different
values of $c$). The relation between couplings is known non-perturbatively
in the region of interest for the $N_{\rm f} =0$ case
\begin{equation} 
  \frac{1}{\bar g_\text{GFT}^2}-\frac{1}{\bar g_{\rm GF}^2} = f(\bar g_{\rm GF})\,,
\end{equation}
see Eq.~(\ref{eq:GFT2GF}).
Neglecting $1/z^2$ terms, we have that
\begin{equation}
  \delta \bar g _\text{GFT}^2 =
  \left( \frac{1}{\bar g_{\rm GF}^4} - f'(\bar g_{\rm GF}) \right) \bar g _\text{GFT}^4
  {0.38}{\bar g _{\rm GF}}\beta_{\rm GF}^{(0)}(\bar g _{\rm GF}) \frac{\delta z}{z} \,. 
\end{equation}
In practice $f'(\bar g_{\rm GF})\approx 0$, and the whole uncertainty $\delta \bar
g _\text{GFT}^2$ is well below the statistical uncertainty of $\bar g
_\text{GFT}^2$. The uncertainty $\delta z$ contributes less
than a 0.2\% to the final error squared in $\Lambda_{\overline{\rm MS}
}^{(3)}$. 

\clearpage

\section{Line of constant physics for $L/a=40,48$}
\label{sec:lcp4048}
For $L/a>32$ we do not have direct simulations to fix our line of
constant physics $\bar g^2_{\rm GF}(\mu_{\rm dec}, a\mu_{\rm dec}) = 3.949$. For
the case $a\mu_{\rm dec}=1/48$, we first determine the inverse lattice step scaling
function from the data available in the
literature~\cite{DallaBrida:2016kgh}, in particular for the fit used
in~\cite{DallaBrida:2016kgh} for the $\beta$-function. 
The result of such determination is
\begin{equation}
  \Sigma^{-1}(3.95,a\mu_{\rm dec}=1/24) = 2.9951(62)\,.
\end{equation}
Now we determine the value of $\beta$ for $a\mu_{\rm dec}=1/24$ such
that the coupling equals 2.9951. This can be determined by
interpolating the data of table~\ref{tab:lcp4048} using as fit ansatze
\begin{equation}
    1/\bar{g}^2_{\rm GF} - \beta/6 = P_3(\beta)\,,
\end{equation}
where $P_3(x)$ is a third degree polynomial. 
The result of this procedure is
\begin{equation}
  \beta =  5.1742(58)\qquad (L/a=48)\,.
\end{equation}
s.t. 
the coupling for $L/a=48$ would be exactly $3.95$. 
Note that this value has to be slightly corrected to the true LCP
$\bar g^2_{\rm GF}(\mu_{\rm dec}) = 3.949$ (see table~\ref{tab:lcp}).

Finally, the values of table~\ref{tab:lcp} with $L/a\le 32$, together
with this last determination can be used to interpolate the value of
$\beta$ s.t. 
$\bar g^2_{\rm GF} = 3.949$ for $L/a=40$. We choose as fit functional
\begin{equation}
  \log(L/a) = P_4(\beta)\,, 
\end{equation}
where $P_4(x)$ is a fourth degree polynomial. 
The resulting value is
\begin{equation}
  \beta =        5.0497(41)\qquad (L/a=40)\,.
\end{equation}

\begin{table}[h!]
  \centering
  \begin{tabular}{lll}
    \toprule
    $\beta$ & $\bar g^2_{\rm GF}$ & $1/g^2_{\rm GF} - \beta/6$ \\
    \midrule
    5.543070 &       2.5043(76)  &       -0.5245(12)\\
    5.242465 &       2.8963(87)  &       -0.5285(10)\\
    4.938726 &        3.403(11)  &      -0.52930(95)\\
    4.634654 &        4.180(14)  &      -0.53319(81)\\
    4.331660 &        5.380(25)  &      -0.53609(85)\\
    4.128217 &        6.785(36)  &      -0.54065(78)\\
    \bottomrule
  \end{tabular}
  \caption{Simulations with $L/a=24$ used to interpolate the vaulue
    $\bar g^2_{\rm GF} = 2.9951$.}
  \label{tab:lcp4048}
\end{table}

\clearpage

\end{appendices}
\addcontentsline{toc}{section}{References}
\bibliographystyle{utphys}
\bibliography{paper}
\end{document}